\pdfoutput=1
\documentclass[pdftex,twocolumn,epjc3]{svjour3}

\RequirePackage[utf8]{inputenc}
\smartqed

\RequirePackage{graphicx}
\RequirePackage{mathptmx}      
\RequirePackage{flushend}
\RequirePackage[numbers,sort&compress]{natbib}
\RequirePackage[colorlinks,citecolor=blue,urlcolor=blue,linkcolor=blue]{hyperref}

\usepackage{amsmath}
\usepackage{amssymb}
\usepackage{subfig}
\usepackage{graphicx}
\usepackage{paralist}
\usepackage{multirow}
\usepackage{amsbsy}
\usepackage{amsfonts}
\usepackage{cuted} 
\usepackage{ulem}
\usepackage{xspace}

\usepackage{widetext}

\usepackage[utf8]{inputenc} 
\PassOptionsToPackage{numbers,sort&compress}{natbib} %
\usepackage{lmodern}%

\journalname{Eur. Phys. J. C}

\newcommand{\ncteqfit}{{\tt nCTEQ15}\xspace}

\newcommand{\gsim}{\gtrsim}

\newcommand{\ord}{{\ensuremath{\cal O}}}
\newcommand{\ppb}{pP\lowercase{b}\xspace}
\newcommand{\pbpb}{P\lowercase{b}P\lowercase{b}\xspace}

\begin{document}
%
%
\def\tabDefone{
%
\begin{table*}[t]
\centering
\renewcommand{\arraystretch}{1.8}		
{
\begin{tabular}{c|c|c|c|c}
& \rule{4ex}{0pt} & Observable & Cuts (GeV) & Figure\\ \hline\hline
\multirow{6}{*}[-1.0em]{\rotatebox{90}{\large{pP\lowercase{b}}}}	&\multirow{3}{*}{\rotatebox{90}{ATLAS}} &$d\sigma(Z\rightarrow \ell^{+}\ell^{-})/dy_Z$~\cite{Aad:2015gta}& $|y_{Z}^{\mathrm{CM}}|<3.5$; $60< m_{\ell^+\ell^-}<120$&Fig.~\ref{fig:atlas_z_pPb_comp}\\\cline{3-5}
&& $d\sigma(W^{+}\rightarrow \ell^{+}\nu)/dy_{\ell^+}$\cite{AtlasWpPb} &$p_T^{\ell^\pm}>25$; $m_T^{\ell^{\pm}}>40$; $|\eta_{lab}^{\ell^{\pm}}|<2.4$ &Fig.~\ref{fig:atlas_wpm_pPb_comp_wp}\\\cline{3-5}
&&  $d\sigma(W^{-}\rightarrow \ell^{-}\bar{\nu})/dy_{\ell^-}$\cite{AtlasWpPb} &$p_T^{\ell^\pm}>25$; $m_T^{\ell^{\pm}}>40$; $|\eta_{lab}^{\ell^{\pm}}|<2.4$ &Fig.~\ref{fig:atlas_wpm_pPb_comp_wm}\\\cline{2-5}
 & \multirow{3}{*}{\rotatebox{90}{CMS}} &$d\sigma(Z\rightarrow \ell^{+}\ell^{-})/dy_Z$\cite{Khachatryan:2015pzs}& $|\eta_{lab}^{\ell^{\pm}}|<2.4$; $60< m_{\ell^+\ell^-}<120$; $p_T^{\ell^{+}(\ell^-)}>20$&Fig.~\ref{fig:cms_z_pPb_comp}\\\cline{3-5}
&& $d\sigma(W^{+}\rightarrow \ell^{+}\nu)/dy_{\ell^+}$\cite{Khachatryan:2015hha}& $p_T^{\ell^{\pm}}>25$; $|\eta_{lab}^{\pm}|<2.4$& Fig.~\ref{fig:cms_wpm_pPb_comp_wp}\\\cline{3-5}
&& $d\sigma(W^{-}\rightarrow \ell^{-}\bar{\nu})/dy_{\ell^-}$\cite{Khachatryan:2015hha}& $p_T^{\ell^{\pm}}>25$; $|\eta_{lab}^{\pm}|<2.4$&Fig.~\ref{fig:cms_wpm_pPb_comp_wm}\\\cline{2-5}
& \rotatebox{90}{LHCb\,}  & $\sigma(Z\rightarrow \ell^{+}\ell^{-})$~\cite{Aaij:2014pvu}&$60<m_{\ell^{+}\ell^{-}}<120$; $p_T^{\ell^+(\ell^-)}>20$; $2.0<\eta^{\ell^{\pm}}<4.5$; $-4.5<\eta_{\ell^{\pm}}<-2.0$ & Fig.~\ref{fig:lhcb_z_pPb_comp}\\\cline{2-5}
& \multirow{2}{*}{\rotatebox{90}{ALICE\ }}  &$\sigma(W^{+}\rightarrow \ell^{+}\nu)$~\cite{Senosi:2015omk}& $p_{T}^{\ell^{\pm}}>10$; $2.03<\eta_{lab}^{\ell^{\pm}}<3.53$; $-4.46<\eta_{lab}^{\ell^{\pm}}<-2.96$&Fig.~\ref{fig:alice_wpm_pPb_comp_wp}\\\cline{3-5}
& & $\sigma(W^{-}\rightarrow \ell^{-}\bar{\nu})$~\cite{Senosi:2015omk}& $p_{T}^{\ell^{\pm}}>10$; $2.03<\eta_{lab}^{\ell^{\pm}}<3.53$; $-4.46<\eta_{lab}^{\ell^{\pm}}<-2.96$&Fig.~\ref{fig:alice_wpm_pPb_comp_wm}\\\hline\hline
\multirow{4}{*}{\rotatebox{90}{\large{P\lowercase{b}P\lowercase{b}}}} &\multirow{2}{*}{\rotatebox{90}{ATLAS\ }} & $1/\sigma_{tot}d\sigma/dy_Z$\cite{Aad:2012ew}&$66 < m_{\ell^{+}\ell^{-}}<116$; $|y_Z|<2.5$ & Fig.~\ref{fig:cmsatlasZPbPbcomparison_atlas}\\\cline{3-5}
& &$A_\ell$~\cite{Aad:2014bha}&$p_T^{\ell} <25$; $|\eta^\ell_{lab}|<2.5$; $m_T>40$; $p_T^{miss}<25$ &Fig.~\ref{fig:cmsatlasWPbPbcomparison_atlas}\\\cline{2-5}
&\multirow{2}{*}{\rotatebox{90}{CMS}} &$1/\sigma_{tot}d\sigma/dy_Z$\cite{Chatrchyan:2014csa}&$60 < m_{\ell^{+}\ell^{-}}<120$; $|y_Z|<2.0$ & Fig.~\ref{fig:cmsatlasZPbPbcomparison_cms}\\\cline{3-5}
& &$A_\ell$~\cite{Chatrchyan:2012nt}&$p_T^{\ell} <25$; $|\eta^\ell_{lab}|<2.1$; $m_T>40$ &Fig.~\ref{fig:cmsatlasWPbPbcomparison_cms}\\\cline{2-5}
\end{tabular}
}
	\caption{LHC data sets considered in this analysis.}
	\label{tab:LHCdatasets}
\end{table*}
} 
\def\tabDeftwo{
\begin{table}[t] 
\renewcommand{\arraystretch}{1.5}
\centering{}
\begin{tabular}{|c||c|c|c|c|}
\hline 
Beam Energy [TeV] & 3.5  & 4  & 6.5 & 7 \\
\hline 
\hline 
$\sqrt{s_{pp}}$ & 7.00 & 8.00 & 13.00 & 14.00\\
\hline 
\hline
$\sqrt{s_{PbPb}}$ & 2.76 & 3.15 & 5.12 & 5.52\\
\hline 
\hline 
$\sqrt{s_{pPb}}$ & 4.40 & 5.02 & 8.16 & 8.79 \\
\hline 
\end{tabular}
\caption{The CM energy per nucleon for pp, pPb and PbPb collisions
{\it vs.} the proton beam energy in TeV units.
}
\label{tab:energy}
\end{table} 
} 
\def\figDefkin{
\begin{figure}[tb]
\centering{}
\includegraphics[width=0.48\textwidth]{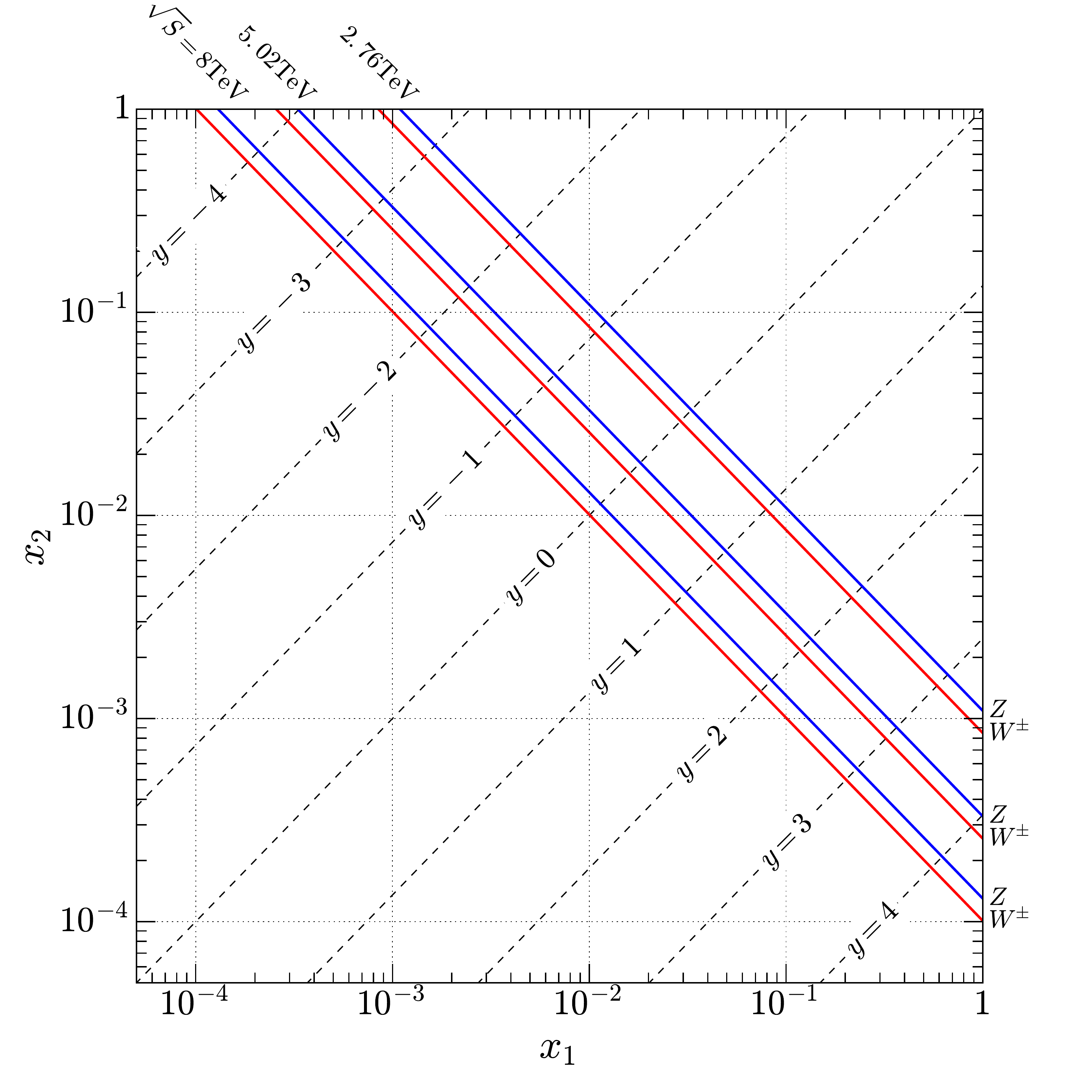}
\caption{The kinematic $(x_1,x_2)$ space explored by the measurements in this study.
We display lines of constant $\tau=M_{V}/\sqrt{s}$ where $M_{V}$ is the invariant
mass of the produced $W^\pm /Z$ vector boson, as well as the center of mass (CM) rapidity $y$.
In case of pPb collisions, we use the standard convention where 
 $x_1$ corresponds to the proton and $x_2$ to the Pb momentum fraction.
\label{fig:xspace}}
\end{figure}
} 
\def\figDefdatarange{
%
\begin{figure}[t]
\centering{}
\includegraphics[width=0.48\textwidth]{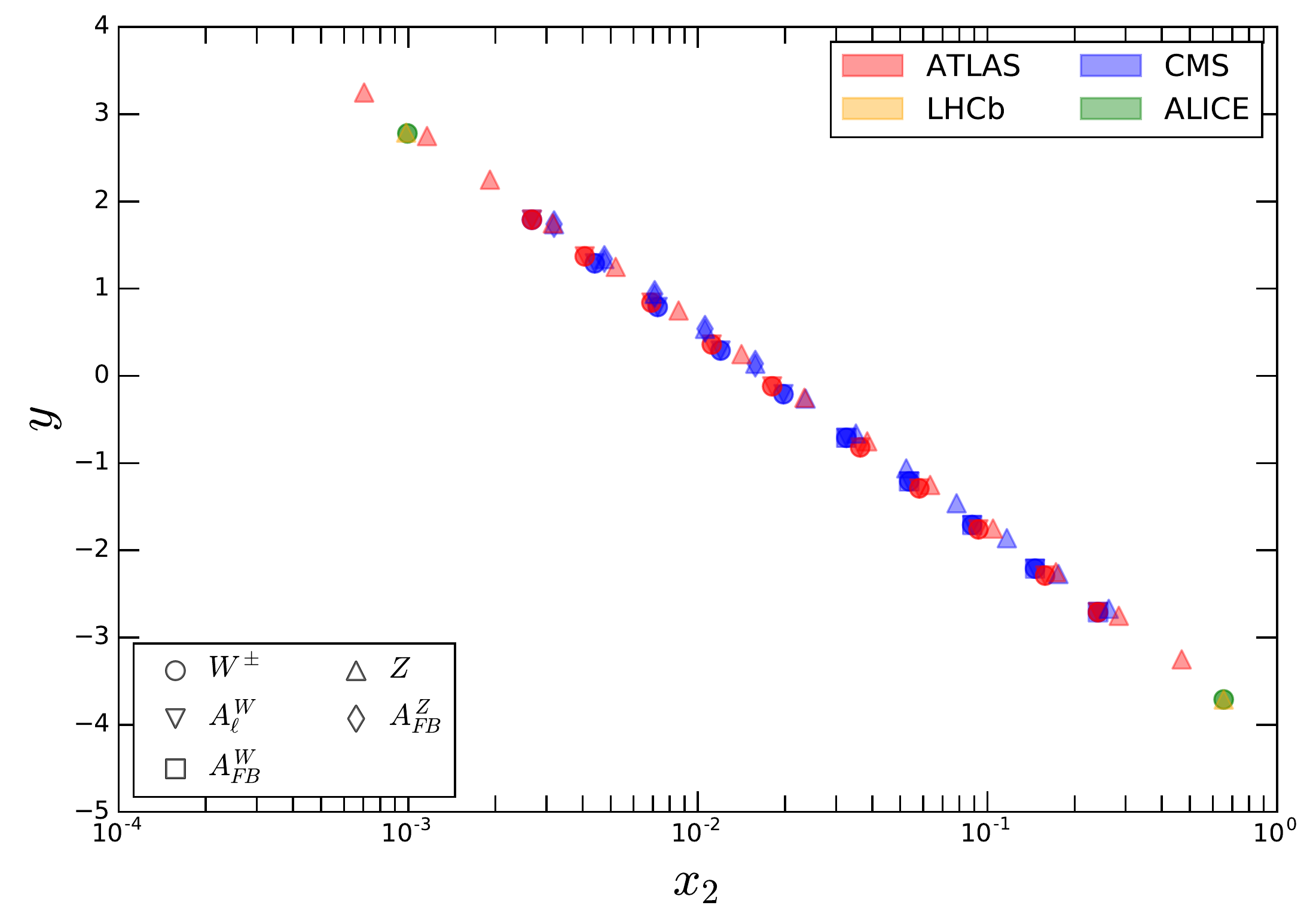}
\caption{Range of the \ppb data used for reweighting.
$y$ is rapidity in the CM frame and $x_2$ is momentum of the parton from the lead beam.
}

\label{fig:data_range}
\end{figure}
} 
\def\figDefatlaspPb{
%
\begin{figure}[!htb]
\centering{}
\includegraphics[width=0.48\textwidth]{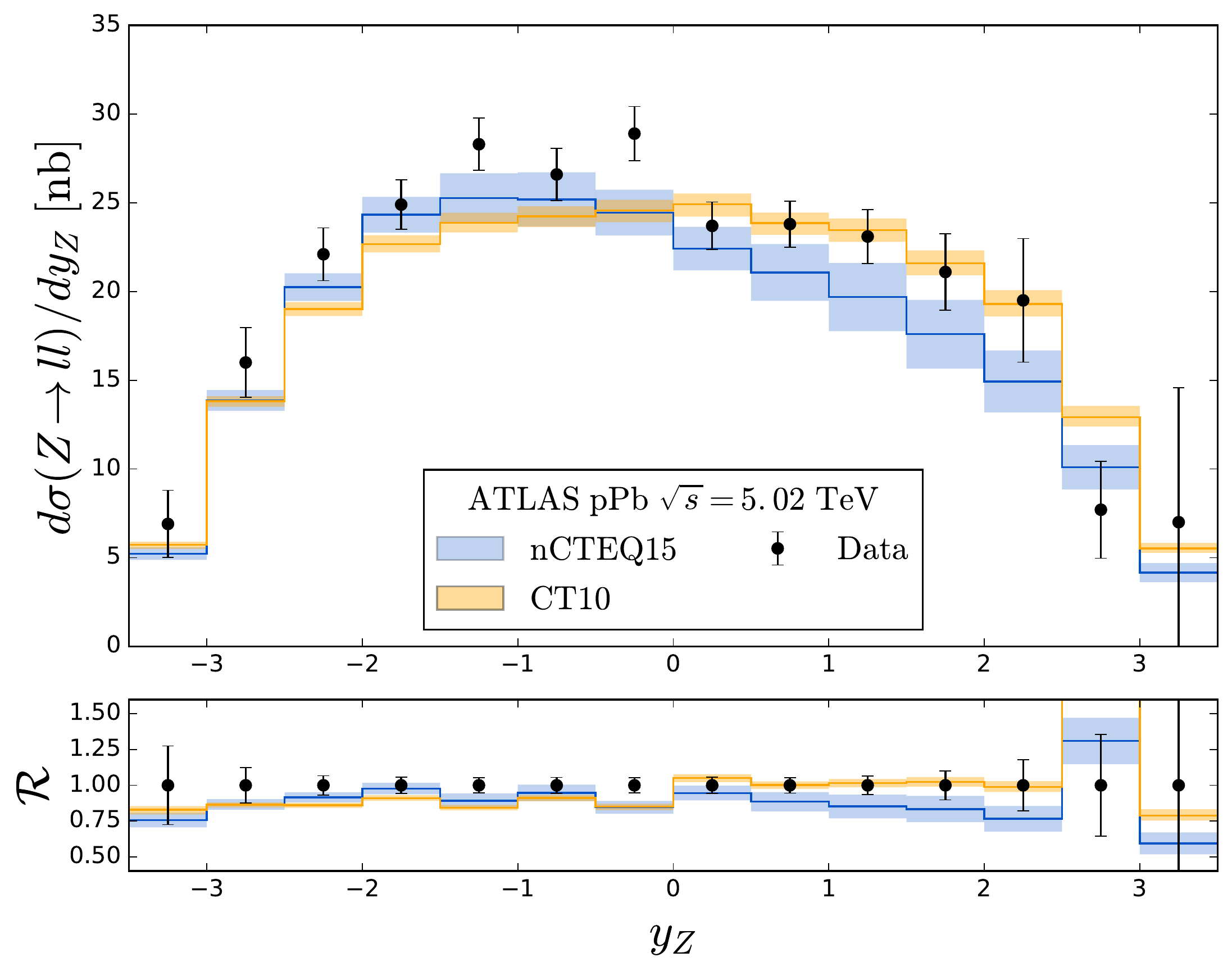}
\caption{ATLAS $Z$ production in \ppb collisions.
}
\label{fig:atlas_z_pPb_comp}
\end{figure}
} 
\def\figDefcmspPb{
%
\begin{figure}[!htb]
\centering{}
\includegraphics[width=0.48\textwidth]{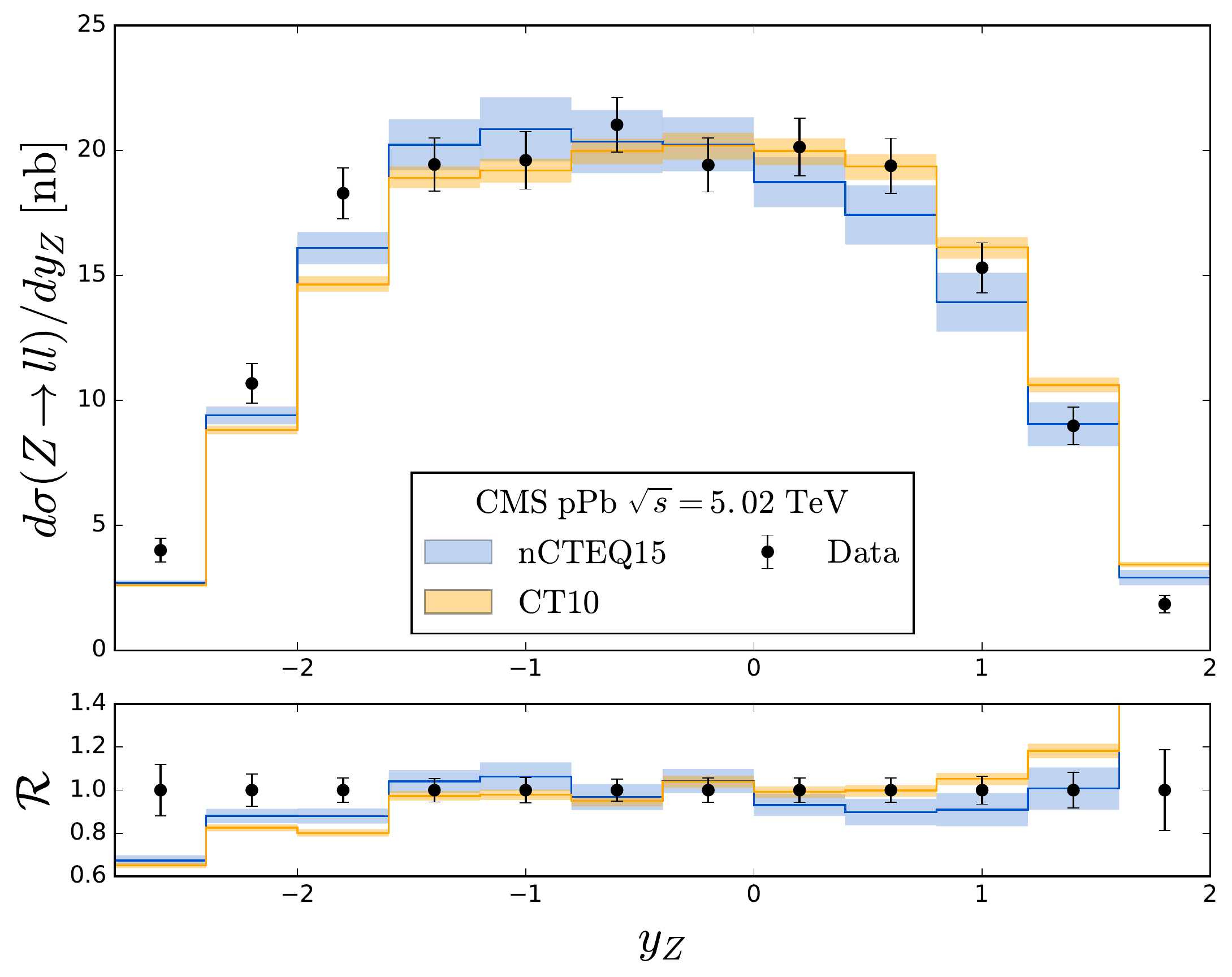}
\caption{CMS $Z$ production in \ppb collisions.
}
\label{fig:cms_z_pPb_comp}
\end{figure}
} 
\def\figDeflhcbpPb{
%
\begin{figure}[!htb]
\centering{}
\includegraphics[width=0.48\textwidth]{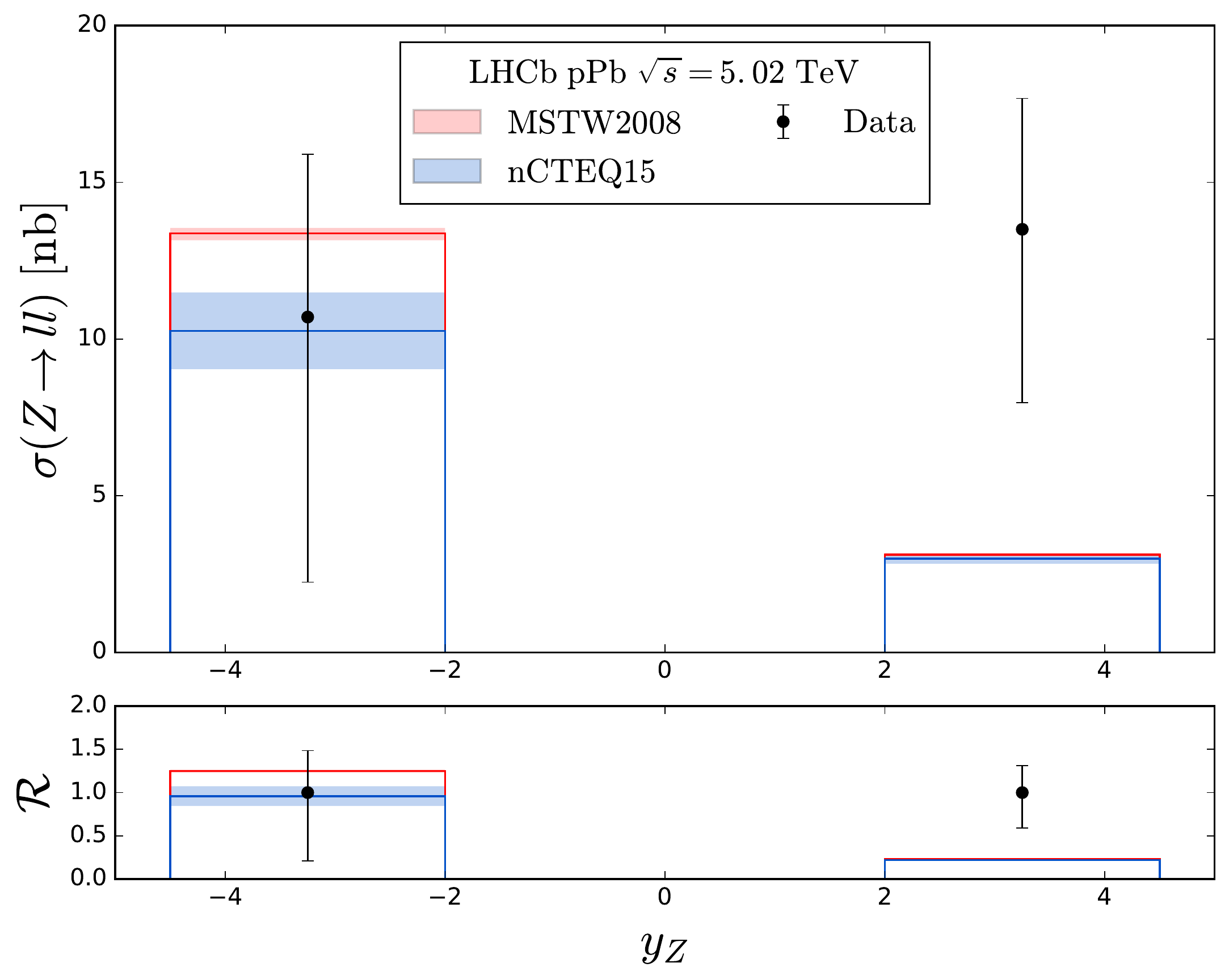}
\caption{LHCb $Z$ production in \ppb collisions.}
\label{fig:lhcb_z_pPb_comp}
\end{figure}
} 
\def\figDefcmsw{
%
\begin{figure*}[!htb]
\centering{}
\subfloat[$W^{+}$\label{fig:cms_wpm_pPb_comp_wp}]{
\includegraphics[width=0.48\textwidth]{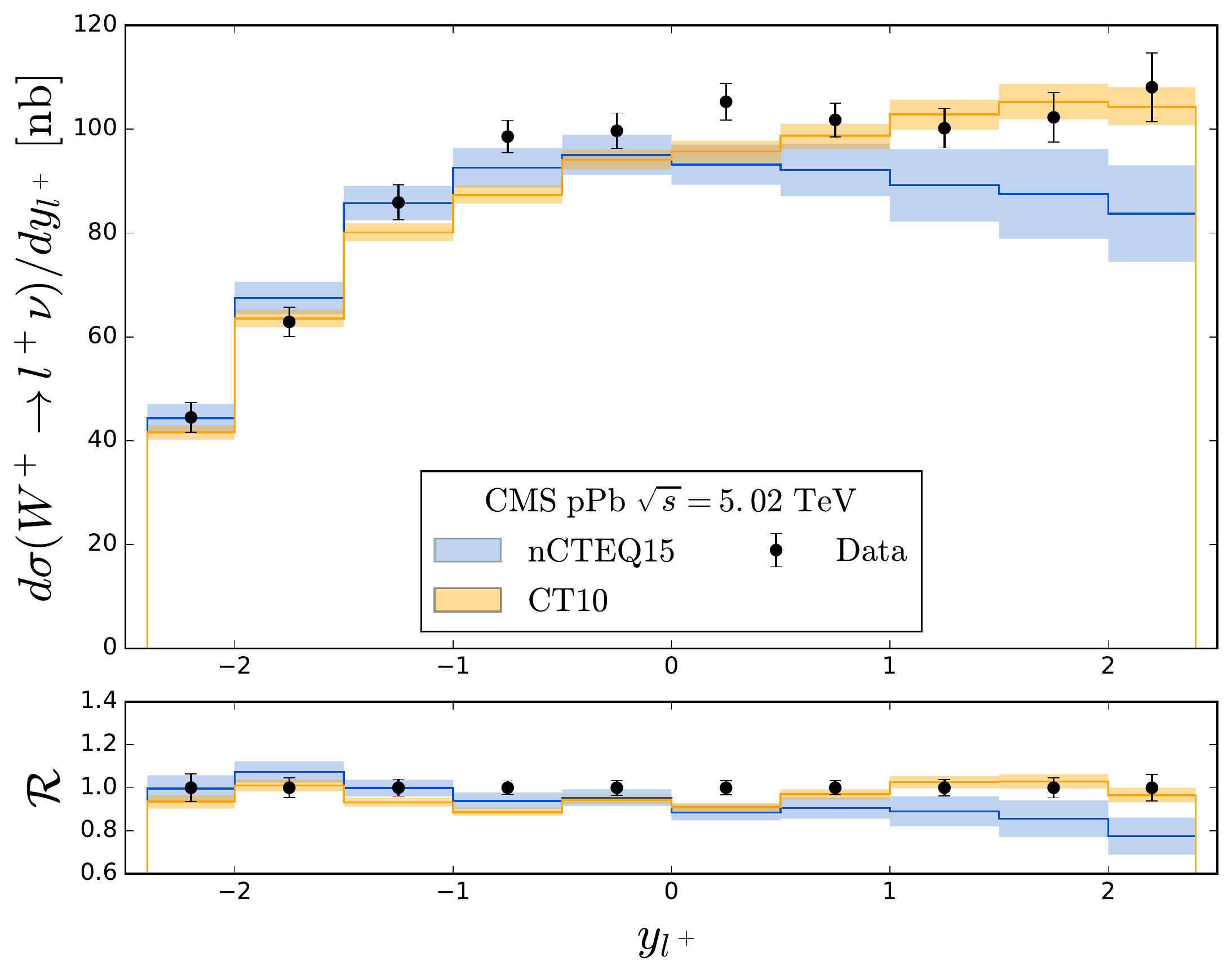}}
\hfil
\subfloat[$W^{-}$\label{fig:cms_wpm_pPb_comp_wm}]{
\includegraphics[width=0.48\textwidth]{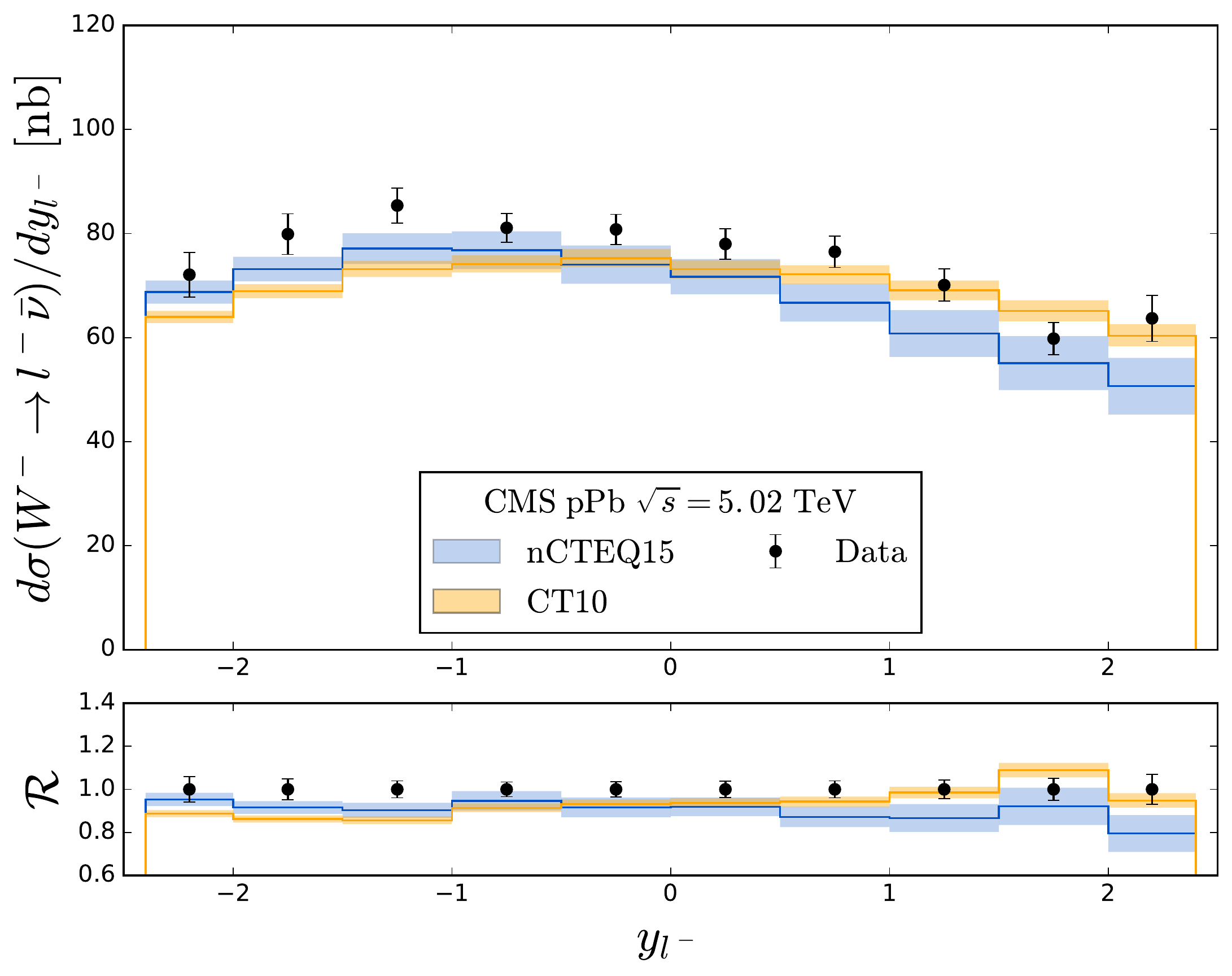}
}
\caption{CMS $W^{\pm}$ production in pP\lowercase{b} collisions at the LHC.}
\label{fig:cms_wpm_pPb_comp}
\end{figure*}
} 
\def\figDefatlasw{
%
\begin{figure*}[!htb]
\centering{}
\subfloat[$W^{+}$\label{fig:atlas_wpm_pPb_comp_wp}]{
\includegraphics[width=0.48\textwidth]{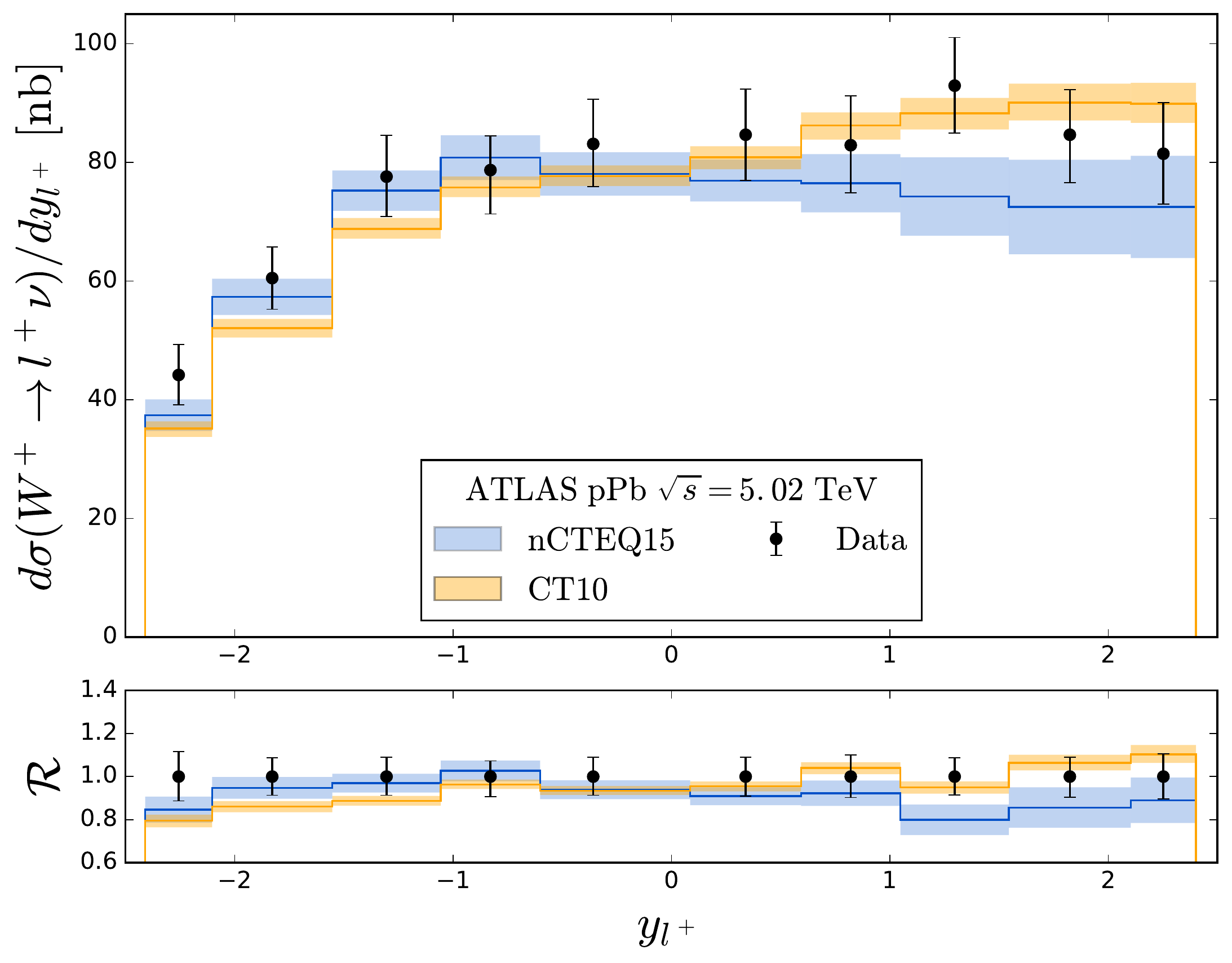}}
\hfil
\subfloat[$W^-$\label{fig:atlas_wpm_pPb_comp_wm}]{
\includegraphics[width=0.48\textwidth]{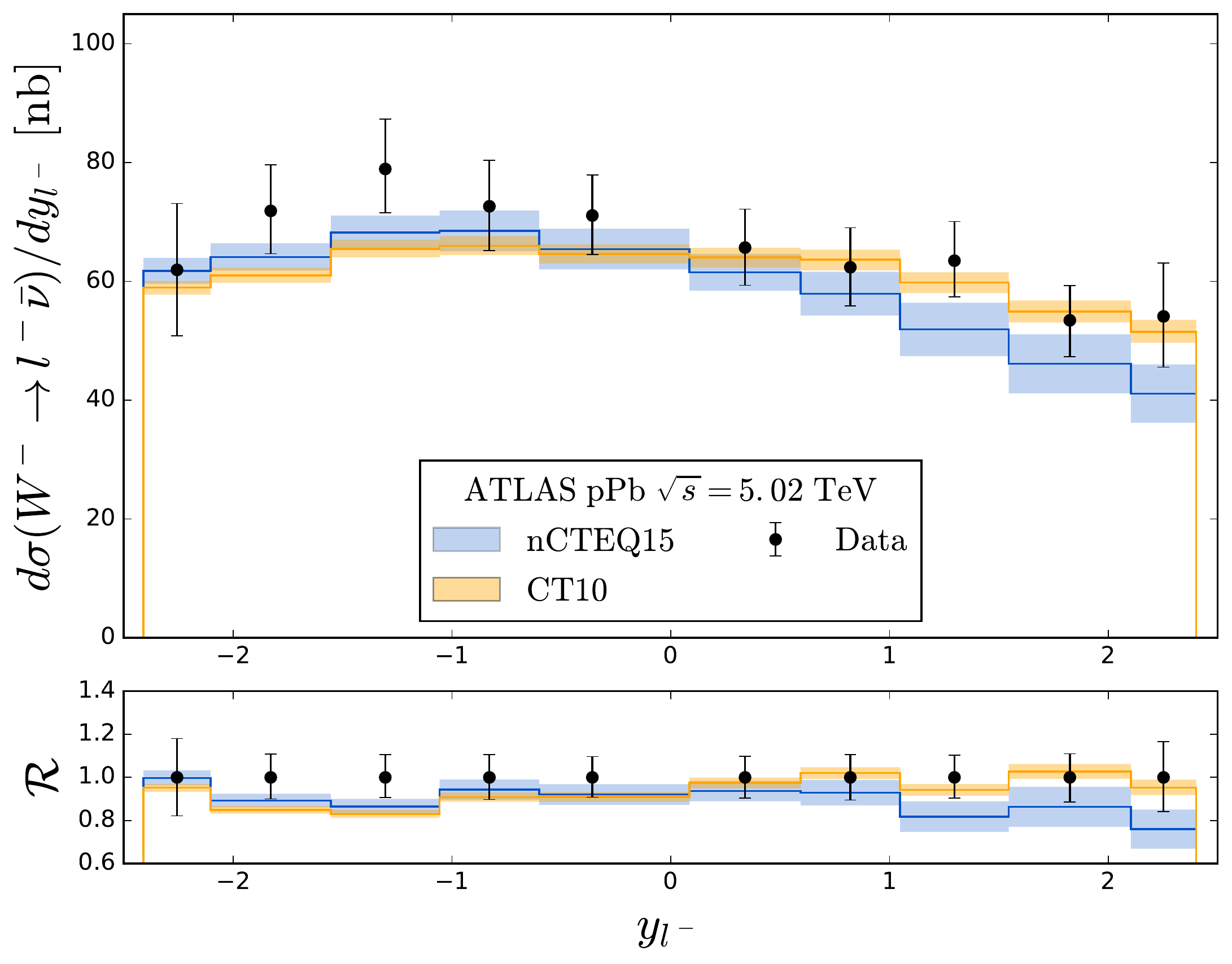}}
\caption{ATLAS $W^{\pm}$ production in pP\lowercase{b} collisions at the LHC.}
\label{fig:atlas_wpm_pPb_comp}
\end{figure*}
} 
\def\figDefalicew{
%
\begin{figure*}[!htb]
\centering{}
\subfloat[$W^{+}$\label{fig:alice_wpm_pPb_comp_wp}]{
\includegraphics[width=0.48\textwidth]{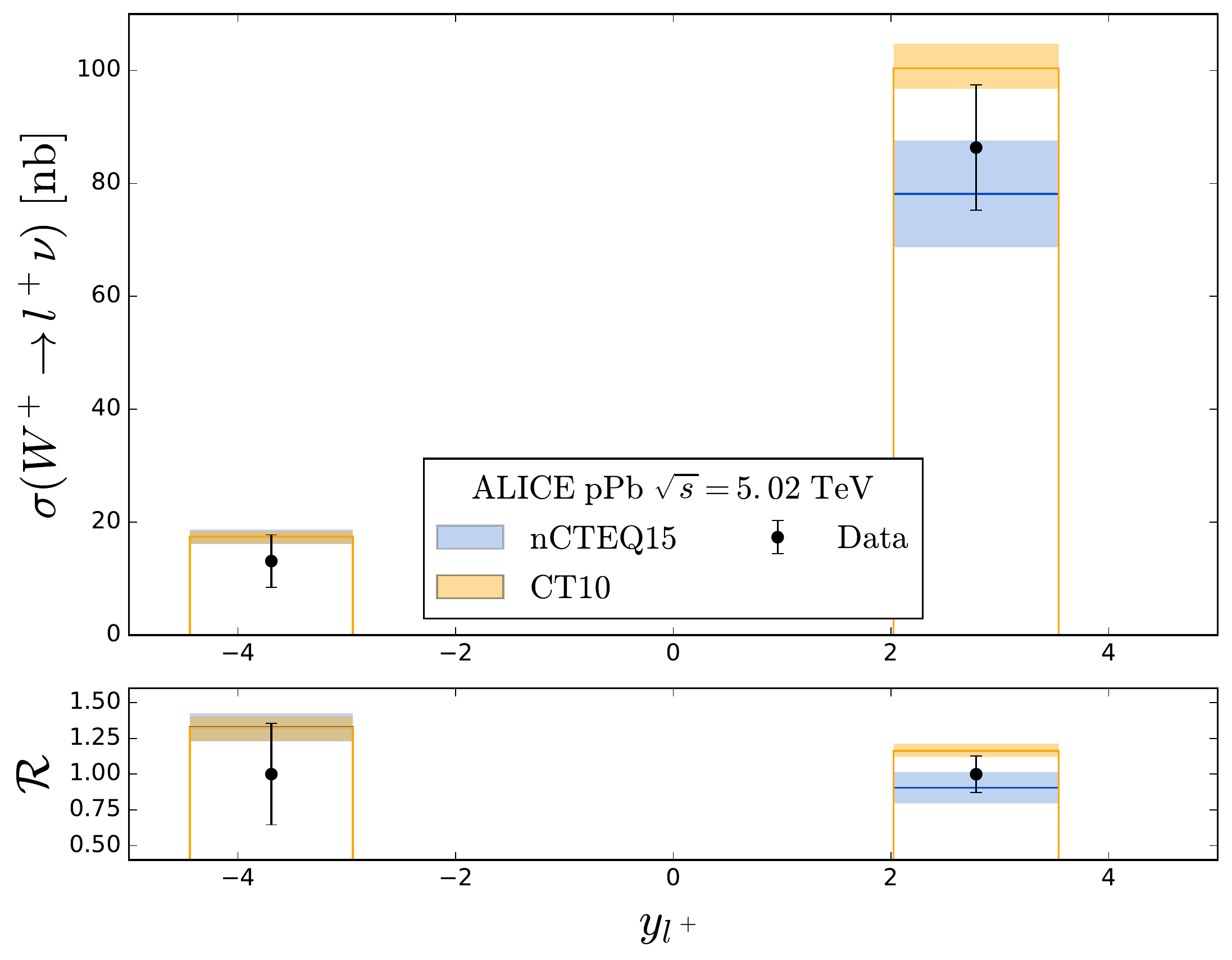}}
\hfil
\subfloat[$W^{-}$\label{fig:alice_wpm_pPb_comp_wm}]{
\includegraphics[width=0.48\textwidth]{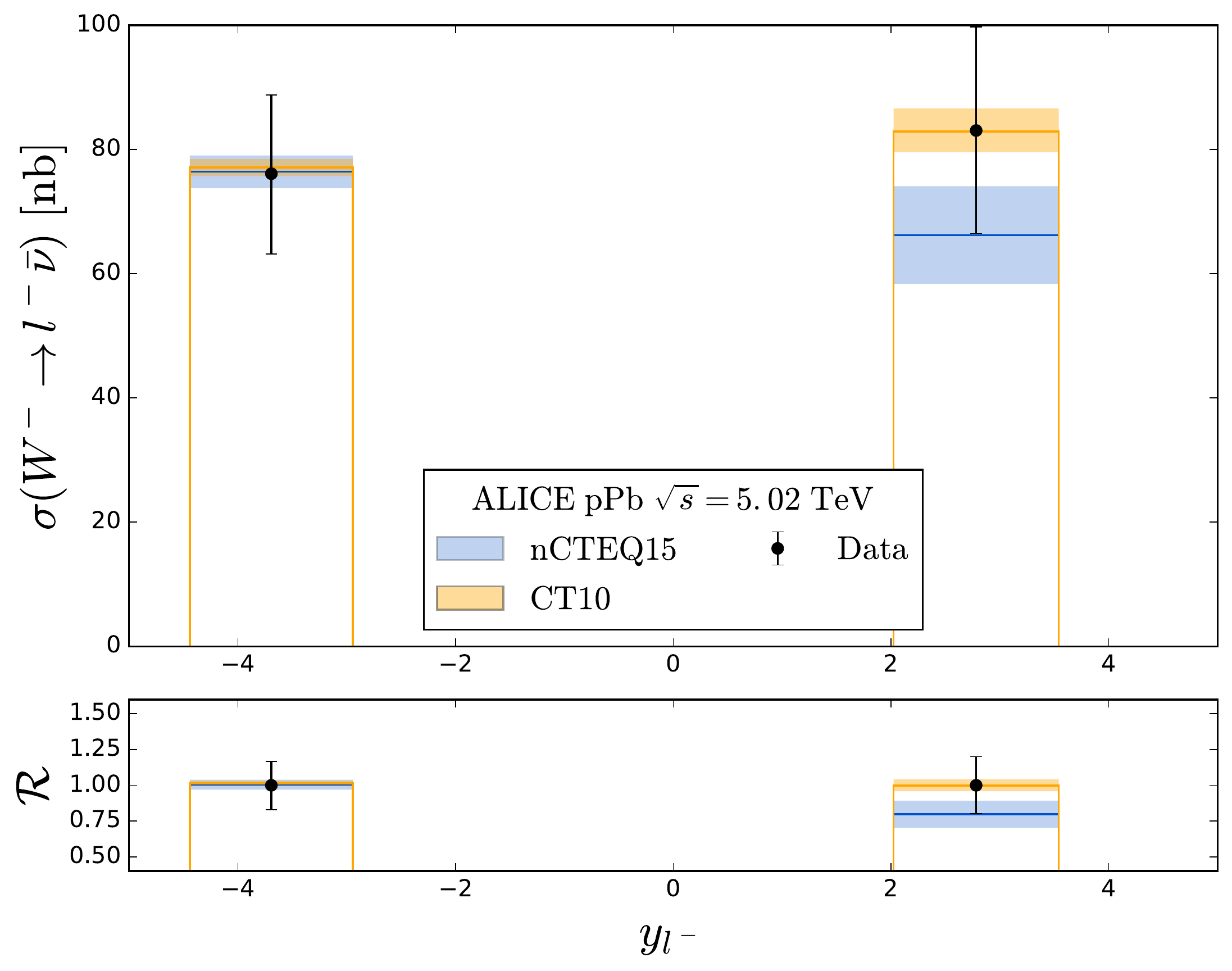}}
\caption{ALICE $W^{\pm}$ production in pP\lowercase{b} collisions at the LHC.}
\label{fig:alice_wpm_pPb_comp}
\end{figure*}
} 
\def\figDefpbpb{
\begin{figure*}[!htb]
\begin{center}
\subfloat[ATLAS\label{fig:cmsatlasZPbPbcomparison_atlas}]{
\includegraphics[width=0.48\textwidth]{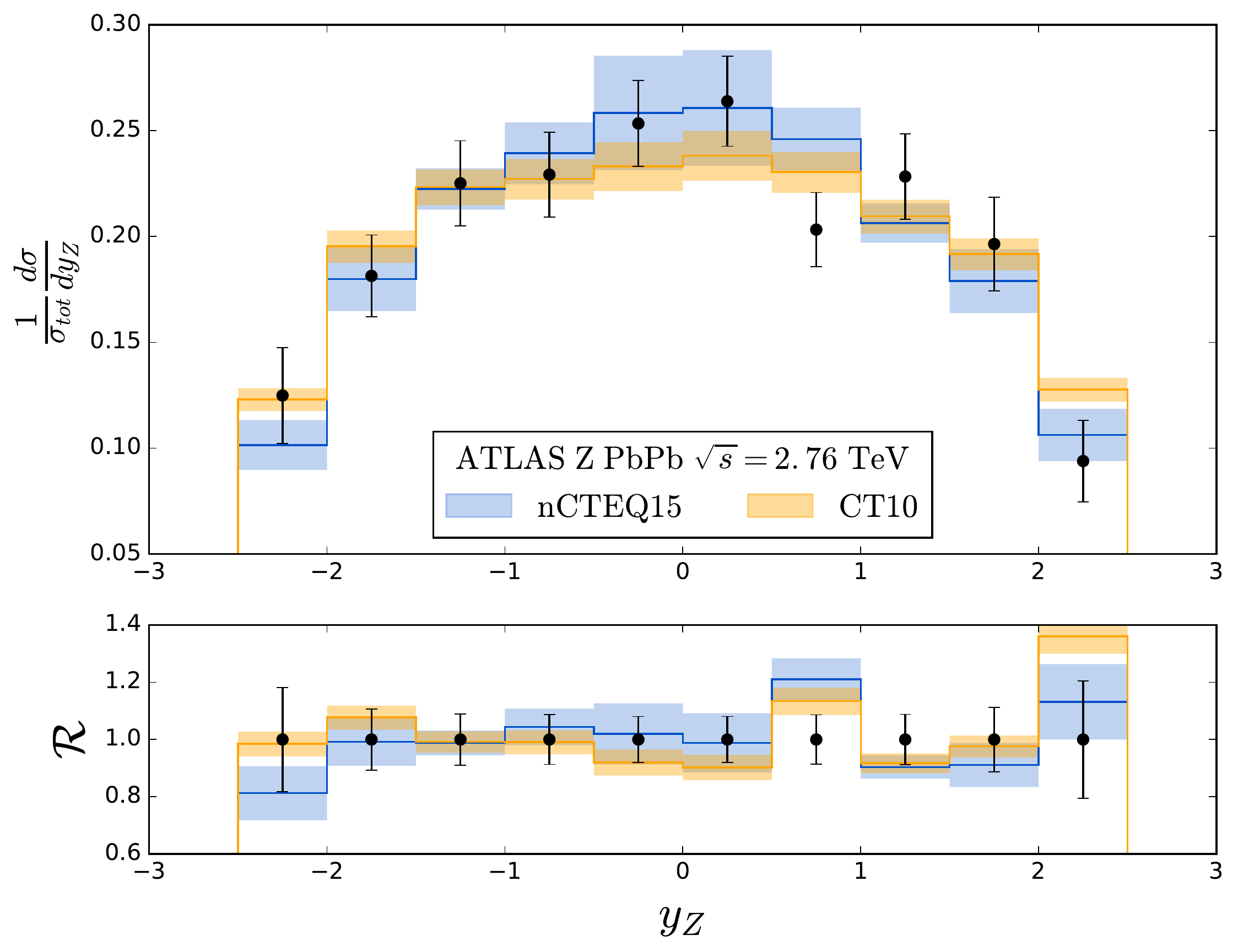}}
\hfil
\subfloat[CMS\label{fig:cmsatlasZPbPbcomparison_cms}]{
\includegraphics[width=0.48\textwidth]{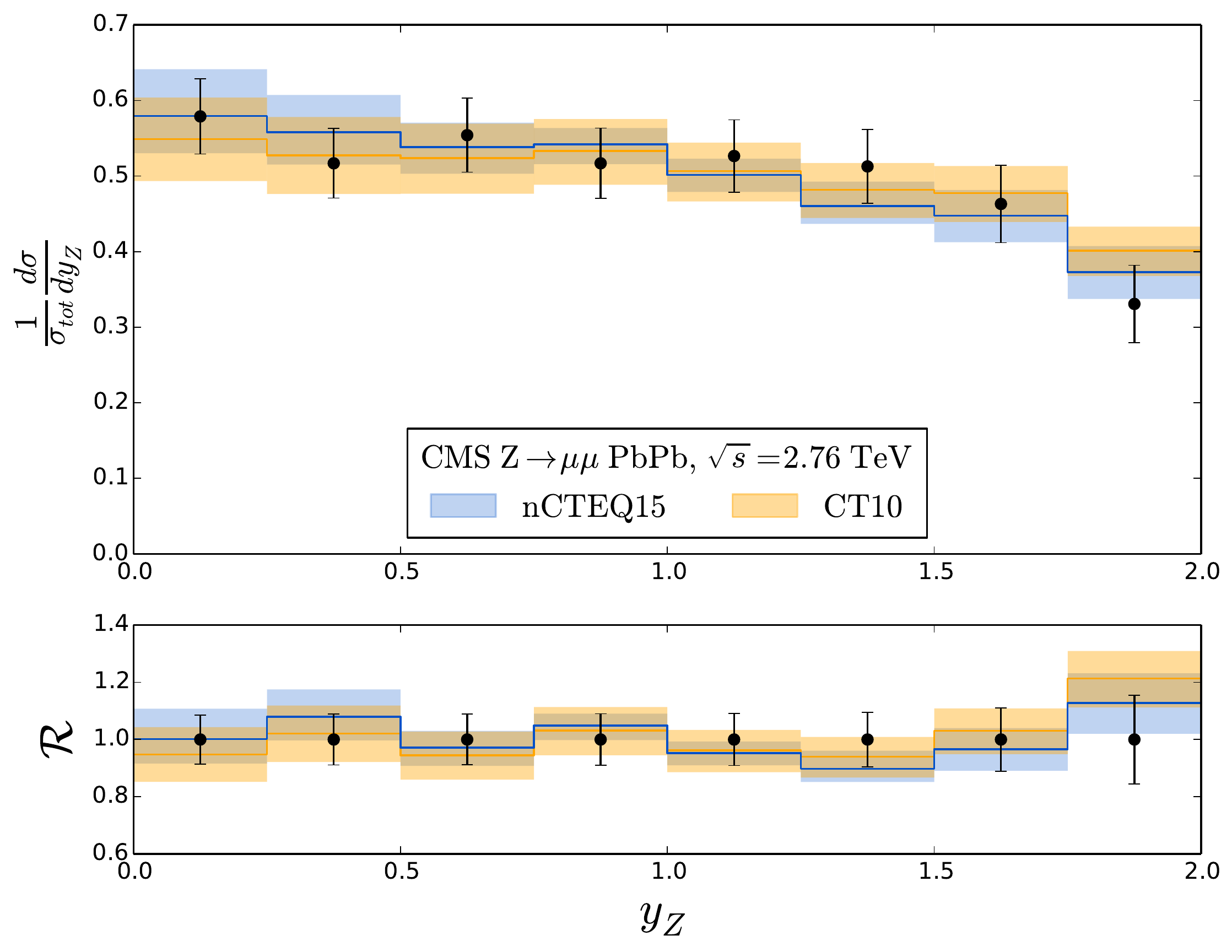}}
\caption{Z boson production cross section normalized to the total cross section
for P\lowercase{b}P\lowercase{b} collisions at the LHC with $\sqrt{s}=2.76$ TeV as measured
by the ATLAS and CMS collaborations. Corresponding
predictions obtained with \ncteqfit and CT10 PDFs are also shown.}
\label{fig:cmsatlasZPbPbcomparison}
\end{center}
\end{figure*}
} 
\def\figDefpbpbw{
\begin{figure*}[!htb]
\begin{center}
\subfloat[ATLAS\label{fig:cmsatlasWPbPbcomparison_atlas}]{
\includegraphics[width=0.48\textwidth]{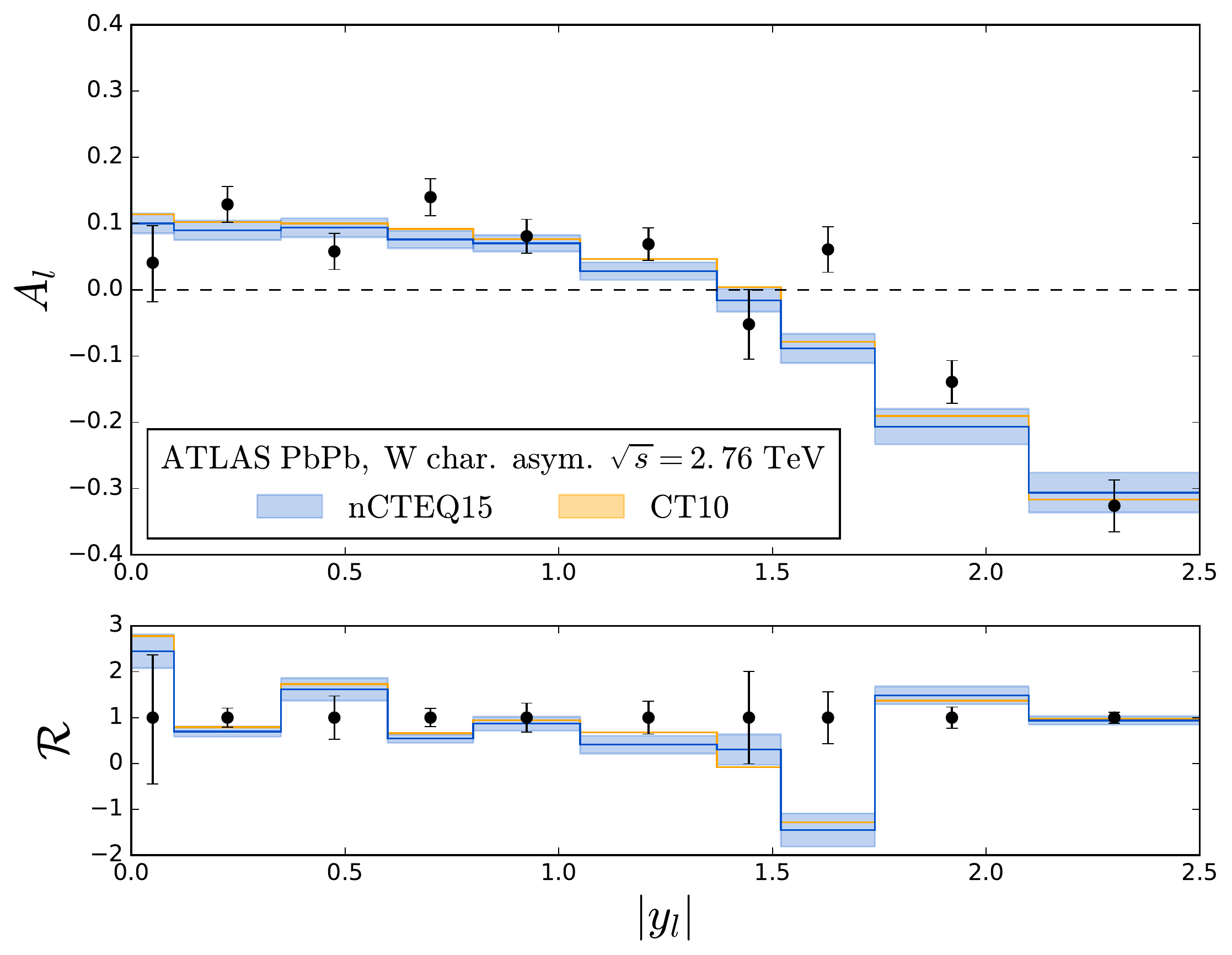}}
\hfil
\subfloat[CMS\label{fig:cmsatlasWPbPbcomparison_cms}]{
\includegraphics[width=0.48\textwidth]{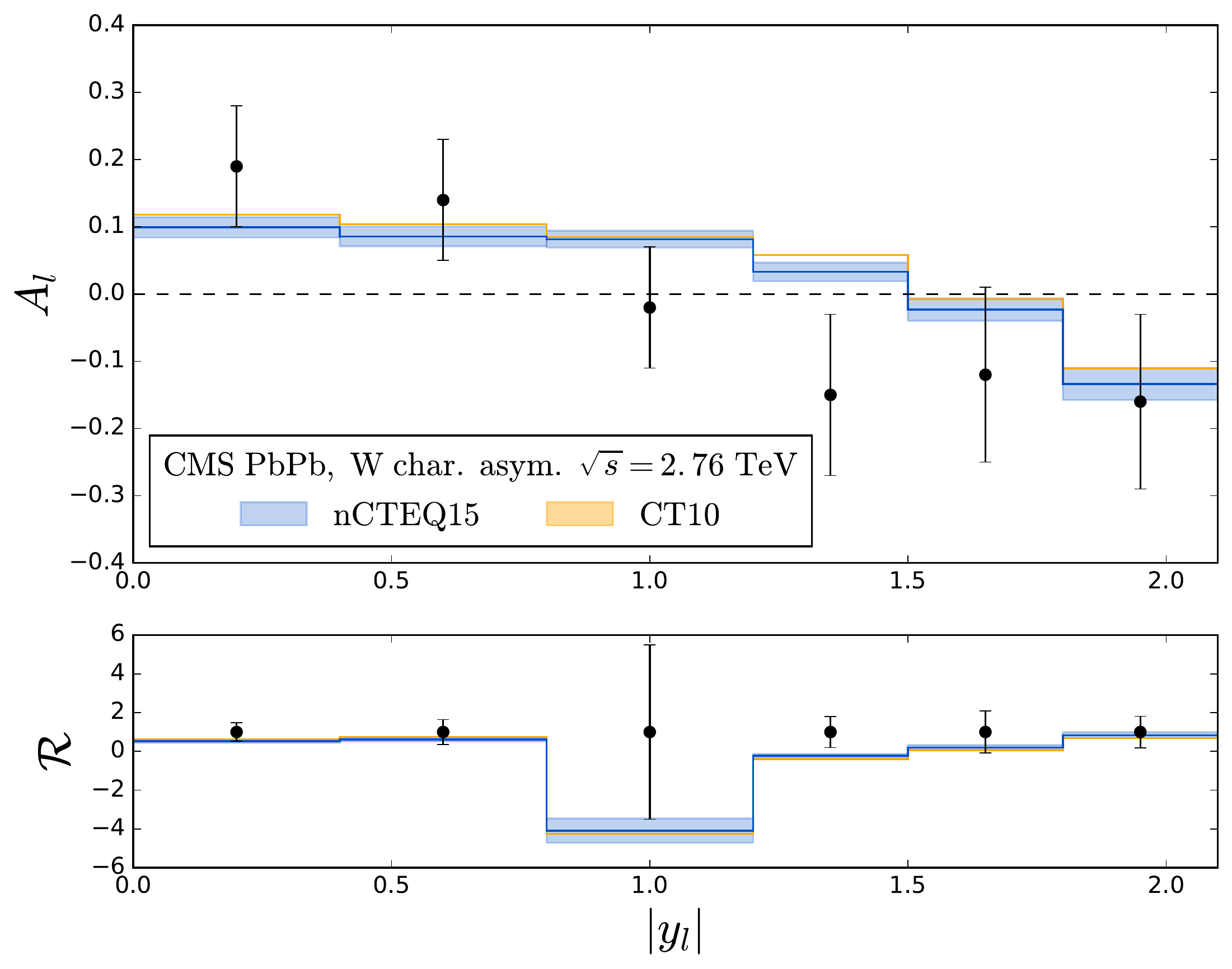}}
\caption{$W$ charge asymmetry for \pbpb collisions at the LHC with $\sqrt{s}=2.76$ TeV as measured
by the ATLAS and CMS collaborations. Corresponding
predictions obtained with \ncteqfit and CT10 PDFs are also shown.}
\label{fig:cmsatlasWPbPbcomparison}
\end{center}
\end{figure*}
%
} 
\def\figDefxsecww{
%
\begin{figure*}[p]
\centering{}
\includegraphics[width=0.97\textwidth]{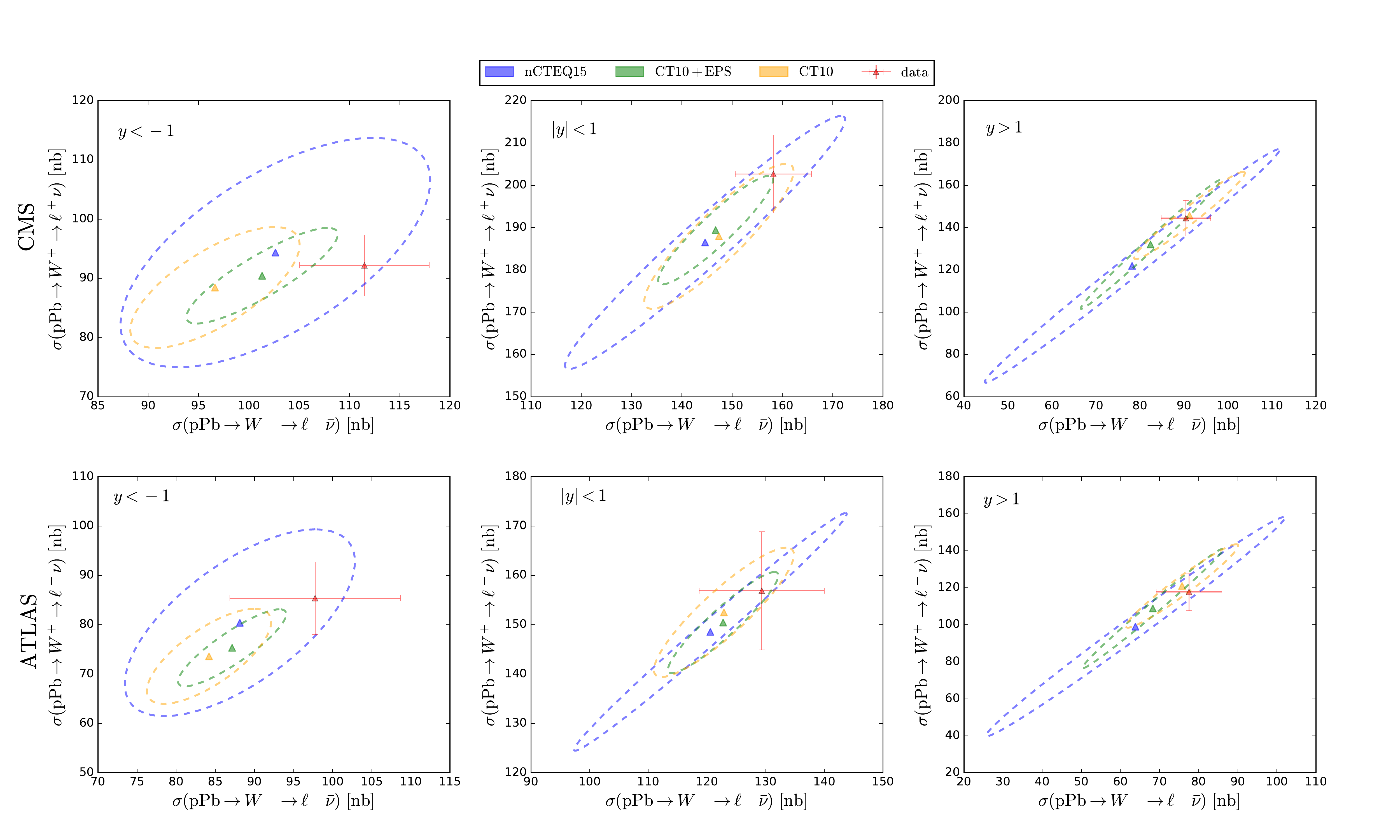}
\caption{Correlations between $W^+$ and $W^-$ cross sections calculated with
different PDFs overlaid with the corresponding LHC data from CMS and ATLAS.}
\label{fig:wpwmCrossSecCorrelations}
\end{figure*}
%
} 
\def\figDefxseczw{
\begin{figure*}[p]
\centering{}
\includegraphics[width=0.97\textwidth]{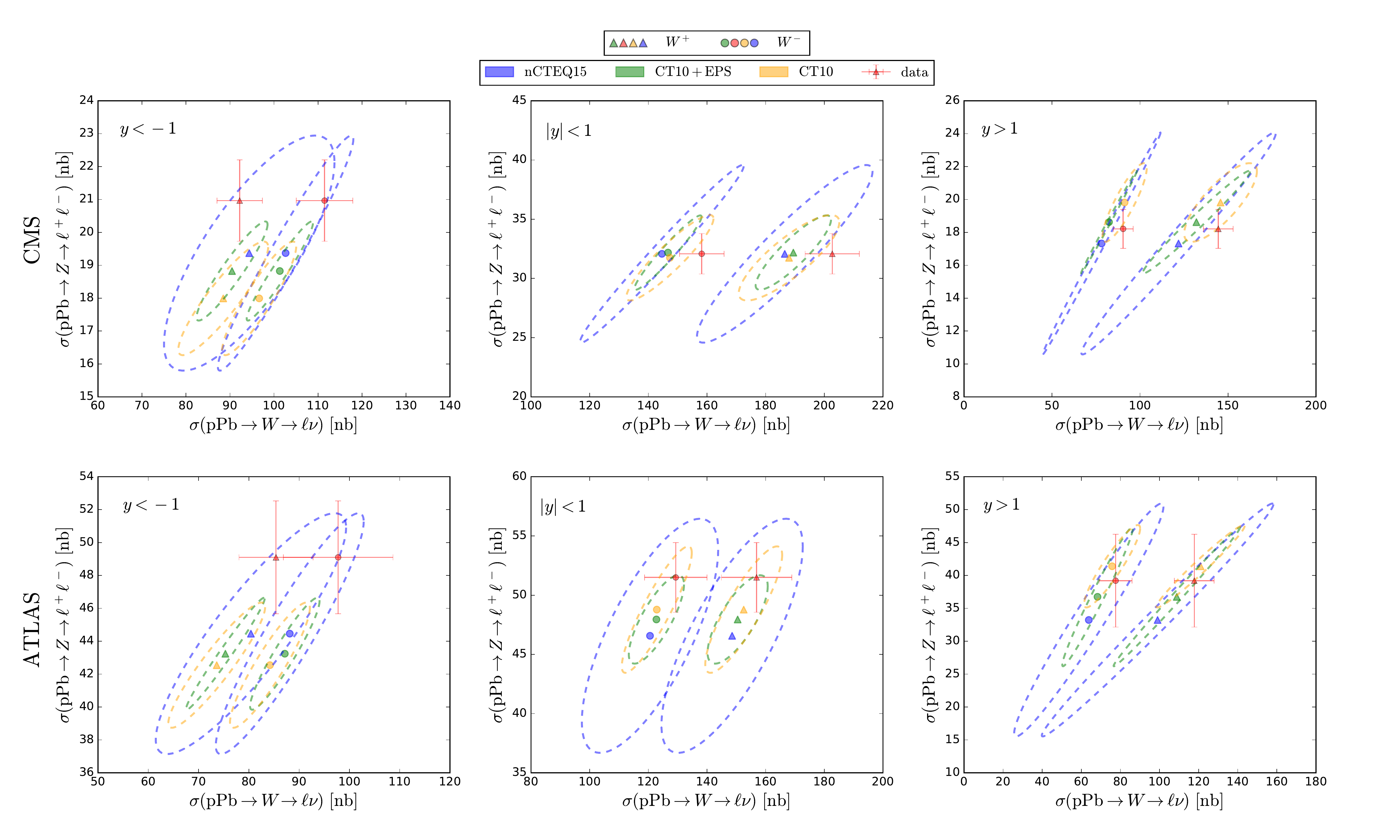}
\caption{Correlations between $Z$ and $W^+/W^-$ cross sections calculated with
different PDFs overlaid with the corresponding LHC data from CMS and ATLAS.}
\label{fig:wzCrossSecCorrelations}
\end{figure*}
%
} 
\def\figDefxsecwwF{
%
\begin{figure}[!htb]
\centering{}
\includegraphics[width=0.5\textwidth]{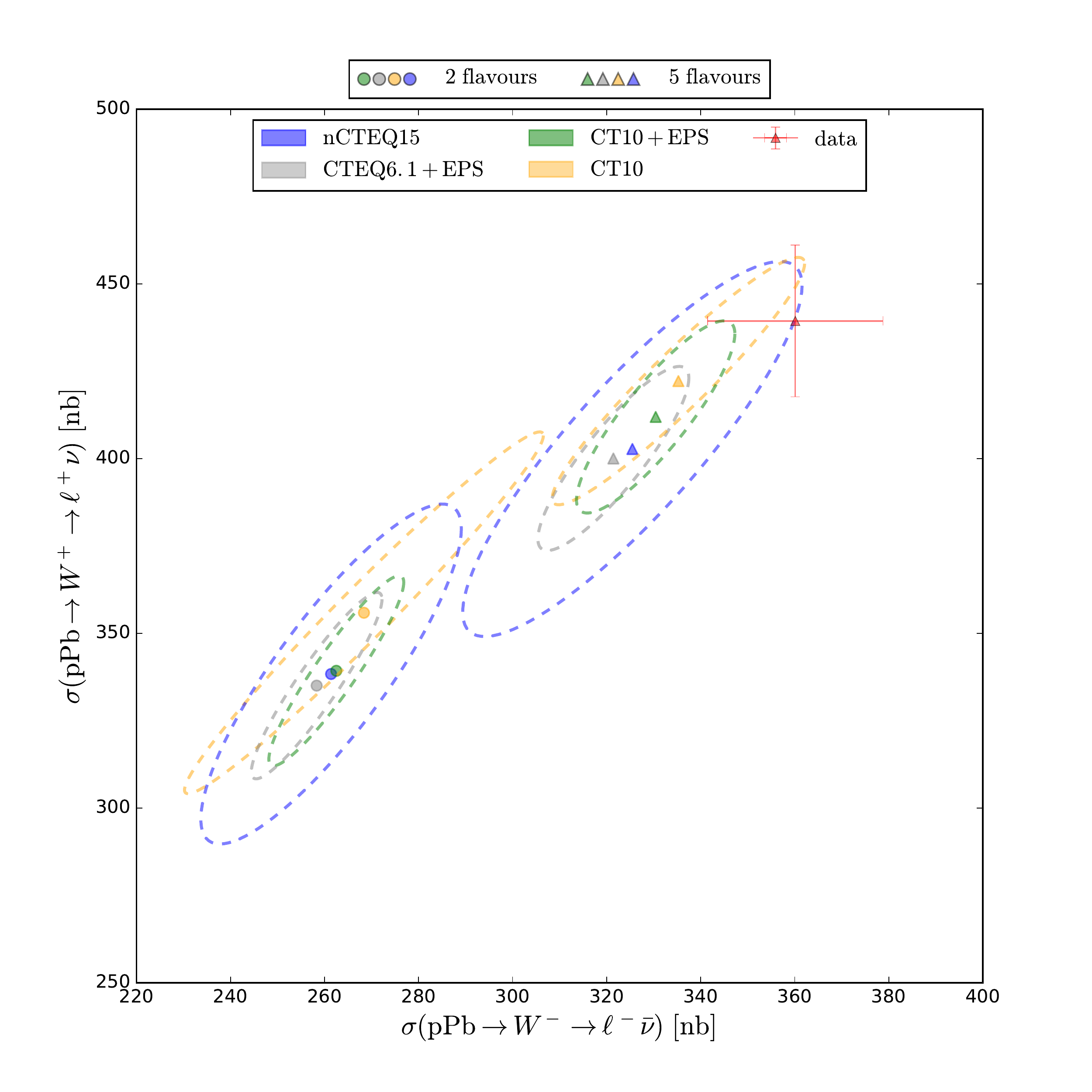}
\caption{Comparison of correlations between $W^+$ and $W^-$ cross sections for the case
when only one family of quarks $\{u,d\}$ is included and when all families are accounted for.
We show here results for \ncteqfit, EPS09+CT10, EPS09+CTEQ6.1 and CT10 PDFs overlaid with the CMS data.}
\label{fig:wzCrossSecCor1F}
\end{figure}
%
} 
\def\figDefxsecwwFF{
%
\begin{figure*}[p]
\centering{}
\includegraphics[width=1.0\textwidth]{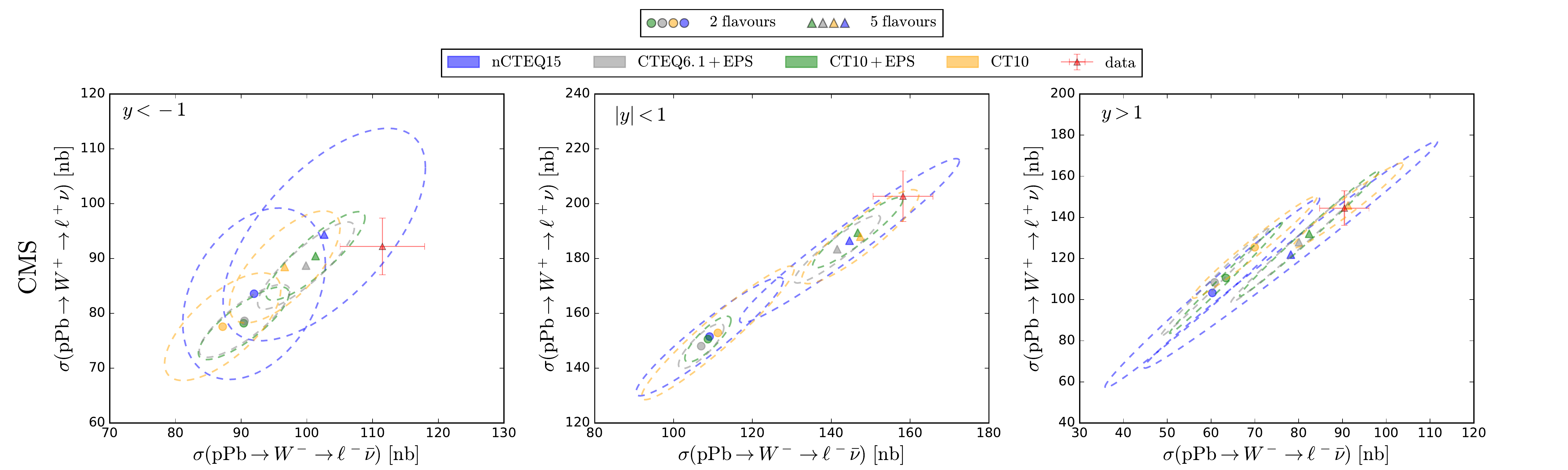}
\caption{Same as Fig.~\ref{fig:wzCrossSecCor1F} but divided into rapidity bins.}
\label{fig:wzCrossSecCor1F_bins}
\end{figure*}
%
} 
\def\figDefcmswrap{
%
\begin{figure*}[p]
\centering{}
\subfloat[$W^{+}$ 5 flavours\label{fig:rap_distr_str_wp_full}]{
\includegraphics[width=0.48\textwidth]{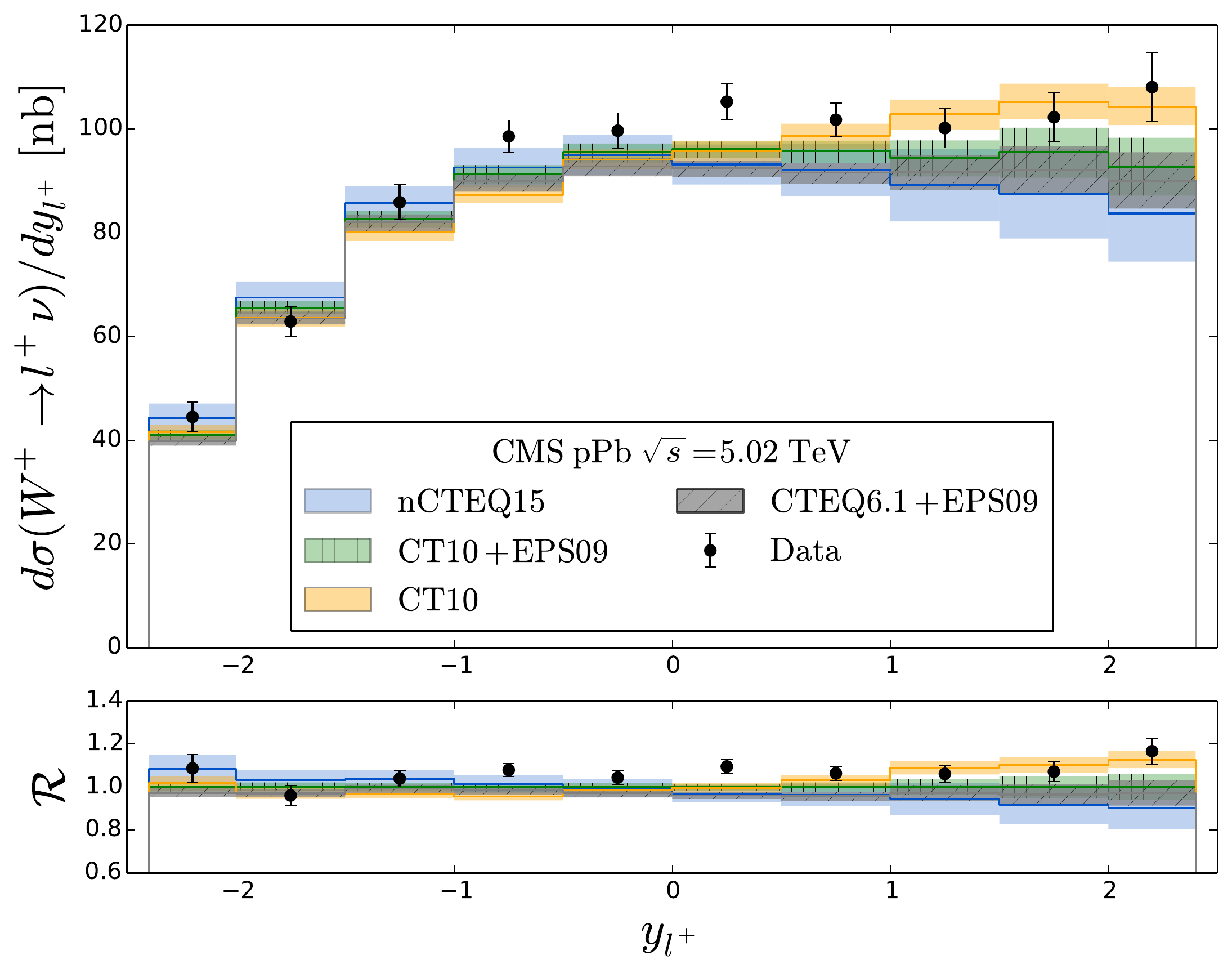}}
\subfloat[$W^{+}$ 2 flavours\label{fig:rap_distr_str_wp_1f}]{
\includegraphics[width=0.48\textwidth]{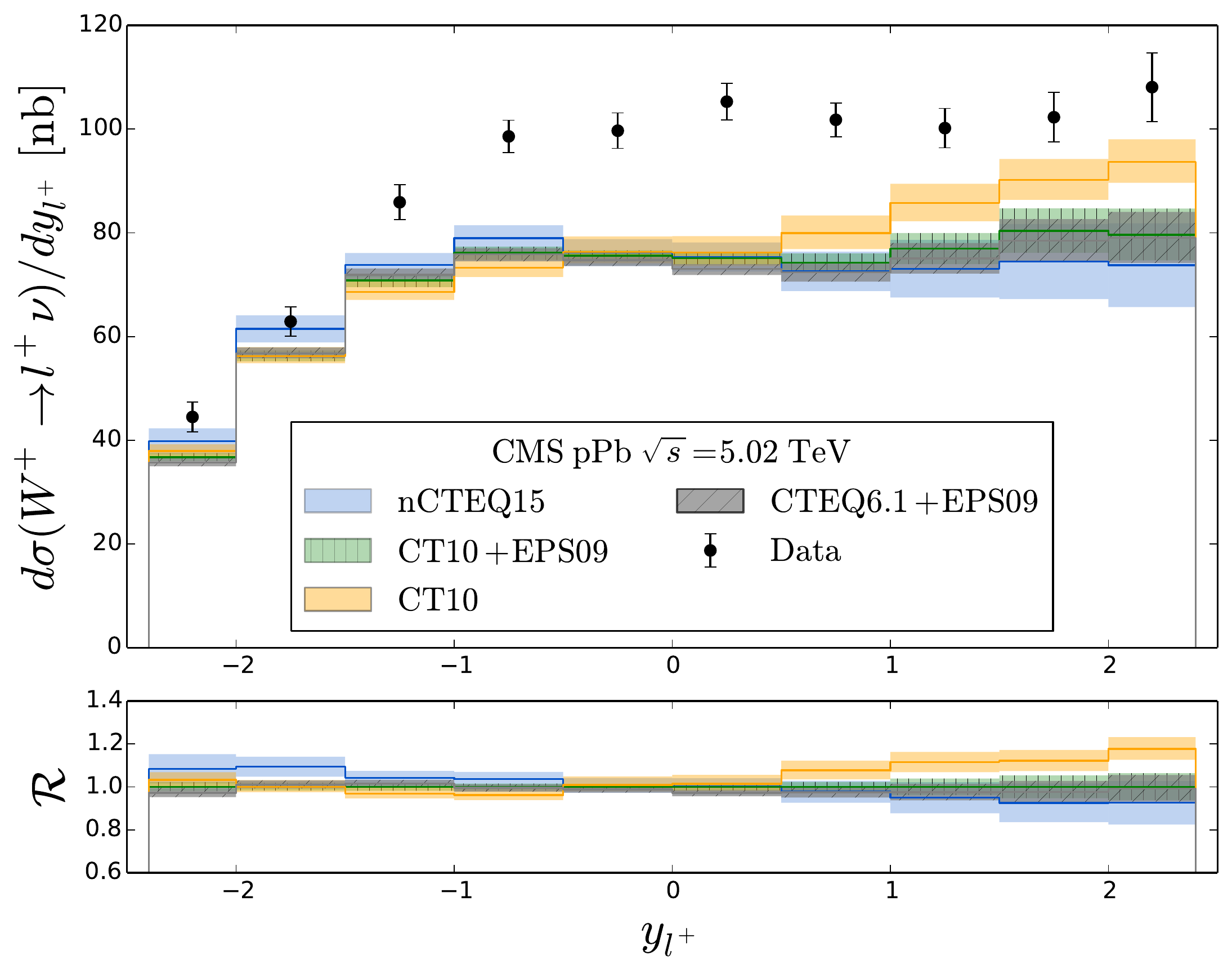}}
\\
\subfloat[$W^{-}$ 5 flavours\label{fig:rap_distr_str_wm_full}]{
\includegraphics[width=0.48\textwidth]{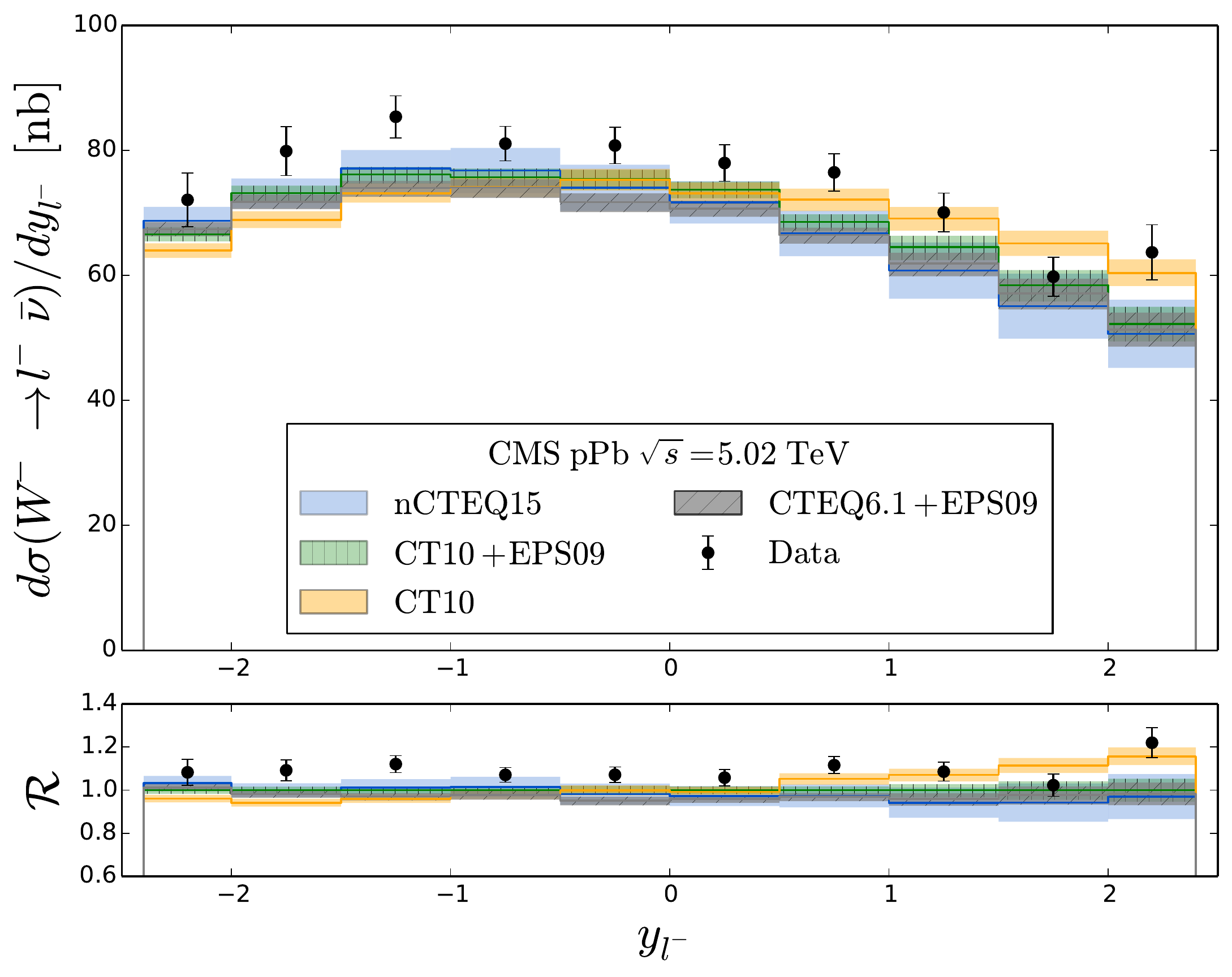}}
\subfloat[$W^{-}$ 2 flavours\label{fig:rap_distr_str_wm_1f}]{
\includegraphics[width=0.48\textwidth]{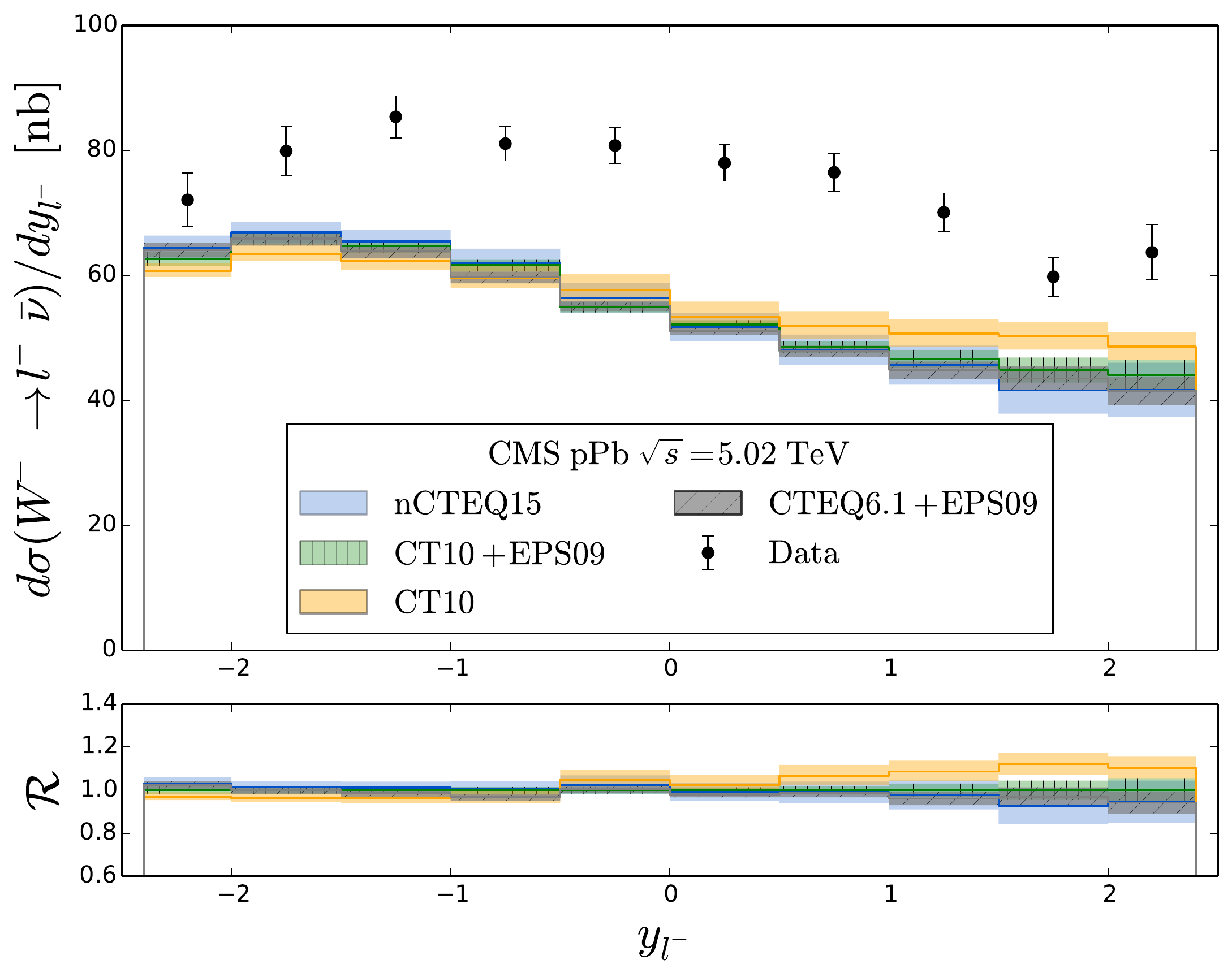}}
\caption{Rapidity distributions for $W^\pm$ cross sections measured by CMS compared with predictions from the \ncteqfit, EPS09+CT10,
EPS09+CTEQ6.1 and CT10 PDFs. 
Figures (a) and (c) show the results for 5~flavors, while 
Figures (b) and (d) show the 2~flavors results.}
\label{fig:rap_distr_str}
\end{figure*}
%
} 
\def\figDefHessGlue{
\begin{figure*}[htb]
\begin{center}
\subfloat[Comparison of the \ncteqfit central gluon PDF and Hessian error bands with the average and variance calculated from $N_{\text{rep}}=10^{4}$ 
replicas.\label{fig:rep-vs-hess-a}]{
\includegraphics[width=0.4\textwidth]{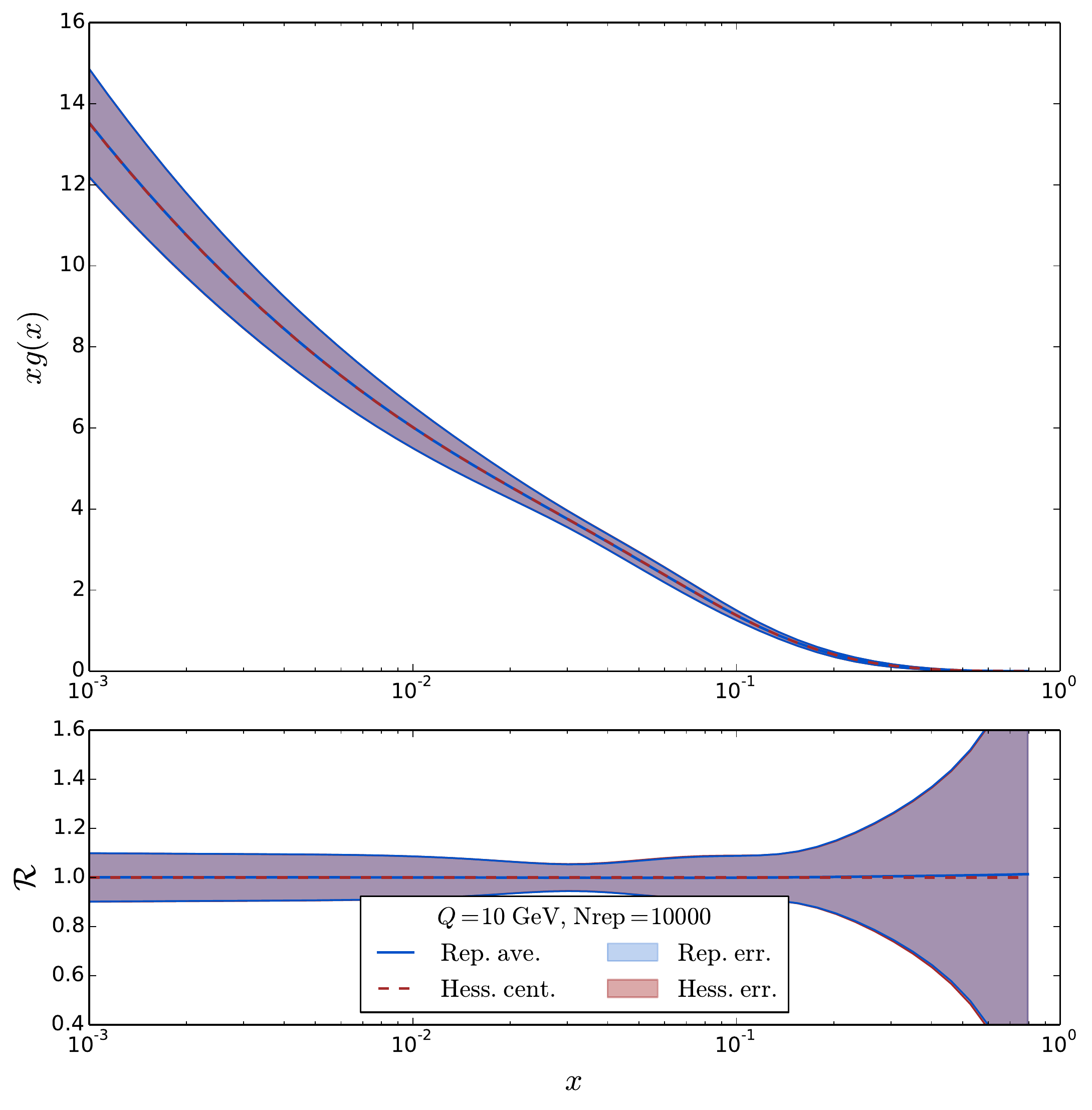}}
\hfil
\subfloat[Ratio of the average gluon PDF calculated using $N_{\text{rep}}=\{10^{2},10^{3},10^{4}\}$ replicas to the \ncteqfit central gluon PDF.\label{fig:rep-vs-hess-b}]{
\includegraphics[width=0.45\textwidth]{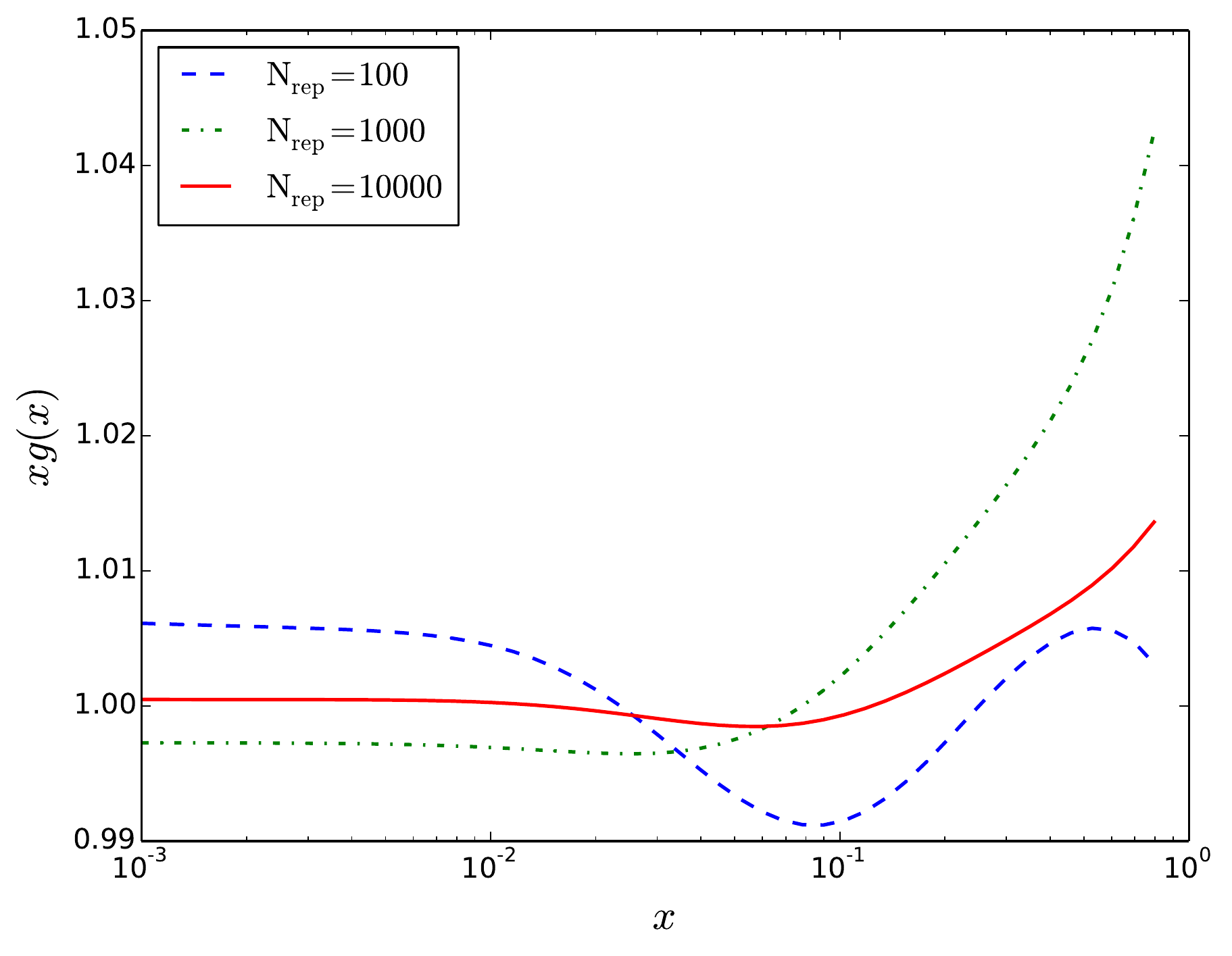}}
\caption{Comparison of the \ncteqfit Hessian gluon distribution and its reproduction in terms of replicas at a scale of $Q=10$ GeV.}
\label{fig:rep-vs-hess}
\end{center}
\end{figure*}
} 
\def\figDefRWcmsW{
\begin{figure}[t]
\centering{}\includegraphics[width=0.45\textwidth]{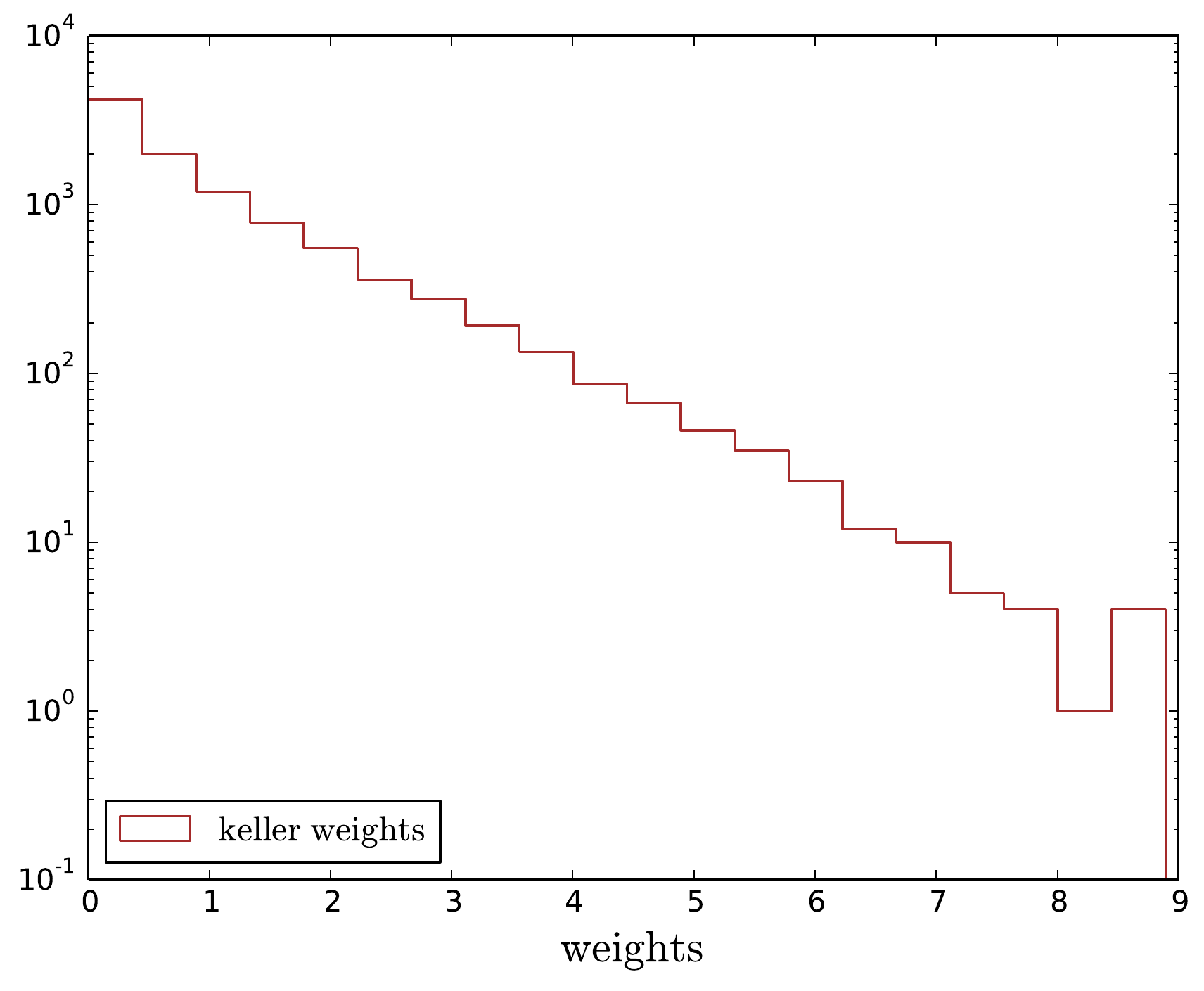}
\caption{Weight distribution after reweighting using rapidity distributions
of charged leptons from CMS $W^{\pm}$ production data. 
\label{fig:weights_cmsW}}
\end{figure}
} 
\def\figDefTheoCMSw{
\begin{figure*}[p]
\subfloat[$W^{+}$\label{fig:theo_after_before_cmsW_wp}]{
\centering{}\includegraphics[width=0.48\textwidth]{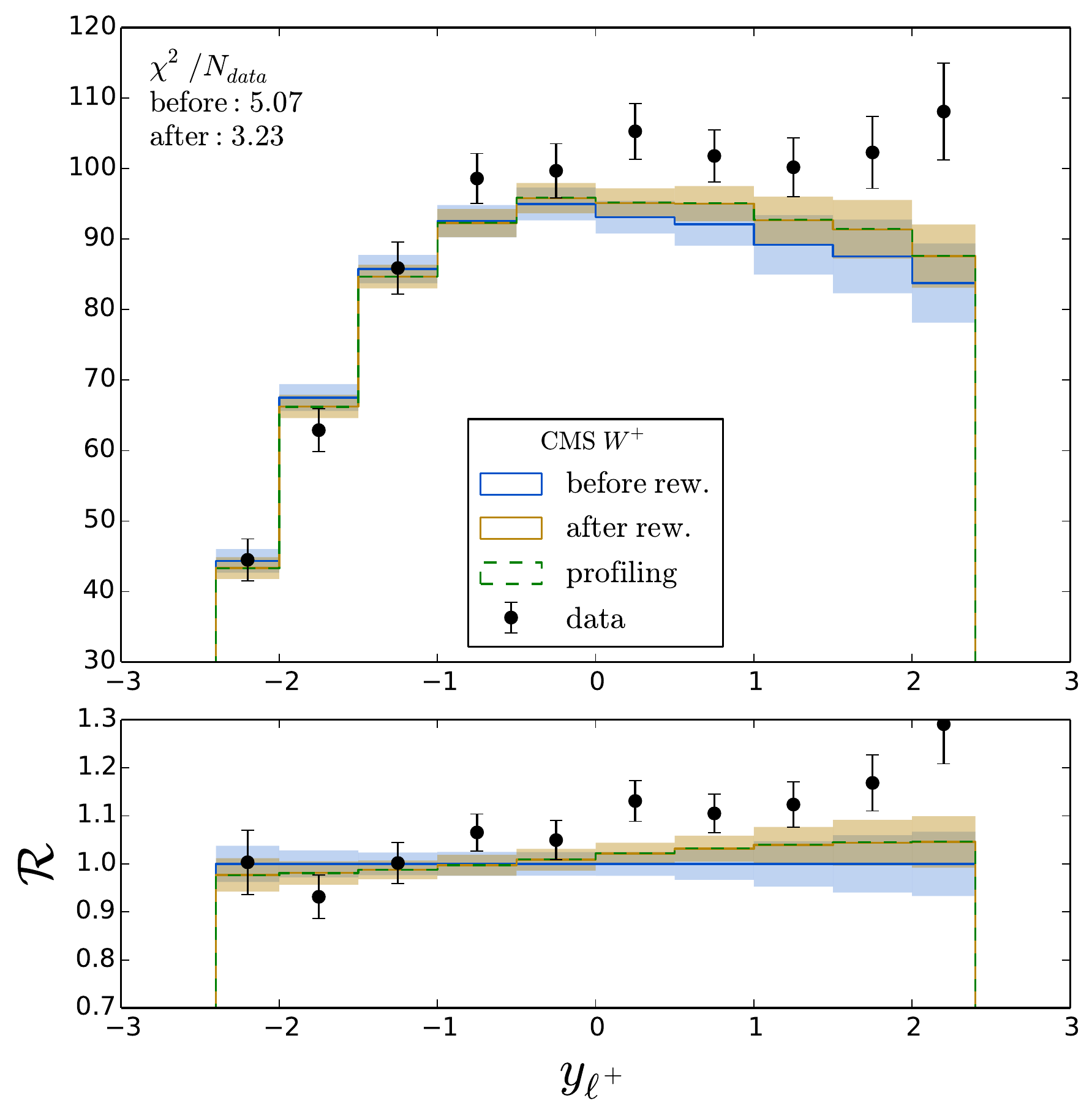}}
\subfloat[$W^{-}$\label{fig:theo_after_before_cmsW_wm}]{
\hfil
\includegraphics[width=0.48\textwidth]{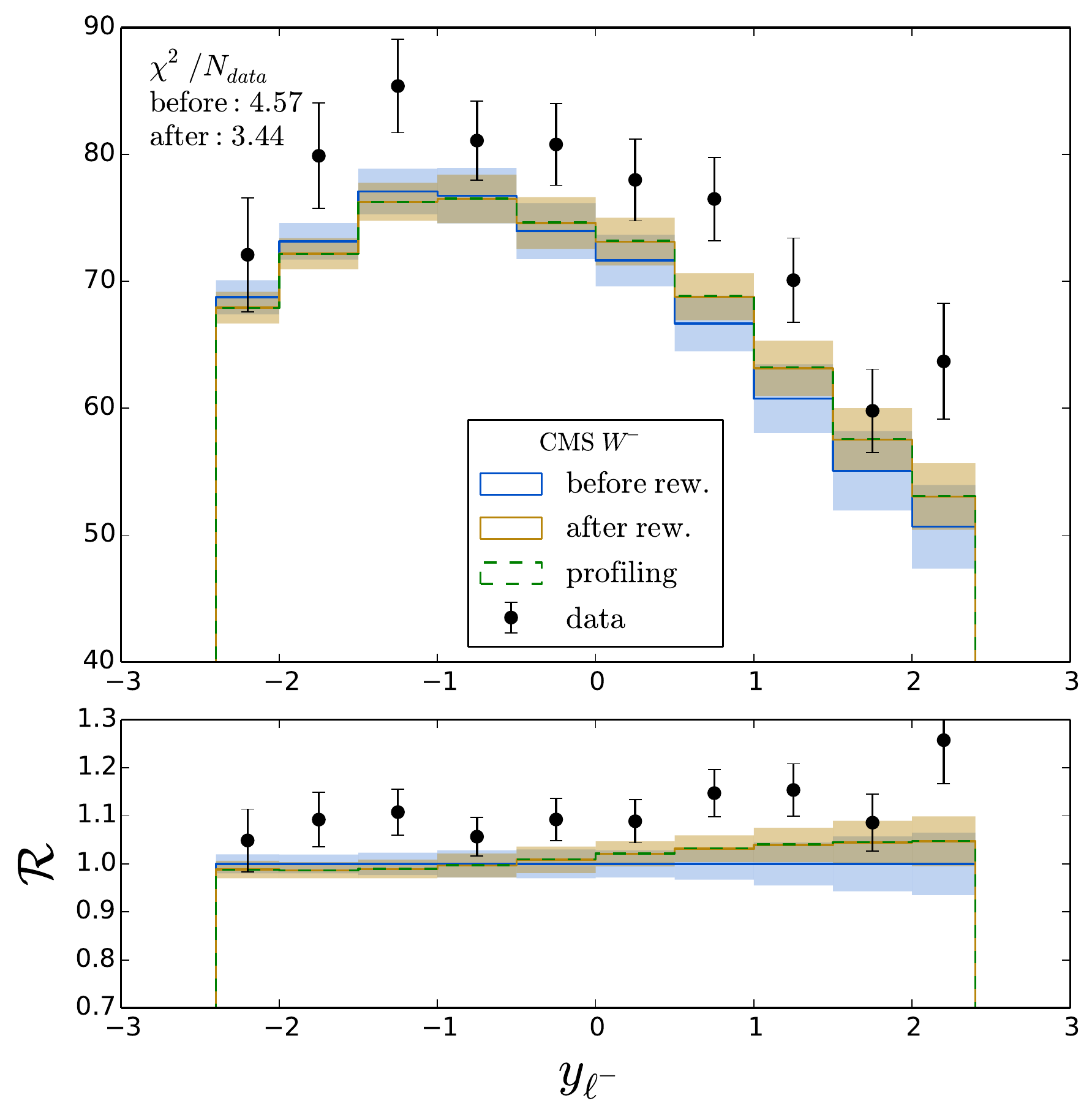}}
\caption{Comparison of data and theory before and after the reweighting procedure
for the rapidity distributions of charged leptons in $W^{\pm}$ production
measured by CMS~\cite{Khachatryan:2015hha}.
}
\label{fig:theo_after_before_cmsW}
\end{figure*}
} 
\def\figDefTheoCMSWii{
\begin{figure*}[p]
\centering{}
\subfloat[$u$ PDF\label{fig:pdfs_after_before_cmsW_10_u}]{
\includegraphics[width=0.48\textwidth]{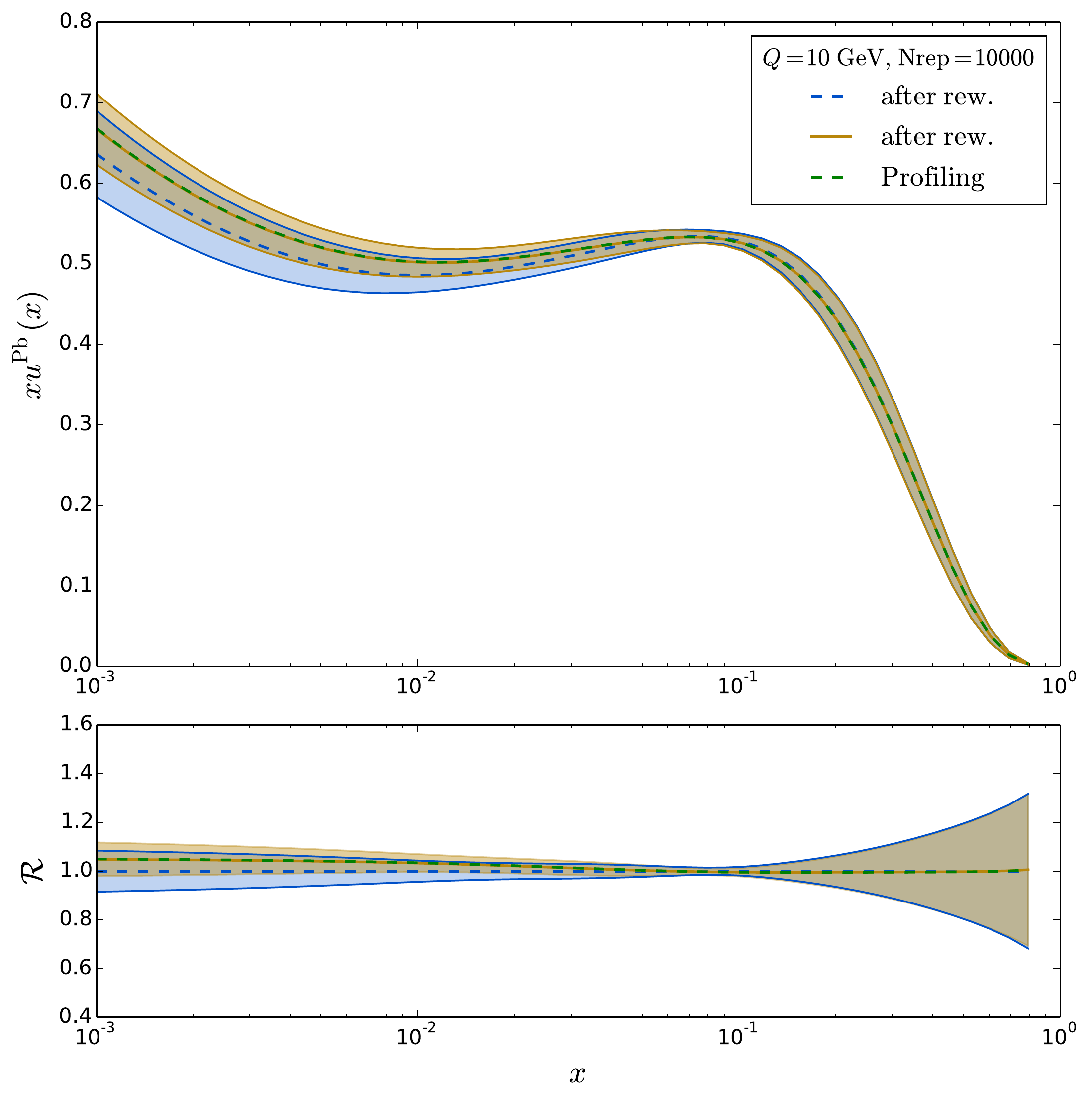}}
\hfil
\subfloat[$g$ PDF\label{fig:pdfs_after_before_cmsW_10_g}]{
\includegraphics[width=0.48\textwidth]{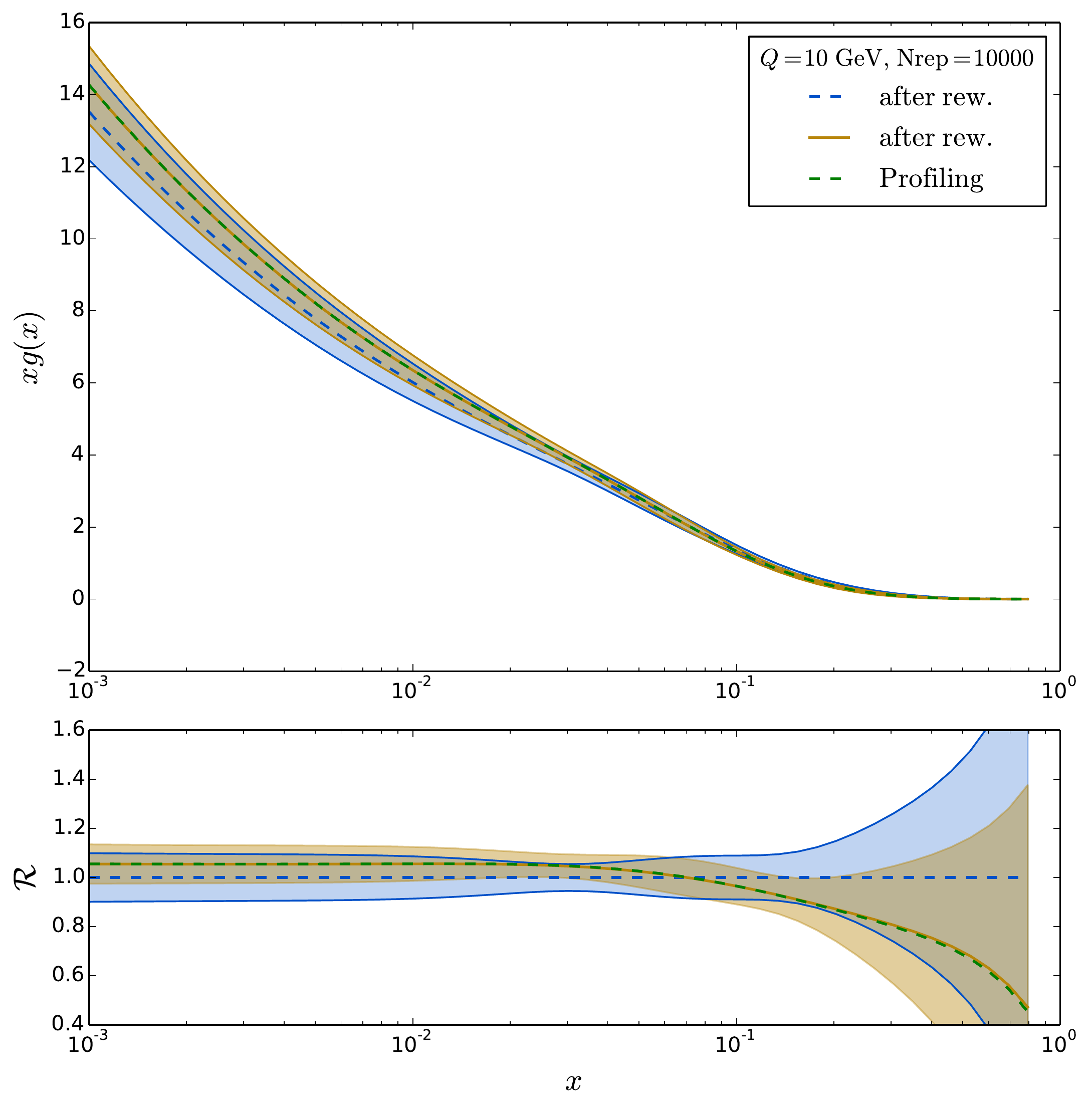}}
\caption{Comparison of PDFs at $Q=10$ GeV before and after the reweighting procedure
using only the CMS rapidity distribution data set. The lower plots show the
ratio compared to the central (average) distribution before the reweighting.
}
\label{fig:pdfs_after_before_cmsW_10}
\end{figure*}
} 
\def\figDefcmsAsym{
\begin{figure}[tb]
\centering{}
\includegraphics[width=0.48\textwidth]{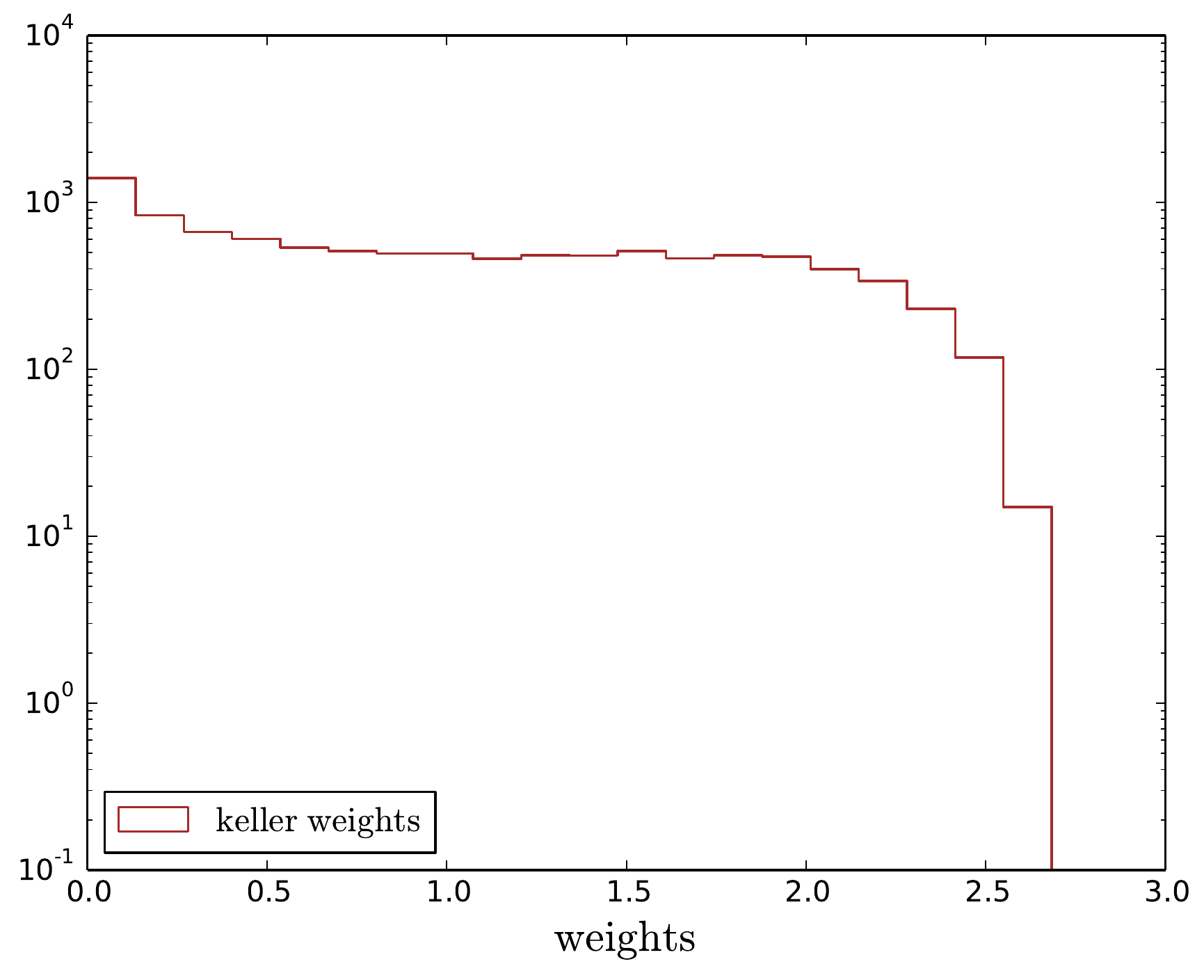}
\caption{Weight distribution after reweighting using the lepton charge asymmetry
and forward-backward asymmetry from CMS $W^{\pm}$ production data.
}
\label{fig:weights_cmsWasym}
\end{figure}
} 
\def\figDefcmsAsymii{
\begin{figure*}[p]
\centering{}
\subfloat[Charge Asymmetry
\label{fig:theo_after_before_cmsWasym_asym}]{
\includegraphics[width=0.48\textwidth]{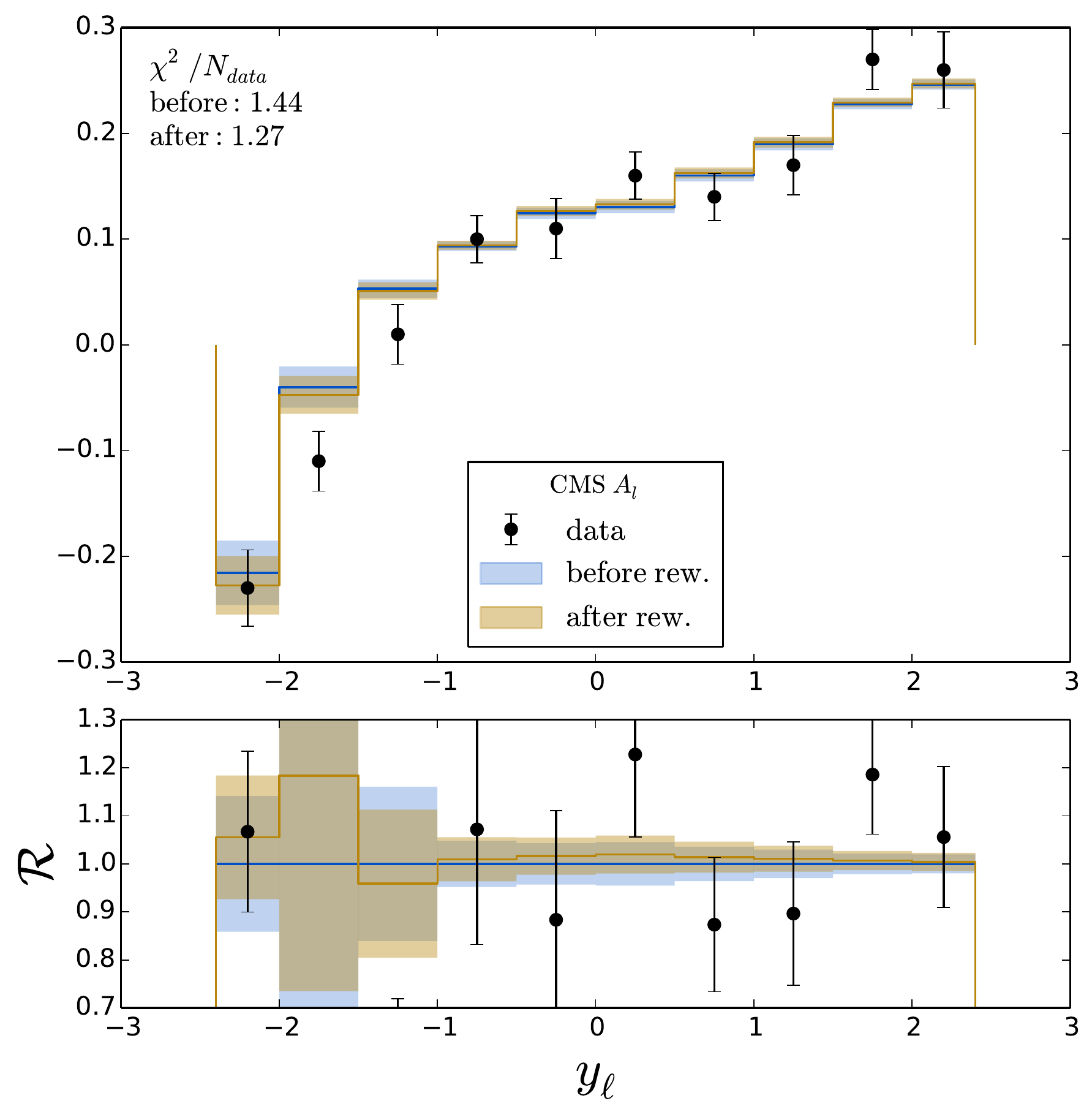}}
\hfil
\subfloat[Forward-backward Asymmetry]{
\includegraphics[width=0.48\textwidth]{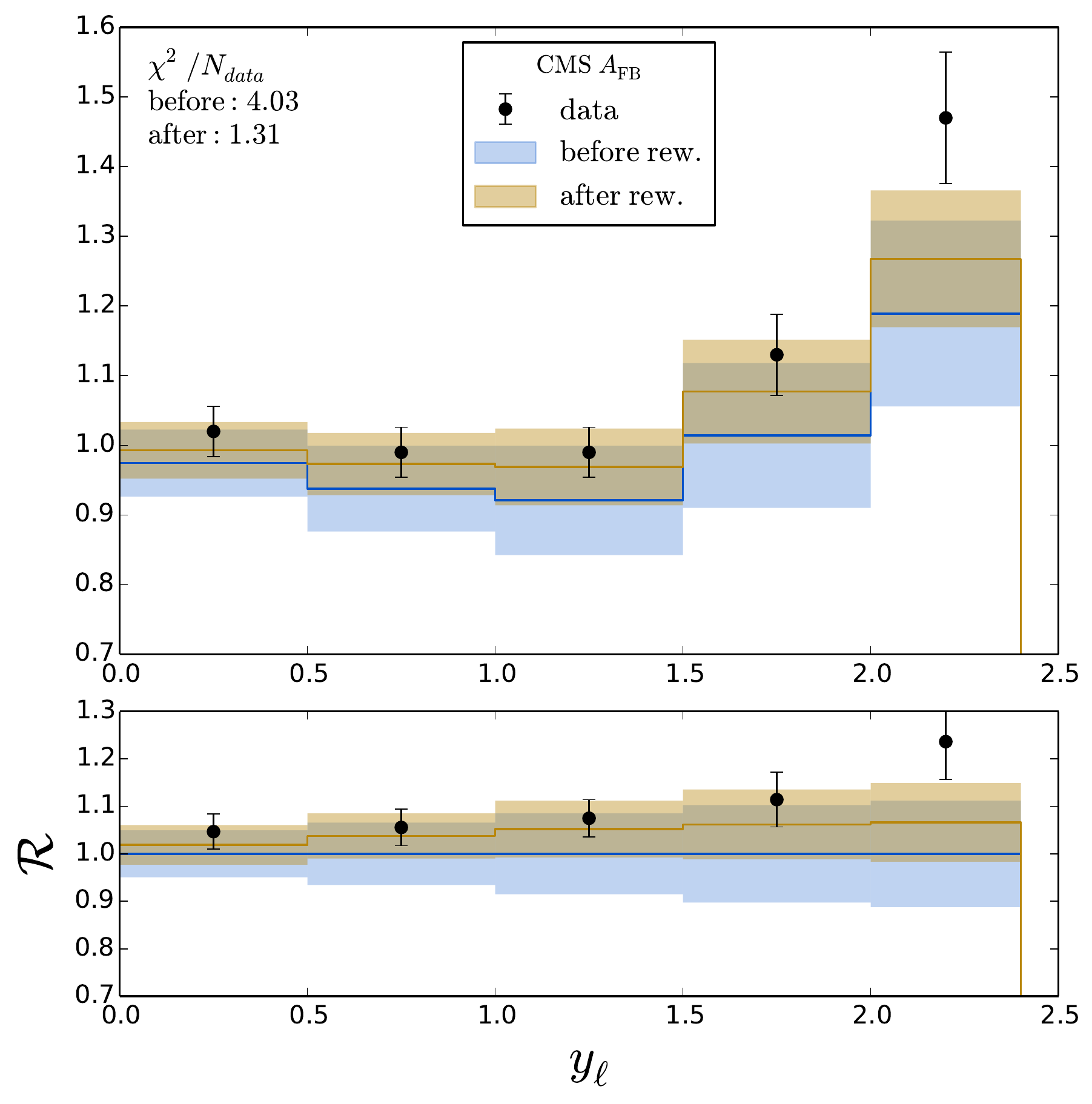}}
\caption{Comparison of data and theory before and after the reweighting procedure
using the charge asymmetry (left) and the forward-backward asymmetry (right)
from $W^{\pm}$ production data by CMS~\cite{Khachatryan:2015hha}.
}
\label{fig:theo_after_before_cmsWasym}
\end{figure*}
} 
\def\figDefcmsAsymiii{
\begin{figure*}[p]
\centering{}
\subfloat[$u$ PDF\label{fig:pdfs_after_before_cmsWasym_10_u}]{
\includegraphics[width=0.48\textwidth]{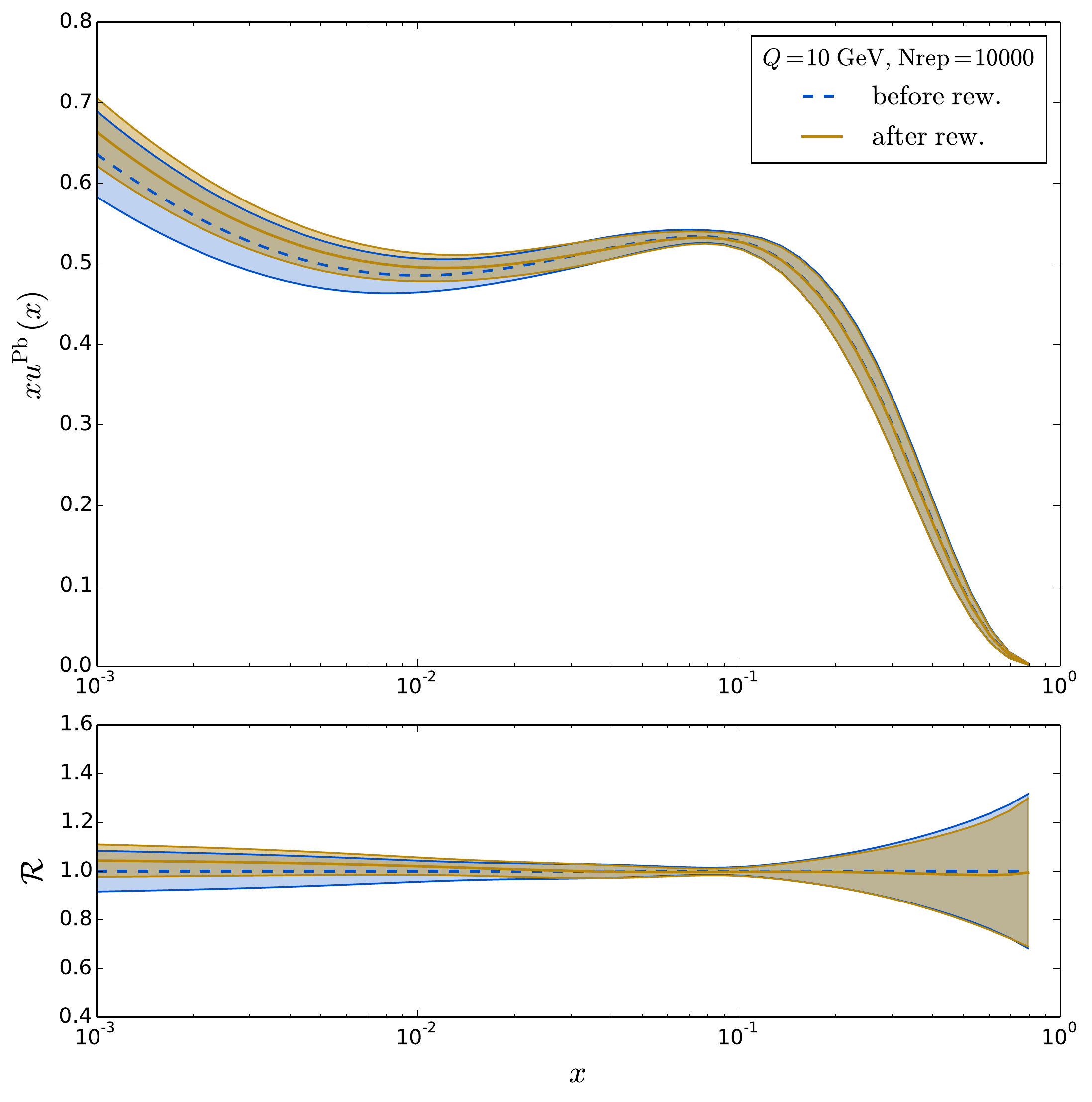}}
\hfil
\subfloat[$g$ PDF\label{fig:pdfs_after_before_cmsWasym_10_g}]{
\includegraphics[width=0.48\textwidth]{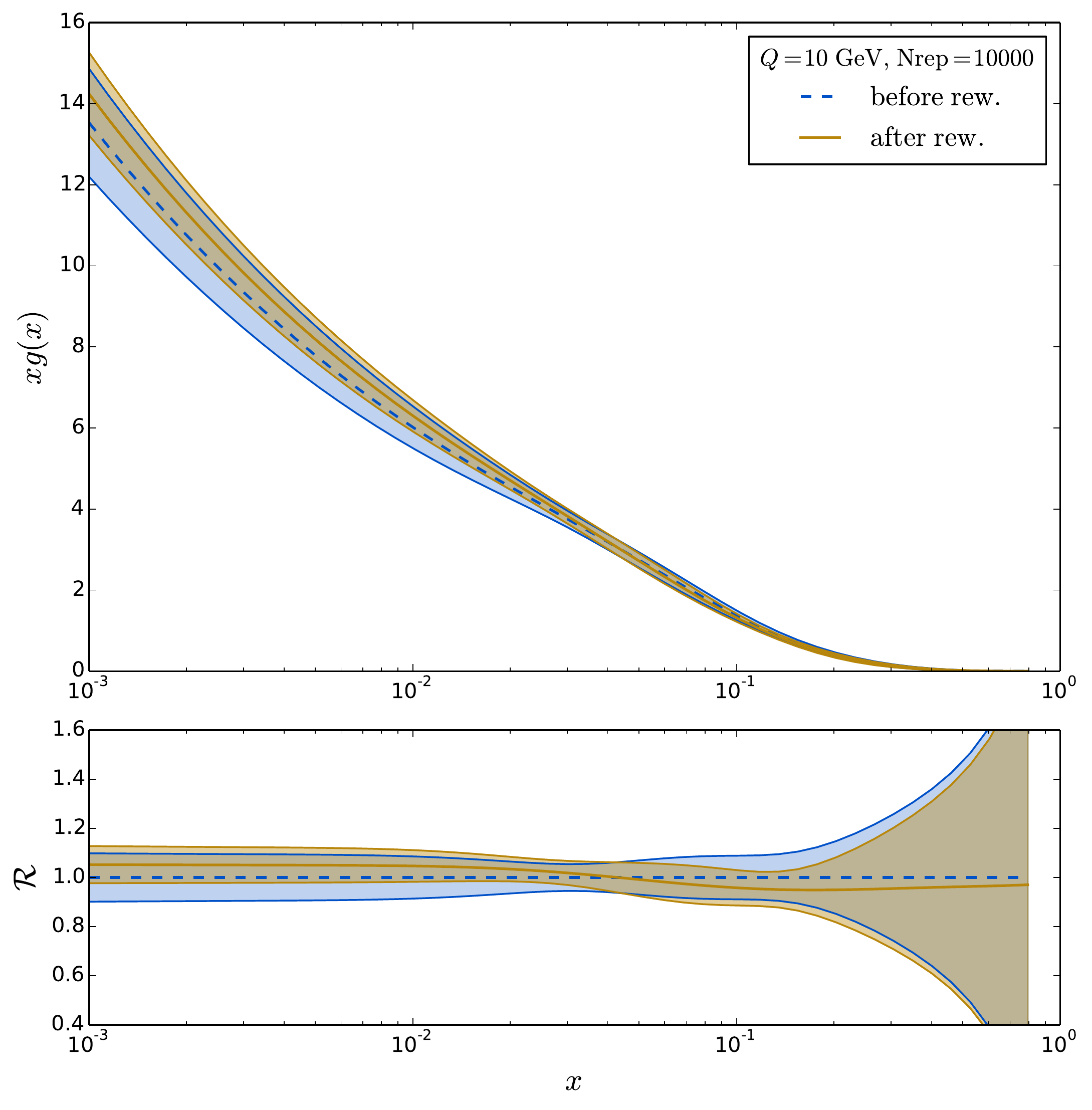}}
\caption{Comparison of PDFs at a scale of $Q=10$ GeV before and after the reweighting procedure using asymmetries measured
by CMS~\cite{Khachatryan:2015hha}. The lower plots show the ratio compared
to the central (average) distribution before the reweighting. 
}
\label{fig:pdfs_after_before_cmsWasym_10}
\end{figure*}
} 
\def\figDefwAll{
\begin{figure}[htb]
\centering{}\includegraphics[width=0.48\textwidth]{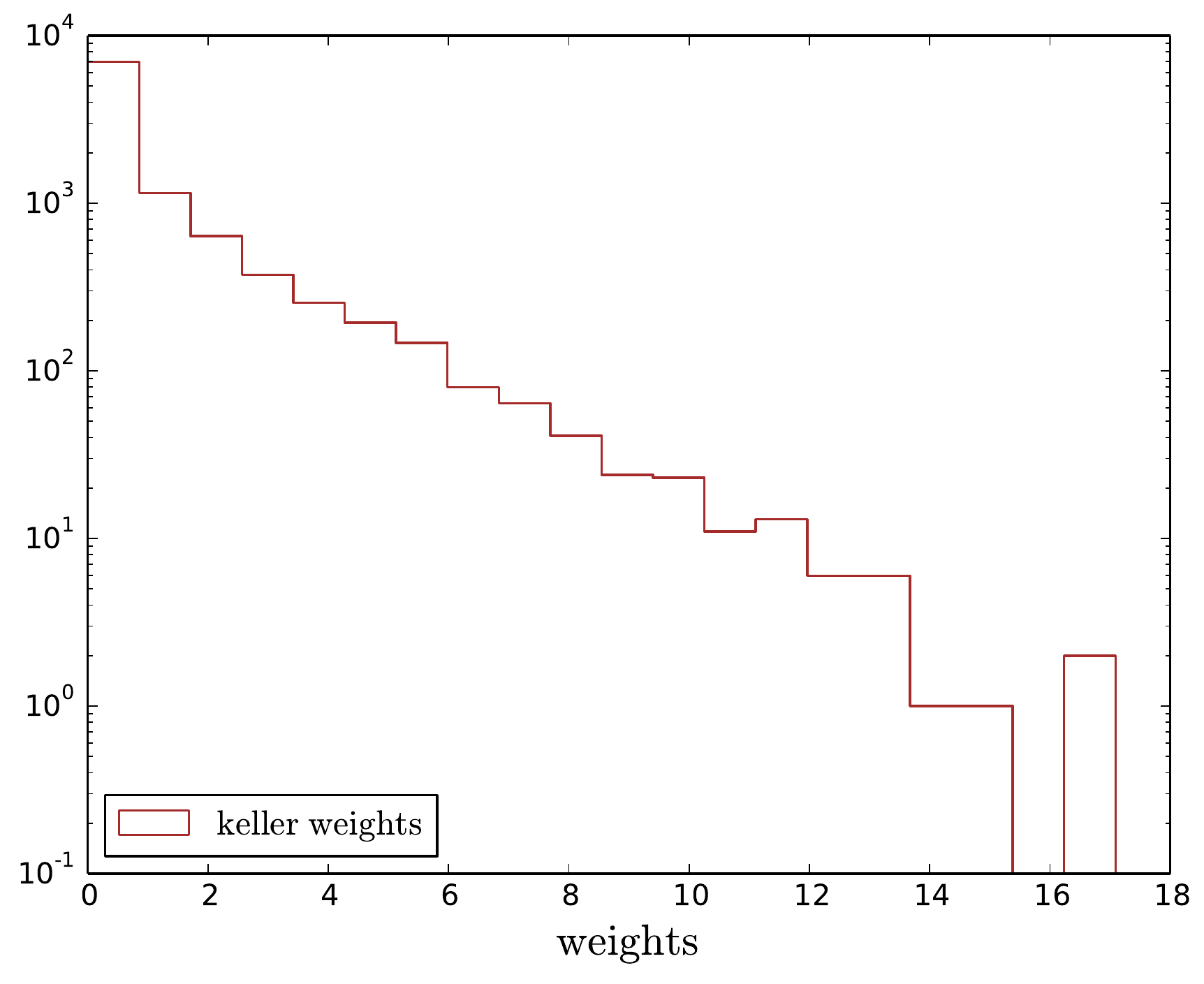}
\caption{Weight distribution after reweighting using all LHC \ppb data on $W/Z$
production. \label{fig:weights_all}}
\end{figure}
} 
\def\figDeffull{
\begin{figure}[t]
\centering{}\includegraphics[width=0.48\textwidth]{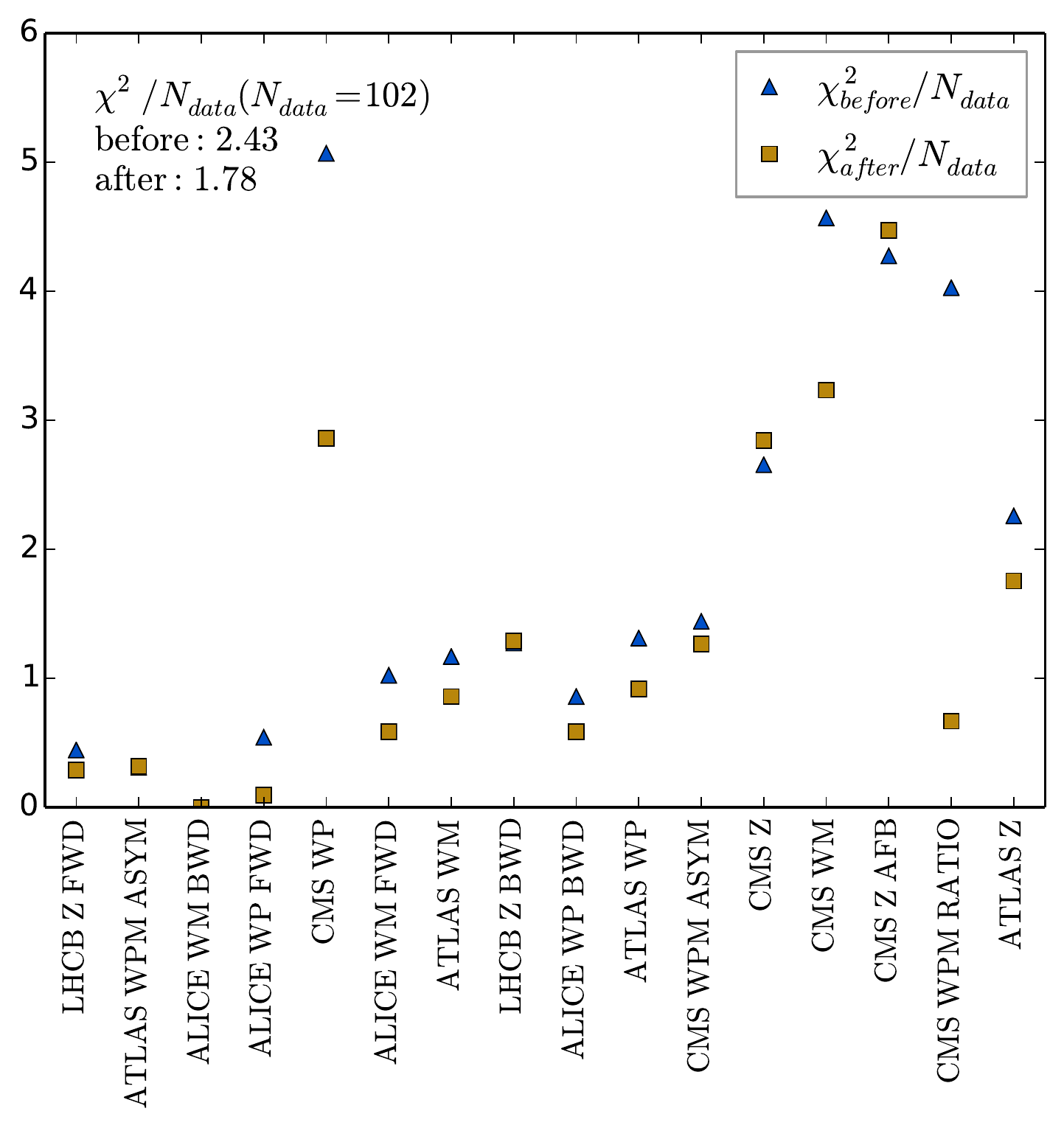}
\caption{$\chi^{2}$ per experiment before and after the reweighting procedure
using all LHC \ppb data.\label{fig:chisqr_full}}
\end{figure}
} 
\def\figDefallCmsW{
\begin{figure*}[p]
\centering{}
\subfloat[$W^{+}$\label{fig:theo_after_before_all_cmsW_wp}]{
\includegraphics[width=0.48\textwidth]{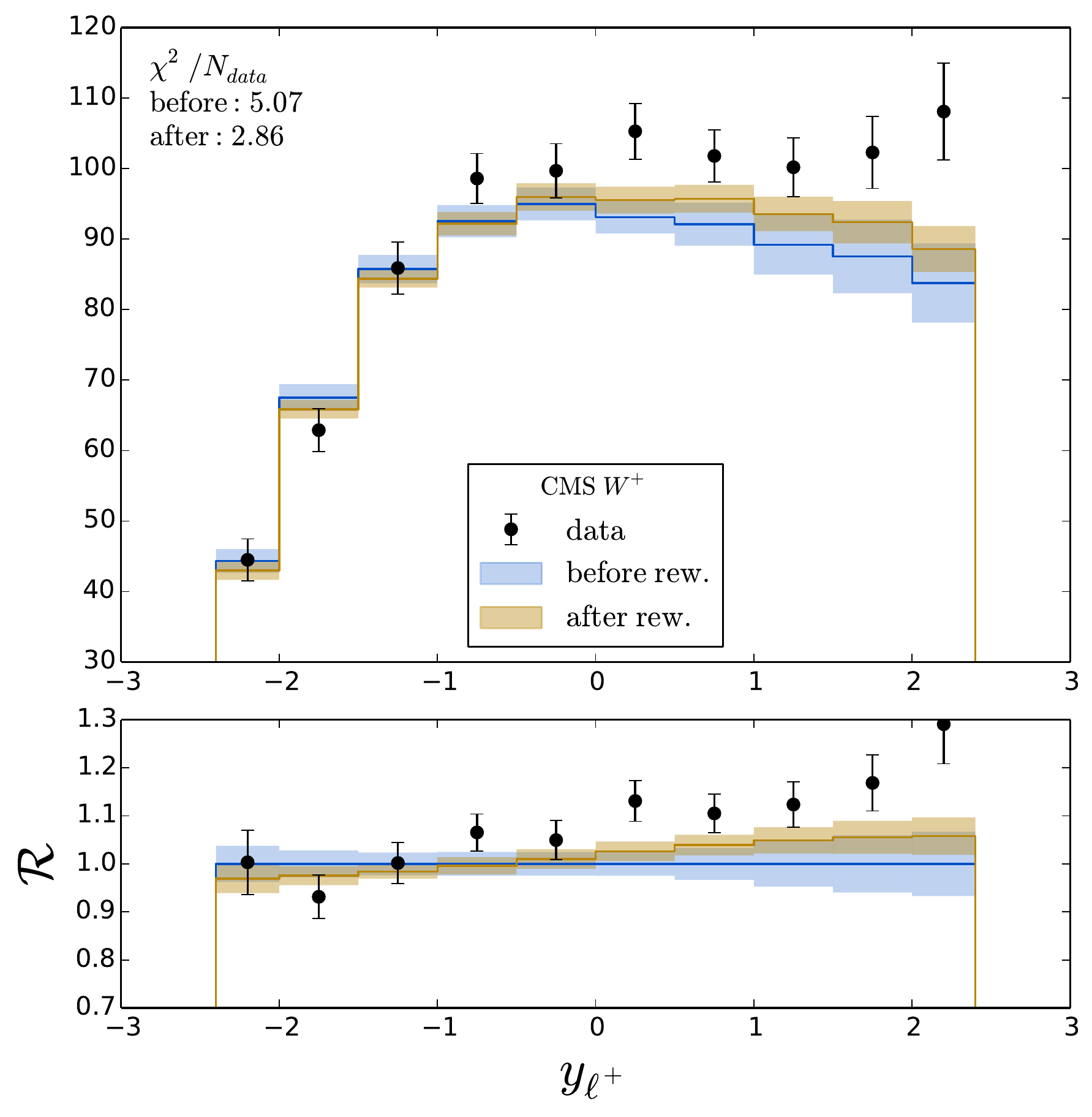}}
\hfil
\subfloat[$W^{-}$\label{fig:theo_after_before_all_cmsW_wm}]{
\includegraphics[width=0.48\textwidth]{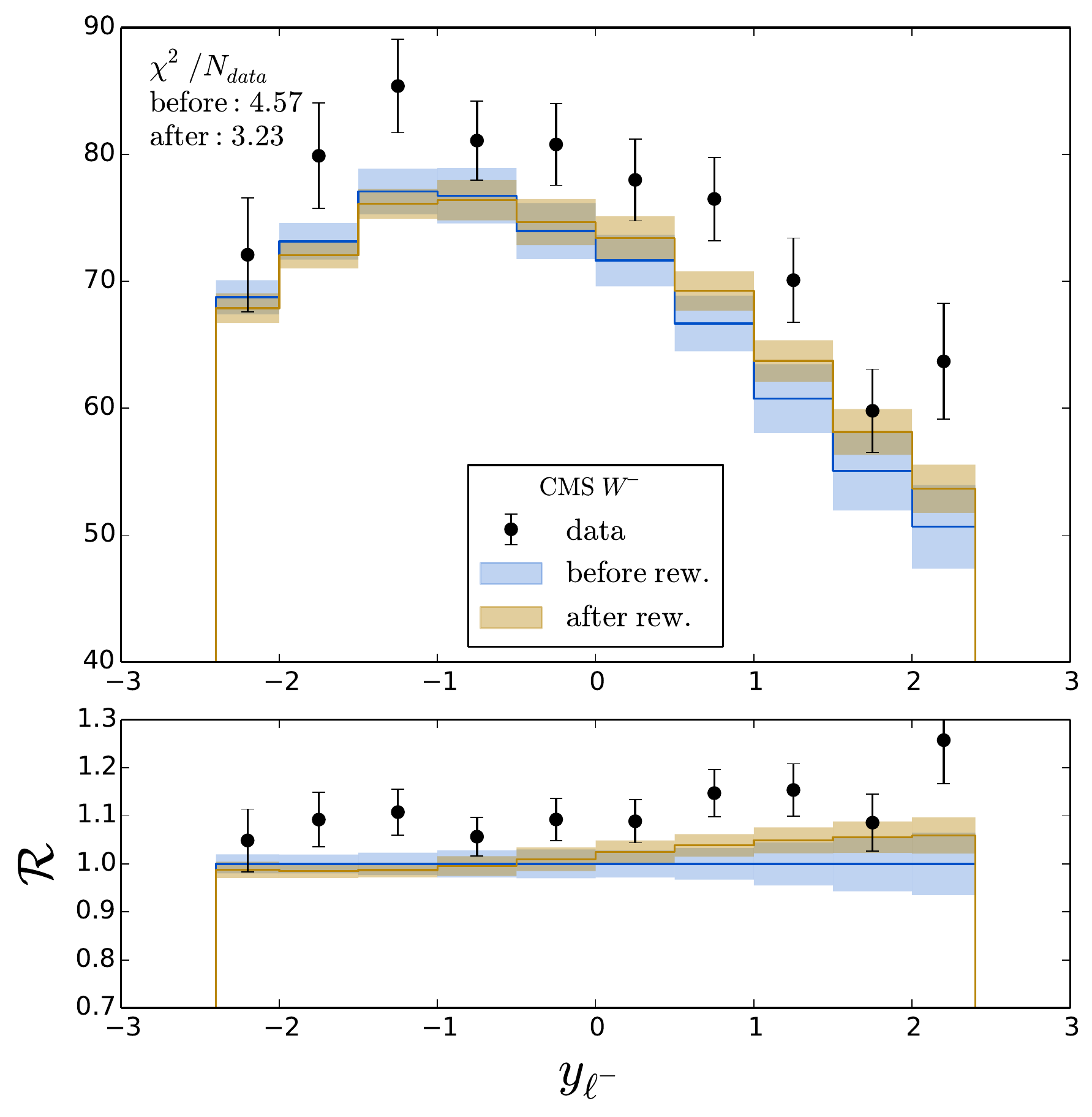}}
\caption{Comparison of data and theory before and after the reweighting procedure
using all LHC \ppb data. 
The  results for the CMS  $W^{+}$ (left) and $W^{-}$ (right) distributions are shown.}
\label{fig:theo_after_before_all_cmsW}
\end{figure*}
} 
\def\figDefallAtlasw{
\begin{figure*}[p]
\centering{}
\subfloat[$W^{+}$\label{fig:theo_after_before_all_atlasW_wp}]{
\includegraphics[width=0.48\textwidth]{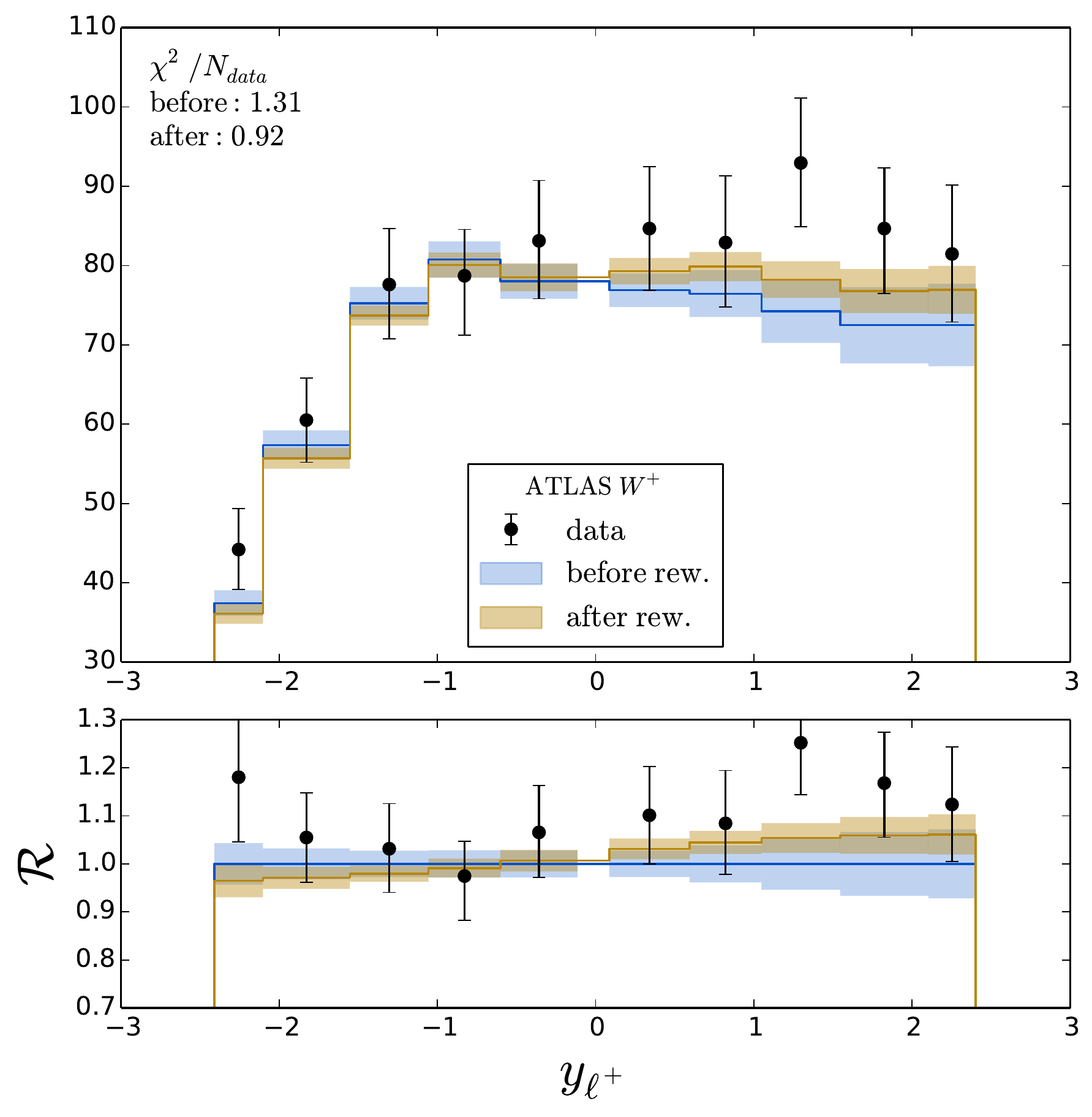}}
\hfil
\subfloat[$W^{-}$\label{fig:theo_after_before_all_atlasW_wm}]{
\includegraphics[width=0.48\textwidth]{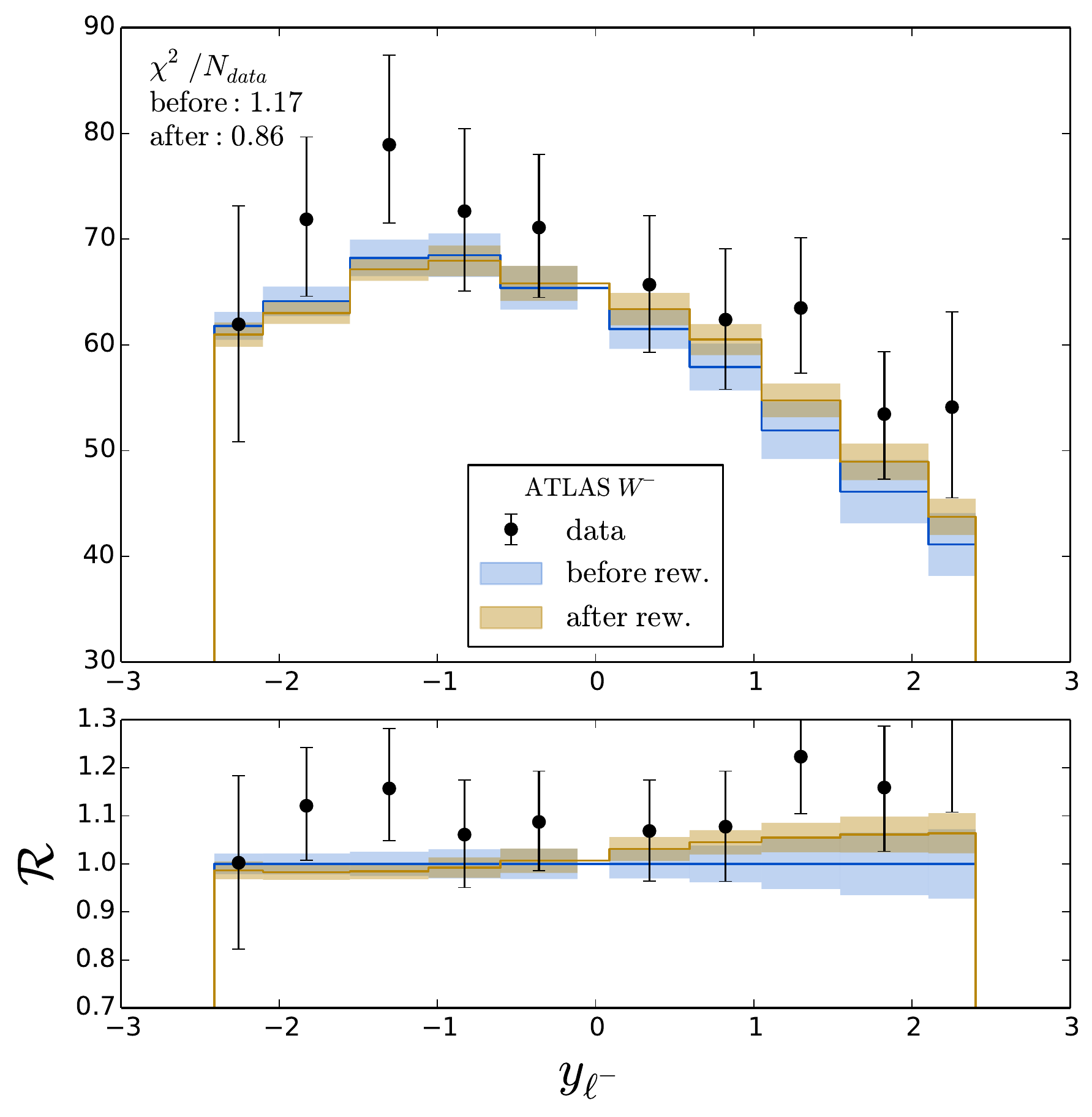}}
\caption{Comparison of data and theory before and after the reweighting procedure
using all LHC \ppb data. 
The  results for the ATLAS  $W^{+}$ (left) and $W^{-}$ (right) distributions are shown.}
\label{fig:theo_after_before_all_atlasW}
\end{figure*}
} 
\def\figDefallud{
\begin{figure*}[p]
\centering{}
\subfloat[$u$ PDF]{
\includegraphics[width=0.48\textwidth]{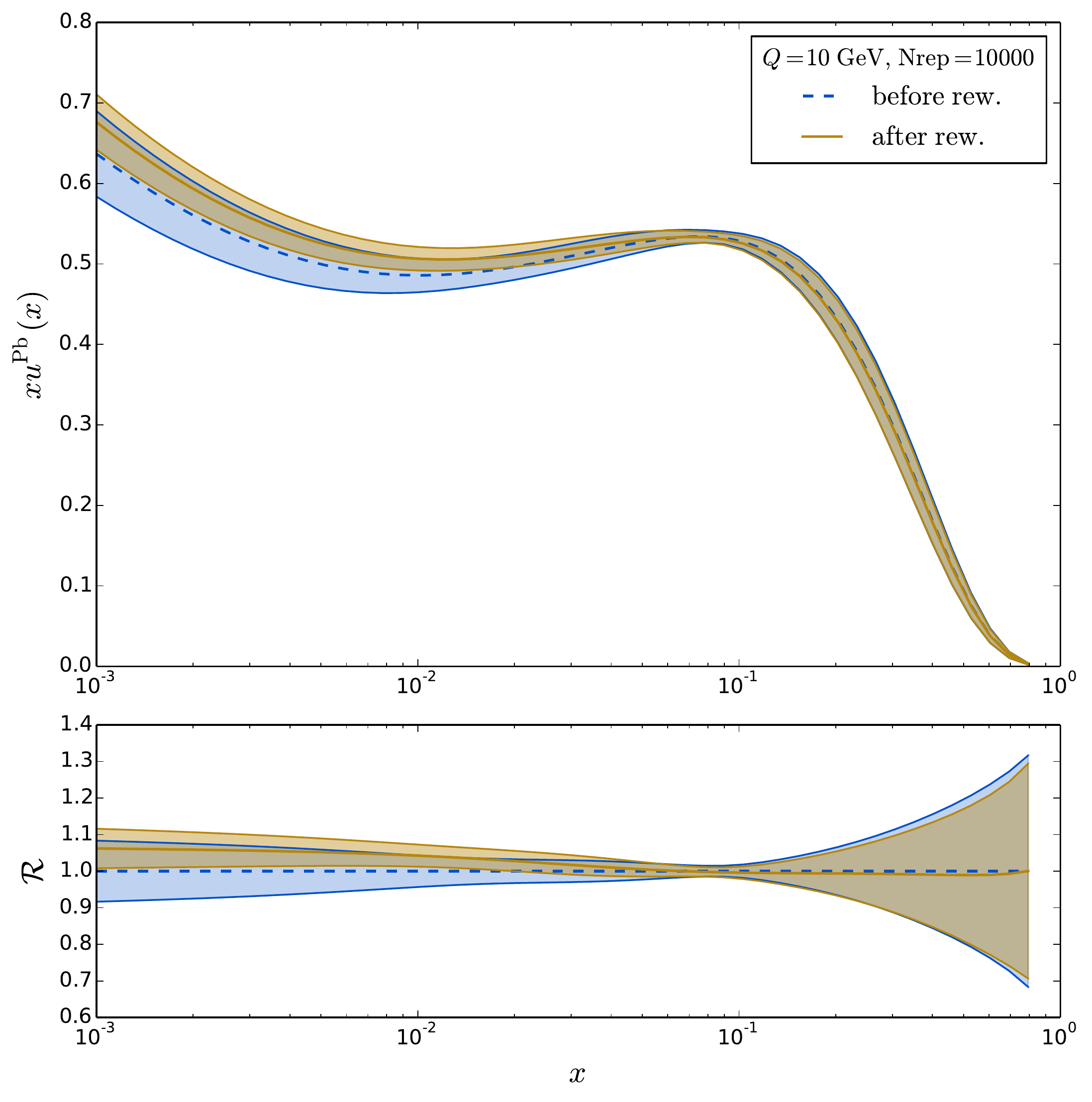}}
\hfil
\subfloat[$d$ PDF]{
\includegraphics[width=0.48\textwidth]{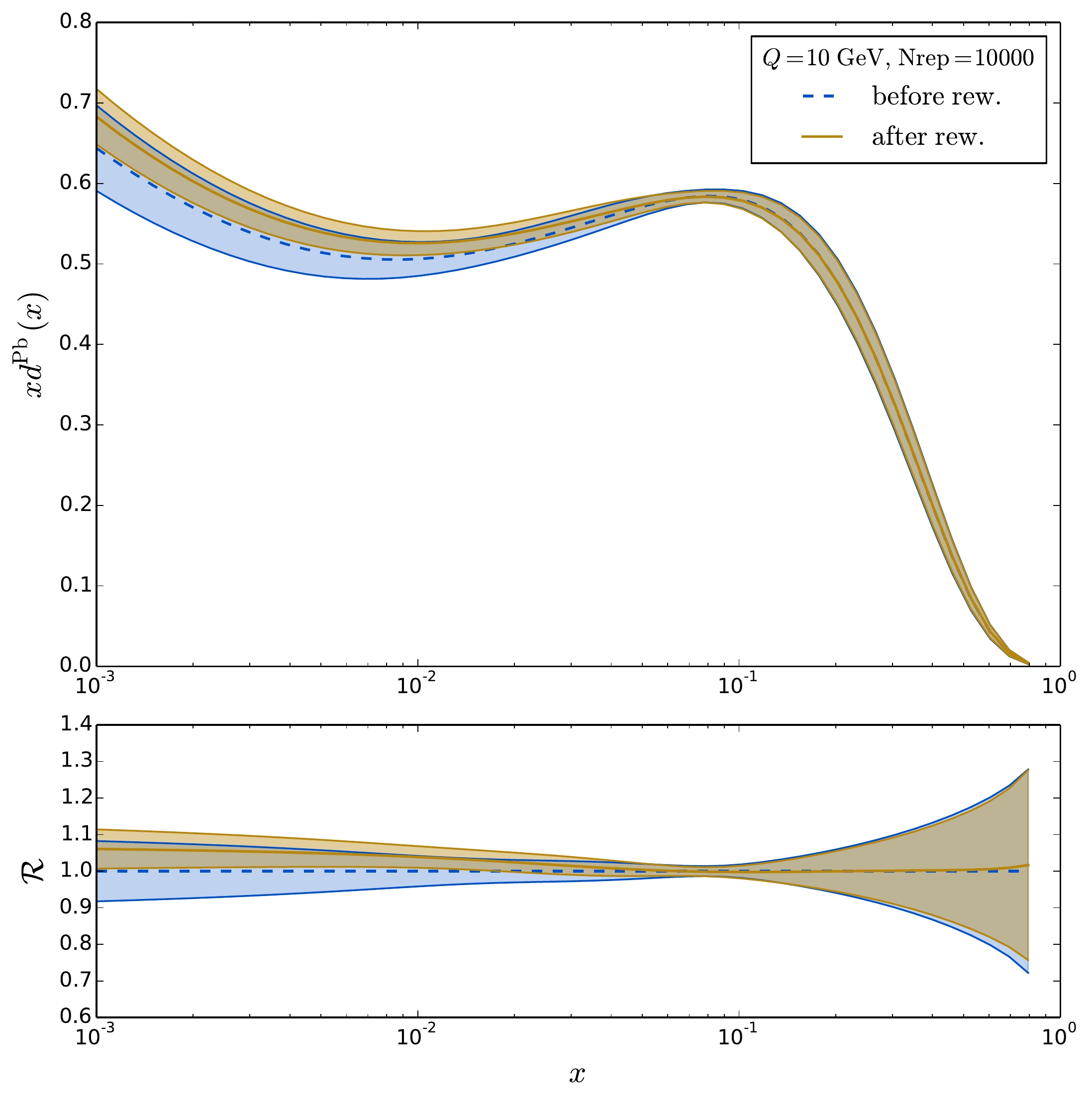}}
\caption{$u$ and $d$ PDFs before and after the reweighting using all LHC \ppb data sets.}
\label{fig:pdfs_after_before_all_ud}
\end{figure*}
} 
\def\figDefallubdb{
\begin{figure*}[p]
\centering{}
\subfloat[$\bar{u}$ PDF]{
\includegraphics[width=0.48\textwidth]{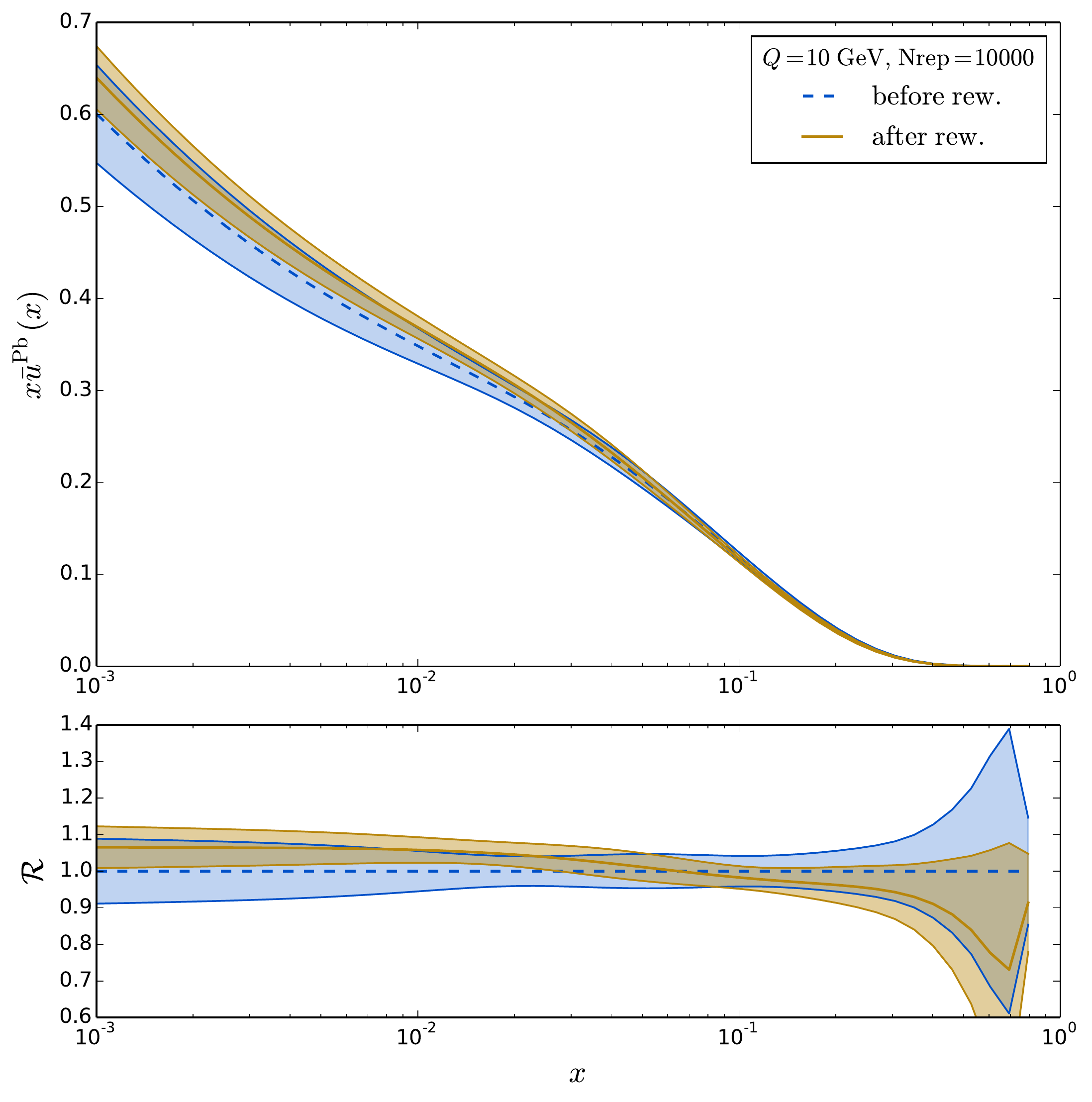}}
\subfloat[$\bar{d}$ PDF]{
\hfil
\includegraphics[width=0.48\textwidth]{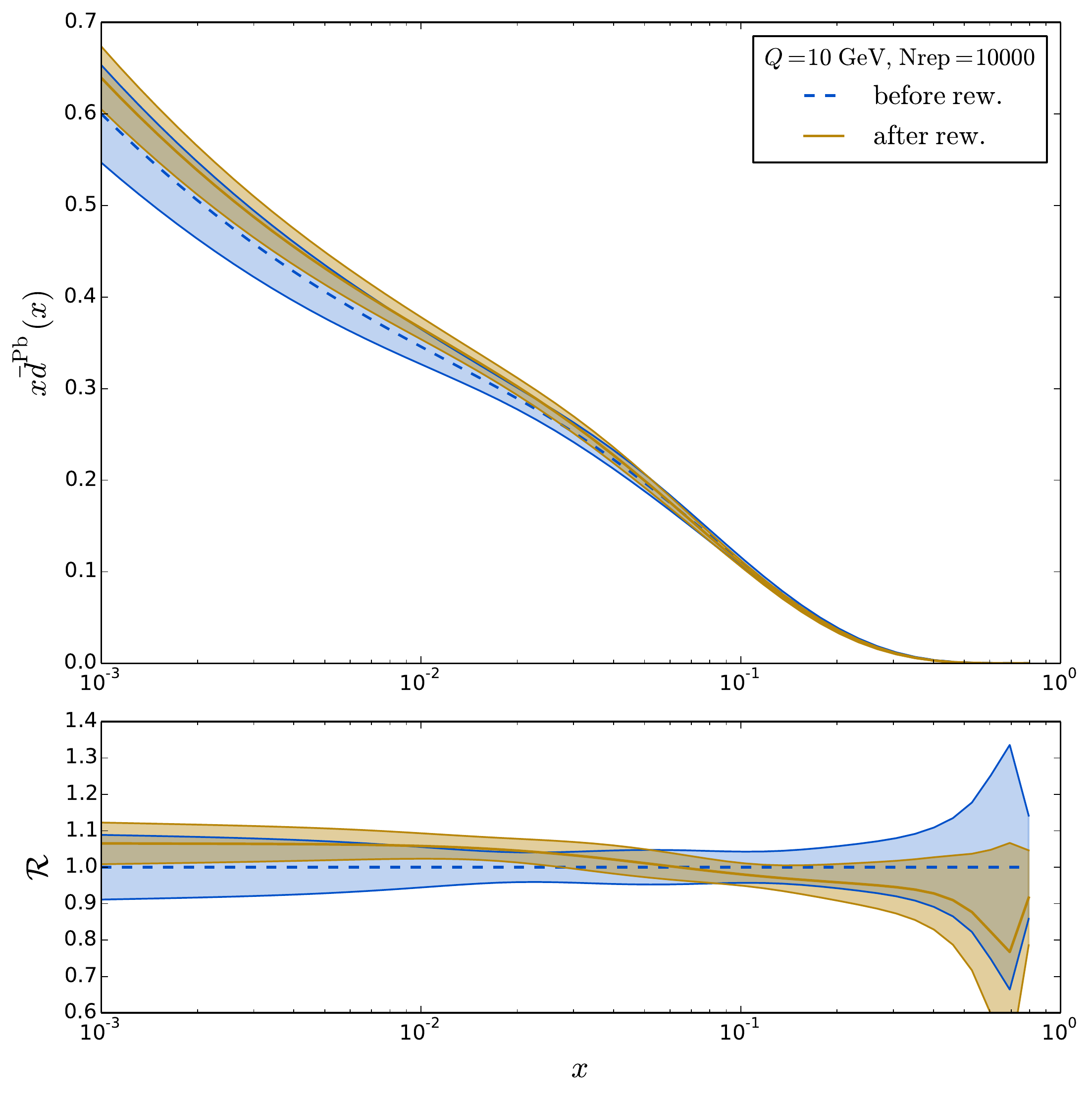}}
\caption{$\bar{u}$ and $\bar{d}$ PDFs before and after the reweighting using all LHC \ppb data sets.}
\label{fig:pdfs_after_before_all_ubardbar}
\end{figure*}
} 
\def\figDefallG{
\begin{figure*}[!htb]
\centering{}
\subfloat[$g$ PDF]{
\includegraphics[width=0.48\textwidth]{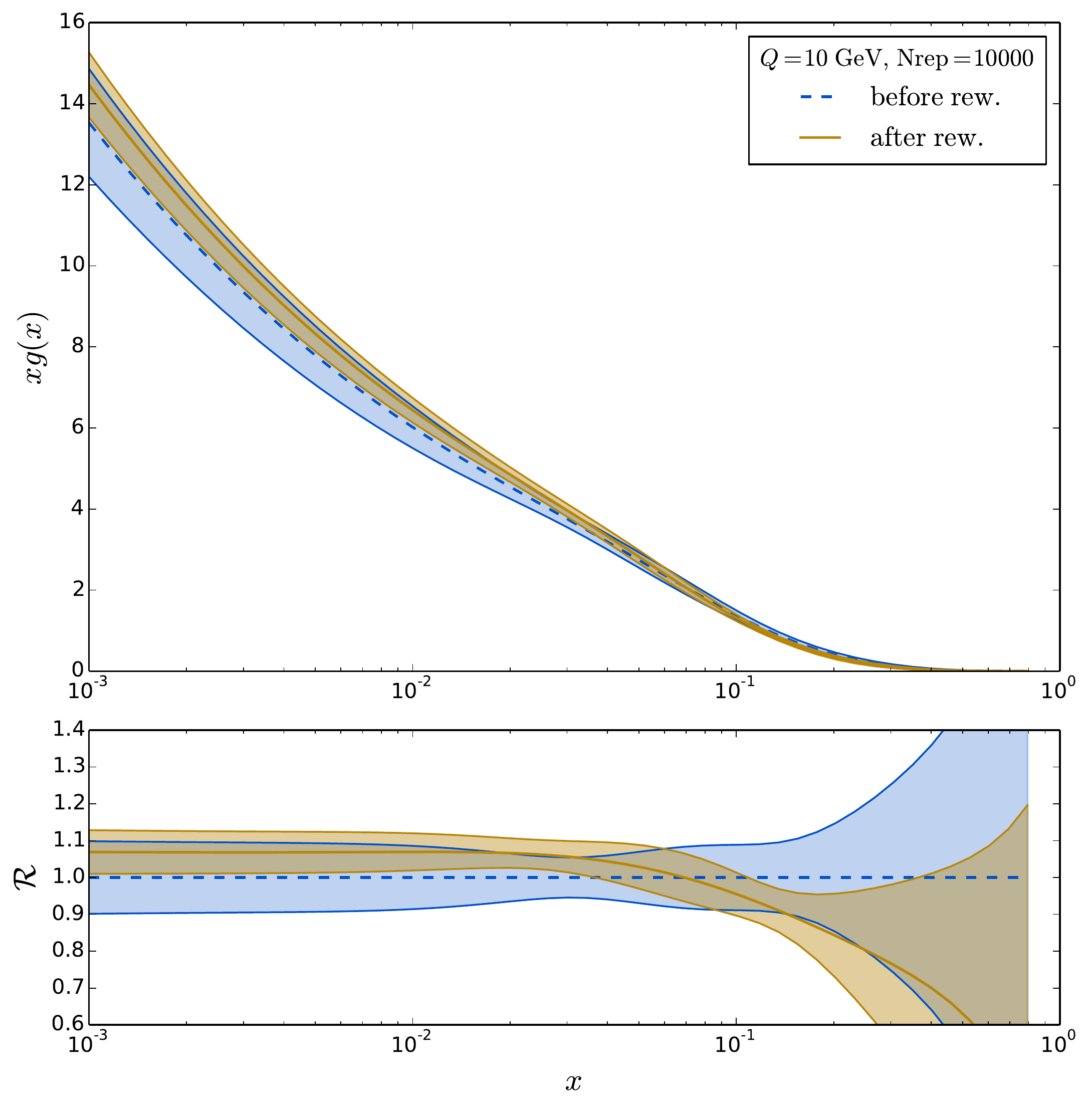}}
\subfloat[$s$ PDF]{
\hfil
\includegraphics[width=0.48\textwidth]{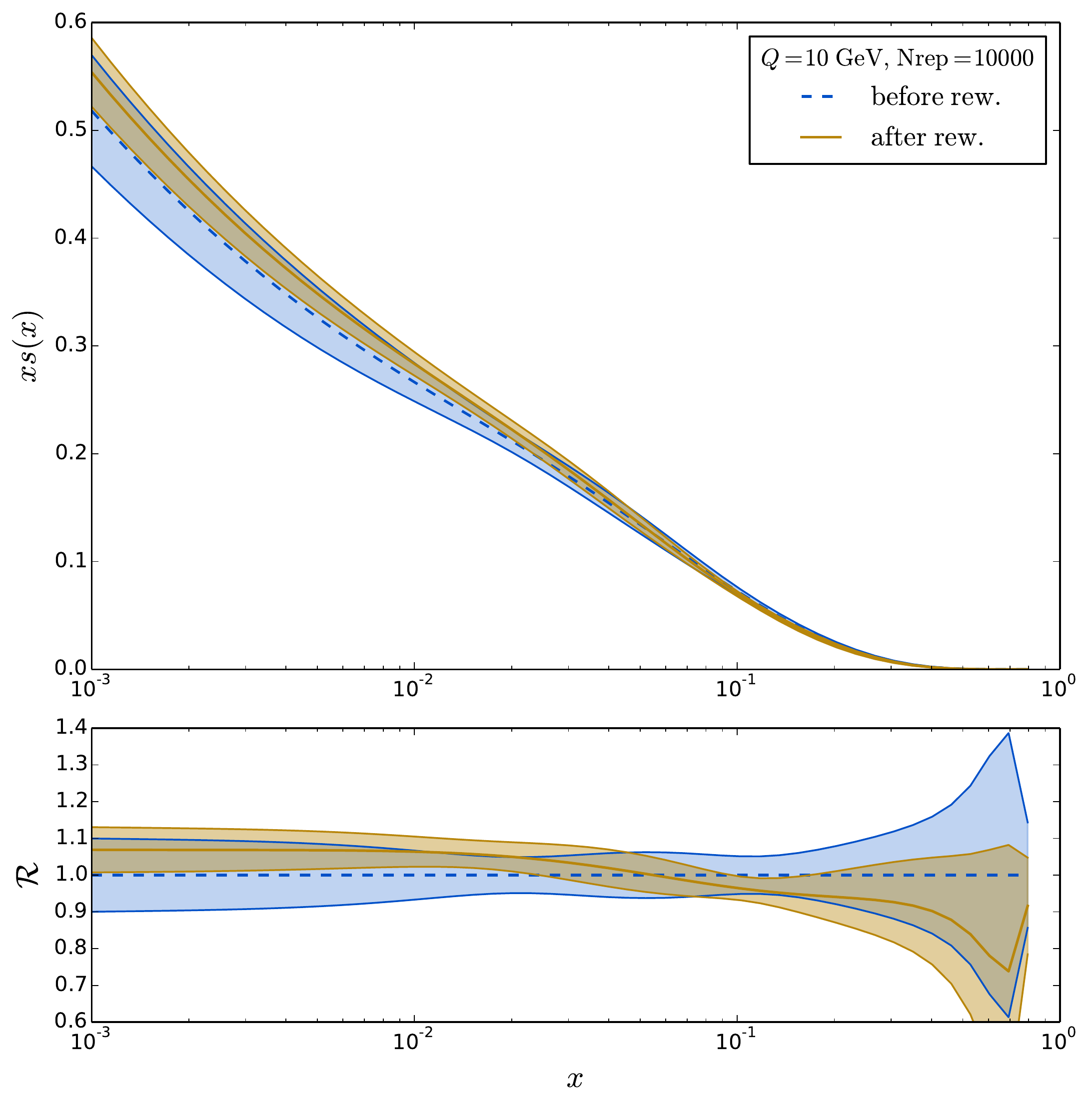}}
\caption{$g$ and $s$ PDFs before and after the reweighting using all LHC \ppb data sets.}
\label{fig:pdfs_after_before_all_g}
\end{figure*}
} 

%
%


\title{Vector boson production in \ppb and PbPb collisions at the LHC and
its impact on \ncteqfit PDFs}

\author{A.~Kusina\thanksref{e1,addr1,addr2}
        \and
        F.~Lyonnet\thanksref{e2,addr3}
        \and
        D.~B.~Clark\thanksref{e3,addr3}
        \and
        E.~Godat\thanksref{e4,addr3}
        \and
        T.~Ježo\thanksref{e5,addr4}
        \and
        K.~Kovařík\thanksref{e6,addr5}
        \and
        F.~I.~Olness\thanksref{e7,addr3}
        \and
        I.~Schienbein\thanksref{e8,addr1}
        \and
        J.~Y.~Yu\thanksref{e9,addr3}
}

\thankstext{e1}{e-mail: kusina@lpsc.in2p3.fr}
\thankstext{e2}{e-mail: flyonnet@smu.edu}
\thankstext{e3}{e-mail: dbclark@smu.edu}
\thankstext{e4}{e-mail: egodat@smu.edu}
\thankstext{e5}{e-mail: tomas.jezo@mib.infn.it}
\thankstext{e6}{e-mail: karol.kovarik@uni-muenster.de}
\thankstext{e7}{e-mail: olness@smu.edu}
\thankstext{e8}{e-mail: ingo.schienbein@lpsc.in2p3.fr}
\thankstext{e9}{e-mail: yu@lpsc.in2p3.fr}

\institute{Laboratoire de Physique Subatomique et de Cosmologie, Université
Grenoble-Alpes, CNRS/IN2P3, 53 avenue des Martyrs, 38026 Grenoble, France\label{addr1}
          \and
          Institute of Nuclear Physics Polish Academy of Sciences,
          PL-31342 Krakow, Poland\label{addr2}
          \and
          Southern Methodist University, Dallas, TX 75275, USA\label{addr3}
          \and
          Physik-Institut, Universit{\"a}t Z{\"u}rich, Winterthurerstrasse 190,
          CH-8057 Z{\"u}rich, Switzerland\label{addr4}
          \and
          Institut f{ü}r Theoretische Physik, Westf{ä}lische Wilhelms-Universit{ä}t
M{ü}nster, Wilhelm-Klemm-Stra{ß}e 9, D-48149 M{ü}nster, Germany\label{addr5}
}

\date{Received: date / Accepted: date}
\setcounter{tocdepth}{4}
\maketitle

\begin{abstract}
We provide a comprehensive comparison of $W^\pm / Z$ vector boson
production data in \ppb and PbPb collisions at the LHC with predictions
obtained using the \ncteqfit PDFs. We identify the measurements
which have the largest potential impact on the PDFs, and estimate
the effect of including these data using a Bayesian reweighting method.
We find this data set can provide information about both the
nuclear corrections and the heavy flavor (strange quark) PDF components.
As for the proton, the parton flavor determination/separation is dependent on nuclear corrections
(from heavy target DIS, for example), this information can also help
improve the proton PDFs.
\end{abstract}
\newpage
\tableofcontents{}
\newpage
%
\section{Introduction \label{sec:intro}}
%
%
%
\tabDefone{}
\figDefkin
\tabDeftwo{}

Vector boson production in hadron collisions is a well understood
process and serves as one of the  ``standard candle'' measurements at the LHC.
$W^{\pm}$ and $Z$ bosons are numerously produced in heavy ion
\ppb and PbPb collisions at the LHC and can be used to gain insight
into the structure of nuclear parton distribution functions (nPDFs).
As the $W^{\pm}$ and $Z$ bosons couple weakly, 
their interaction with the nuclear medium is negligible which makes these
processes one of the cleanest probes of the nuclear structure available
at the LHC. The possibility of using vector boson production data
to constrain nPDFs was previously considered~\cite{Paukkunen:2010qg}, 
and this demonstrated the strong potential for the proton-lead data (especially the asymmetries) 
to constrain the nuclear PDFs. 
The current LHC measurements for  $W^{\pm}$ and $Z$ production include 
rapidity and transverse momentum distributions for both proton-lead (\ppb) and lead-lead (\pbpb)
collisions~\cite{Aad:2015gta,Khachatryan:2015pzs,Aaij:2014pvu,Khachatryan:2015hha,AtlasWpPb,Senosi:2015omk,Aad:2012ew,Chatrchyan:2014csa,Aad:2014bha,Chatrchyan:2012nt}.
Some of these data were already used (along with jet and charged particle
production data) in a recent analysis~\cite{Armesto:2015lrg,Ru:2016wfx} employing a
reweighting method to estimate the impact of these data on EPS09~\cite{Eskola:2009uj}
and DSSZ~\cite{deFlorian:2011fp} nPDFs.%
    \footnote{During the publication process of this study a new global analysis
    including pPb LHC data has been presented~\cite{Eskola:2016oht}.}

The LHC heavy ion $W^\pm /Z$ data set is especially interesting
as it can help to resolve the long-standing dilemma regarding the 
heavy flavor components of the proton PDFs.
Historically, this has been an important issue as
nuclear target data (especially $\nu$-DIS) have been essential in
identifying  the individual parton flavors~\cite{Ball:2014uwa,Harland-Lang:2014zoa,Dulat:2015mca,Khanpour:2016pph};
however, this means that the uncertainties of the heavy flavors are
intimately tied to the (large) nuclear uncertainties. 
The LHC heavy ion $W^\pm /Z$ data has the potential to improve
this situation due to the following two key features. 
First, this data is in a kinematic regime where the heavier quark flavors
(such as strange and charm) contribute substantially.
Second, by comparing  the proton $W^\pm /Z$ data with the heavy ion results
we have an ideal environment to precisely characterize the nuclear corrections.
The combination of the above can not only improve the nuclear PDFs, but also
the proton PDFs which are essential for any LHC study.

In this work we present predictions for vector boson production in
\ppb and PbPb collisions at the LHC obtained using \ncteqfit nuclear
parton distributions, and perform a comprehensive comparison
to the available LHC data. We also identify the measurements which
have the largest potential to constrain the nPDFs, and perform a reweighting
study which allows us to estimate the effects of including these data in
an nPDF fit. 

The rest of the paper is organized as follows.
Sec.~\ref{sec:comp} is devoted to predictions of vector boson production 
at the LHC in nuclear collisions. In particular, 
we provide an overview of the kinematic range probed by the $W^\pm /Z$ 
data and discuss the tools we will use for the calculation. 
Then we present our predictions for \ppb and \pbpb collisions at the LHC and compare them with the experimental data
and other theoretical predictions.
In Sec.~\ref{sec:reweighting} we perform a reweighting  using \ncteqfit distributions
to assess the impact of the nuclear data on the nPDFs. 
Finally, Sec.~\ref{sec:conclusions} summarizes our results and observations.

%
%
%
%

%
\section{$W^\pm / Z$ Production at the LHC \label{sec:comp}}

We begin by presenting our predictions for $W^\pm$ and $Z$ boson production in nuclear collisions at the LHC 
using the recently published \ncteqfit PDFs~\cite{Kovarik:2015cma}.

\subsection{Experimental data and theoretical setup}

For the theoretical calculations in our study we use 
the FEWZ (Fully Exclusive W, Z production)~\cite{Gavin:2010az,Gavin:2012sy}  program version~2.1.
Even though FEWZ can compute $W$ and $Z$ production with decays up to next-to-next-to-leading order,
we work at next-to-leading order (NLO) to be consistent with the order of 
evolution of the nPDFs.\footnote{The CT10 proton PDFs used in the theoretical calculations are also at NLO.}

As FEWZ is designed to handle $pp$ or $p\bar{p}$ collisions, we have extended it so that two different PDF sets can be used 
for the two incoming beams as required for the \ppb collisions.

For the lead PDFs we use the \ncteqfit nPDFs~\cite{Kovarik:2015cma}, 
while we use the CT10 distributions~\cite{Lai:2010vv} for the free protons; the only
exception is the use of MSTW2008 PDFs~\cite{Martin:2009iq} for 
the LHCb $Z$ boson measurement~\cite{Aaij:2014pvu} in order to match the original LHCb publication.
Additionally, we compare these results with predictions calculated using
nuclei made out of free proton PDFs, and in some cases free proton PDFs
supplemented with EPS09 nuclear corrections~\cite{Eskola:2009uj}.

We will consider 
LHC data on $W^{\pm}$ and $Z$ boson production  from the ALICE, ATLAS, CMS, and LHCb experiments. 
The exhaustive list of data sets that we use is provided in Table~\ref{tab:LHCdatasets} along with the experimental
kinematical cuts implemented in the analysis.
While there are measurements for both the rapidity and transverse momentum distributions,
for this study we will focus only on the rapidity measurements. 

Using the transverse momentum ($p_T$) distributions to study the PDFs is more intricate
as it requires resummations in the low $p_T$ region where the cross section is maximal; 
we reserve this for a future study.

In Fig.~\ref{fig:xspace} we display the kinematic space probed by the $W^{\pm}/Z$ production process~\cite{Kusina:2012vh}.
We translate between the $\{x_1, x_2 \}$  and the $\{y,\tau \}$ variables for three values of the collider center of mass (CM) energy,  
$\sqrt{s}$. 
Table~\ref{tab:energy} lists the CM energy per nucleon as a function
of the nominal proton beam energy which is determined from the relation:
\begin{equation}
\sqrt{s_{N_1 N_2}} = \sqrt{s_{p p}} \ \sqrt{\frac{Z_{N_1}}{A_{N_1}}}  \  \sqrt{\frac{Z_{N_2}}{A_{N_2}}} \; ,
\end{equation}
where in case of lead we have $A = 208$ and $Z = 82$.
Additionally for asymmetric collisions there is a rapidity shift,
$\delta y$, between the CM and the laboratory (LAB) frame:
\begin{equation}
\delta y = 
\frac{1}{2} \  
\ln \left[ \frac{E_{N_1}}{E_{N_2}} \right]\, ,
\label{eq:shift}
\end{equation}
and in particular for the case of pPb collisions, $E_{Pb}=(Z_{Pb}/A_{Pb})E_p$
giving $\delta y_{pPb}=\frac{1}{2}\ln\big(\frac{82}{208}\big) \simeq -0.465$, i.e. $y_{CM}=y_{LAB}-0.465$. 

For the asymmetric case of pPb, we use the convention 
where $x_1$ is the proton momentum fraction, and 
 $x_2$ is the lead momentum fraction.
Thus, for pPb at large $y_{CM}$ we have a large proton $x_1$ and 
a small lead $x_2$; conversely, 
at small $y_{CM}$ we have a small proton $x_1$ and 
a large lead $x_2$.

In Fig.~\ref{fig:xspace}, the pair of lines with 
$\sqrt{s}$=2.76~TeV corresponds to PbPb collisions with a beam energy of 3.5~TeV per proton,
and $\sqrt{s}$=5.02~TeV corresponds to \ppb collisions with a beam energy of 4~TeV per proton.

\subsection{Comparison to Proton-Lead (\lowercase{p}P\lowercase{b}) data\label{sec:ppb}}
\figDefdatarange{}
\figDefatlaspPb
\figDefcmspPb
\figDeflhcbpPb

We first consider the LHC \ppb collisions at $\sqrt{s}=5.02$~TeV. 
The distributions are shown in the CM frame, and include the appropriate rapidity shift according to Eq.~\eqref{eq:shift}.
In Fig.~\ref{fig:data_range}, we display the kinematic range of the
\ppb data bins (central values) in the plane $(y, x_2)$ where $y$ is
the rapidity in the CM frame of the relevant vector boson or lepton, and
$x_2$ the lead parton momentum fraction. As expected, there is little
data below $x\sim 10^{-3}$ and most of the constraints from these LHC
data are in the low- to mid-$x$ region.

Figs.~\ref{fig:atlas_z_pPb_comp},~\ref{fig:cms_z_pPb_comp} and
\ref{fig:lhcb_z_pPb_comp} show our predictions for the
ATLAS~\cite{Aad:2015gta}, CMS~\cite{Khachatryan:2015pzs} and
LHCb~\cite{Aaij:2014pvu} $Z$ boson production measurements, 
respectively. In all three cases, results obtained with the \ncteqfit
nPDFs are shown along with those obtained with
a lead nucleus composed by $Z$ protons and $A-Z$ neutrons,
assuming isospin symmetry and using CT10 PDFs;
the ratio of predictions over the data is shown in the lower panel. 
Note that the errors shown
for the \ncteqfit predictions are for nuclear uncertainties only
(and only for the beam with momentum fraction $x_2$) 
which means that the PDF error of the proton beam is not accounted 
for.\footnote{For the symmetric case of PbPb collisions the errors on both 
	beams are taken into account.}
Furthermore, the errors shown for the pPb predictions using lead nuclei constructed from 
CT10 and MSTW2008 proton PDFs are only for the beam with momentum fraction $x_2$. 
By comparing the proton uncertainties (CT10 and MSTW2008) to  
the  nuclear uncertainties, we see that the nuclear uncertainties are much larger.

Examining Figs.~\ref{fig:atlas_z_pPb_comp},~\ref{fig:cms_z_pPb_comp} and
\ref{fig:lhcb_z_pPb_comp},
it is interesting to note the following. 
\begin{enumerate}

\item The data and theory are generally compatible (without significant tension) 
both  with and without nuclear corrections; this situation may change as the 
experimental errors and nuclear uncertainties are reduced. 

\vspace{6pt}

\item 
Focusing on the ATLAS and CMS comparison of Figs.~\ref{fig:atlas_z_pPb_comp} and \ref{fig:cms_z_pPb_comp},
we observe that the distributions peak at negative rapidities
$y_Z \sim -1$.
Referring to Fig.~\ref{fig:xspace},
this corresponds to an enhancement of the $q\bar{q}$ proton-lead luminosity
over the pure proton one in the
$x_2$ region $\sim 0.05$.

\vspace{6pt}

\item 
Focusing on the LHCb data  of Fig.~\ref{fig:lhcb_z_pPb_comp},
we find good agreement for negative $y$, but large differences at positive $y$.
Despite these differences, the large uncertainties will yield a reduced impact 
in our subsequent reweighting procedure. 

\end{enumerate}

We now turn our attention to $W^+$ and $W^-$ production at the LHC.
In Figs.~\ref{fig:cms_wpm_pPb_comp},~\ref{fig:atlas_wpm_pPb_comp} and
\ref{fig:alice_wpm_pPb_comp} we compare the data obtained by CMS~\cite{Khachatryan:2015hha},
ATLAS~\cite{AtlasWpPb} and ALICE~\cite{Senosi:2015omk} for $W^\pm$ production with
theoretical predictions obtained with \ncteqfit and CT10 PDFs.

We find the $W^{-}$
CMS and ATLAS data are adequately described in the negative rapidity range ($y_{\ell^-}<0$),
but the tensions grow as we move to larger rapidity.
This effect is magnified for the case of  $W^{+}$  where we see substantive deviations
at large rapidity ($y_{\ell^+}>1$).
Referring to Fig.~\ref{fig:xspace}, these  deviations are in the smaller 
$x_2$ region ($\sim 3\times 10^{-3}$) where we might expect
nuclear shadowing of the $u\bar{d}$ and $d\bar{u}$ 
luminosities.\footnote{The nuclear correction factors are typically
defined as the ratio of the nuclear quantity to the proton or isoscalar quantity. 
At large $x$ ($\gsim 0.2$) in the EMC region the nuclear quantities are suppressed relative to the proton.
In the intermediate region $x\sim 0.1$ we find ``anti-shadowing'' where the nuclear 
results are enhanced. 
Finally, at smaller $x$ (a few $\times \, 10^{-2}$) we have the ``shadowing'' region
where the nuclear results are suppressed.}
However, this low $x_2$ range is unconstrained by the data
currently used in nPDF fits, so these results
come from an extrapolation of the larger $x_2$ region.
It is interesting to observe that a delayed shadowing (which shifts the
shadowing down to smaller $x_2$ values) would improve the comparison of the
data with the theory in the larger $y_{\ell^\pm}$ region; this type of behavior
was observed in the nuclear corrections extracted from the neutrino-DIS
charged current data~\cite{Kovarik:2010uv,Schienbein:2009kk,Nakamura:2016cnn}.
Taking into account the errors from both the experimental data  and the
theoretical predictions, no definitive conclusions can be drawn at the present time.
Notwithstanding,  this  data  has the  potential to strongly influence the  nPDF fits,
especially in the small $x_2$ region, 
if the uncertainties could be reduced.

Finally,  the ALICE data (Fig.~\ref{fig:alice_wpm_pPb_comp}) currently have large uncertainties,
and we expect they will have  a minimal impact on the reweighting.  

\figDefcmsw
\figDefatlasw
\figDefalicew
\subsection{Comparison to Lead-Lead data\label{sec:pbpb}}
\figDefpbpb
\figDefpbpbw

We now consider the LHC \pbpb collisions at $\sqrt{s}=2.76$~TeV. 
As these beams are symmetric, we now have $y_{CM}=y_{lab}$. Again, we will use \ncteqfit\cite{Kovarik:2015cma} and CT10~\cite{Lai:2010vv} PDFs for the theoretical predictions.
Results from  ATLAS and CMS collaborations are available in the form of either event yields ($Z$ boson production) or charge asymmetries ($A_\ell$).

In Fig.~\ref{fig:cmsatlasZPbPbcomparison_atlas} and \ref{fig:cmsatlasZPbPbcomparison_cms}
we present the comparison of the ATLAS~\cite{Aad:2012ew} and CMS~\cite{Chatrchyan:2014csa}
data with  theoretical predictions with \ncteqfit and CT10 PDFs.
Note that the differential cross sections have been normalized to the total
cross section.
The \pbpb data generally exhibits no tension as the distributions are well described across the kinematical range;
however, this is in part due to the large uncertainties due to two nuclei in the initial state.

The measurement  of charge asymmetries can provide strong constraints 
on the PDF fits as many of the  systematic uncertainties cancel in such ratios.
In Fig.~\ref{fig:cmsatlasWPbPbcomparison} we compute the 
lepton ($\ell=[\mu,e]$) 
charge asymmetry $A_{\ell}(y_{\ell})$:
\begin{equation}
A_{\ell}(y_{\ell}) = 
\frac{
dN(W^+\to \ell^+ \nu_\ell )
-
dN(W^-\to \ell^- \bar{\nu}_\ell )
}
{
dN(W^+\to \ell^+ \nu_\ell )
+
dN(W^-\to \ell^- \bar{\nu}_\ell )
}
\end{equation}
for $W^{+}$ and $W^{-}$ bosons as measured by the ATLAS~\cite{Aad:2014bha} and CMS~\cite{Chatrchyan:2012nt} experiments.
Unfortunately, it appears that the dependence on the nuclear corrections largely cancels  out
in the ratio as the nuclear  \ncteqfit result is indistinguishable from the CT10 proton result. 
Hence, these charge asymmetry ratios cannot constrain the nuclear corrections at the present time.

\subsection{$W^\pm/Z$ Cross Section Correlations}
\label{sec:corr}

\figDefxsecww
\figDefxseczw
\figDefxsecwwF
\figDefxsecwwFF
\figDefcmswrap

In order to analyze our results more quantitatively, it is very useful to look at PDF correlations. In particular, we are interested
in assessing the importance of the strange quark in our results. We first review some standard definitions before presenting our
analysis.

The definition of the correlation cosine of two PDF-dependent
observables $X$ and $Y$ is~\cite{Nadolsky:2008zw} 
\begin{equation}
\begin{split}
\label{eq:cosPhi}
\cos\phi & =\frac{\vec{\nabla}X\cdot\vec{\nabla}Y}{\Delta X\Delta Y}\\
 & =\frac{1}{4\Delta X\Delta Y} \sum_i \left(X_{i}^{(+)}-X_{i}^{(-)}\right)\left(Y_{i}^{(+)}-Y_{i}^{(-)}\right)\, ,
\end{split}
\end{equation}
where $\Delta X$ is the PDF error of the corresponding observable. 
For the \ncteqfit\ PDFs this corresponds to the symmetric error given by
\begin{equation}
\label{eq:Delta}
\Delta X=\frac{1}{2}\sqrt{\sum_{i}^{N}\left(X_{i}^{(+)}-X_{i}^{(-)}\right)^{2}}\quad .
\end{equation}
$X_{i}^{(\pm)}$ is the observable evaluated along the $\pm$ error PDF
eigenvector $i$, and the summation runs over all eigenvector directions. 

In our case we are interested in observables $X,Y\in\{\sigma_{Z},\sigma_{W^+},\,\sigma_{W^-}\}$. 
Here, we focus on the planes formed by the
($W^+$, $W^-$) and the ($Z$, $W^\pm$) boson production cross sections
to visualize the correlations.

Fig.~\ref{fig:wpwmCrossSecCorrelations} shows the correlations of the
$W^{+}$ and $W^{-}$ production cross sections for \ppb collisions at
the LHC in comparison with the CMS and ATLAS measurements. 
Similarly, in Fig.~\ref{fig:wzCrossSecCorrelations} we display the results for
$Z$ and $W^{\pm}$ bosons.
The results are shown
for three different rapidity regions, $y<-1,\ |y|<1,\ y>1$, and 
for several PDFs sets.
For the proton side we always use the CT10 PDFs
and for the lead side we examine three  results:
i) \ncteqfit, 
ii) CT10,
and  
iii) CT10 PDFs
supplemented by the nuclear corrections from EPS09
(CT10+EPS09). 
Finally, the central predictions are supplemented with
uncertainty ellipses illustrating correlations between the cross
sections.
The ellipses are calculated in the following way~\cite{Nadolsky:2008zw},  
\begin{equation}
\begin{split}
X &= X_0 + \Delta X \cos\theta,\\
Y &= Y_0 + \Delta Y \cos(\theta+\phi),
\end{split}
\end{equation}
where $X$, $Y$ represent PDF-dependent observables,
$X_0$ ($Y_0$) is the observable calculated with the central PDF, $\Delta X$ ($\Delta Y$)
is defined in Eq.~\eqref{eq:Delta}, $\phi$ is the correlation angle defined in
Eq.~\eqref{eq:cosPhi}, and $\theta$ is a parameter ranging between $0$ and $2\pi$.

From Figs.\ \ref{fig:wpwmCrossSecCorrelations} and \ref{fig:wzCrossSecCorrelations}
one can generally observe that the ellipses for the different PDF sets overlap.
Furthermore, the central predictions for all three PDF sets lie in the overlapping area
of the three ellipses.
However, a trend can be observed as a function of the rapidity:
\begin{enumerate}

\item For negative rapidities ($y<-1$), the central predictions from the
nuclear PDFs (nCTEQ15, EPS09) are closer to the experimental data as they yield larger cross sections than the  uncorrected (proton) CT10 PDFs.
This can be understood because the lead $x_2$ values probed in this rapidity bin lie in the 
region $x_2 \sim 10^{-1}$  where the nPDFs are enhanced due to anti-shadowing (cf., Fig.~9 in Ref.~\cite{Kovarik:2015cma}).
Due to the larger uncertainties associated with the \ncteqfit predictions, the ATLAS and CMS 
cross sections lie within the  1$\sigma$ ellipse. 
Conversely, the measured data lie outside the uncorrected (proton) CT10 error ellipsis.

\vspace{6pt}

\item For the central rapidity bin ($|y|<1$), the  predictions
from all three PDF sets lie generally very close together.
In this case, the probed $x_2$ values lie in the range $0.007 \le x_2 \le 0.05$
which is in the transition zone from the anti-shadowing to the shadowing 
region.
We find the LHC  $W^+$ and $W^-$ cross sections in Fig.~\ref{fig:wpwmCrossSecCorrelations}
tend to lie above  the theory predictions. 
Examining the $Z$ cross section of Fig.~\ref{fig:wzCrossSecCorrelations},  we  find 
the CMS data agrees closely with  the theory predictions, 
while the ATLAS data is larger by approximately 1$\sigma$.

\vspace{6pt}

\item For the positive rapidity bin ($y>1$),
we find  the central predictions from CT10 match the $W^\pm$ data very closely, 
but slightly overshoot the $Z$ data. 
The nuclear PDFs (nCTEQ15, EPS09) undershoot the  $W^\pm$ data by a bit more than 1$\sigma$, 
but agree with the  $Z$ cross section within 1$\sigma$. 
Here, the probed $x_2$ values are $\lesssim 0.007$; in this region the lead PDFs 
are poorly constrained and  the corresponding
cross sections are dependent on  extrapolations of the PDF parameterization in this region.
\end{enumerate}
Interpreting the above set of results appears  complicated,
so we will  try and break the problem down in to smaller components. 
We now compute the same results as above, but using only 
2 flavors (one family) of quarks: $\{u,d\}$;  
specifically, these plots are produced by zeroing the heavy
flavor components $(s,c,b)$, but keeping $(u,d)$ and the gluon.
For the $Z$ production this eliminates the $s\bar{s}$ and (the
smaller) $c\bar{c}$ contributions, while for $W^+/W^-$ production it is the
$\bar{s}c/s\bar{c}$ contribution which drives the change.
While the charm PDF does play a role in the above (the bottom contribution is minimal),
$c(x)$ is  generated radiatively by the process $g\to c \bar{c}$  (we assume no intrinsic component);
thus, it is essentially determined by the charm mass value and the gluon PDF.
In contrast, the ``intrinsic'' nature of the strange PDF leads to its comparably large uncertainties.
For example, if we compare the free-proton PDF baselines (CTEQ6.1, CT10), 
the strange quark exhibits substantial differences while the 
charm (and bottom) distributions are quite similar; 
this pattern then feeds into the nPDFs. 
Therefore, the  strange quark PDF will be the primary focus of the following discussion.

In  Figs.~\ref{fig:wzCrossSecCor1F} and \ref{fig:wzCrossSecCor1F_bins}
we compare the 5~flavor and 2~flavor results using
nCTEQ15, CT10+EPS09,  CTEQ6.1+ EPS09, and CT10.
We have added  CTEQ6.1+EPS09 as CTEQ6.1 was the baseline used for the EPS09 fit.

Examining  Fig.~\ref{fig:wzCrossSecCor1F}, 
the shift of the 2~flavor results compared to the 5~flavor results
can be as large as $\sim$30\% and reflects,
the contributions of the strange and charm quarks.

For the 5~flavor case ($\blacktriangle$), the calculations are scattered to the low side of the data 
in both $W^+$ and $W^-$.
The CT10 result is the closest to the data, but due to the larger uncertainties of nCTEQ15, 
the data point is  within range of both of their ellipses.
We also observe that the  CT10+EPS09 and CTEQ6.1+EPS09 results bracket the nCTEQ15 value;
again, this is due to the very different strange PDF associated with CT10 and CTEQ6.1.

For the 2~flavor case ($\bullet$), all the nuclear results (nCTEQ15,
CT10+EPS09, CTEQ6.1+EPS09) coalesce, and they are distinct from the
non-nuclear result (CT10).
This pattern suggests that the nuclear corrections of nCTEQ15 and
EPS09 for the $\{u,d \}$ flavors are quite similar, and the spread
observed in the 5~flavor case comes from differences of $s(x)$ in
the underlying base PDF.
Thus we infer that the difference between the nuclear results and the proton result
 accurately represents the nuclear corrections for the 2~flavor case (for  $\{u,d\}$), 
but  for the 5~flavor case it is a mix of nuclear corrections and variations of the underlying sea quarks.

Fig.~\ref{fig:wzCrossSecCor1F_bins} displays the  same information as Fig.~\ref{fig:wzCrossSecCor1F} 
except it is divided into rapidity bins.
As we move from negative $y$ to positive $y$ we move from
high $x$ where the nPDFs are well constrained to
small $x$ where the nPDFs have large uncertainties
({\it cf.},  Fig.~\ref{fig:data_range}).
Thus, it is encouraging that at $y < -1$ we uniformly find the nuclear predictions
yield larger cross sections than the  proton results (without nuclear corrections)
and thus
lie closer to the LHC data.

Conversely, for   $y > 1$ we find the  nuclear predictions
yield smaller cross sections than the  proton results. The comparison
with the LHC data varies across the figures, but
this situation suggests a number of possibilities. 

First, the large nPDF uncertainties in this small $x_2$ region could be improved
using the LHC data.

Second, the lower nPDF cross sections are partly due to the nuclear shadowing
in the small $x$ region; if, for example, this shadowing region were shifted to even
lower $x$ values, this would increase the nuclear results. 
Such a shift was  observed in Refs.~\cite{Kovarik:2010uv,Schienbein:2009kk,Nakamura:2016cnn}
using charged current neutrino-DIS data, and this would move the nuclear predictions
of Fig.~\ref{fig:wpwmCrossSecCorrelations} at $y>1$ toward the LHC data. 

Finally, we note that  measurements of the strange quark asymmetry~\cite{Mason:2007zz}
indicate that $s(x)\not=\bar{s}(s)$ which is unlike what is used in the current nPDFs;
this would influence the $W^\pm/Z$ cross sections separately
as (at leading-order)~\cite{Kusina:2012vh} 
$W^+\sim \bar{s}c$,
$W^-\sim s\bar{c}$,
and 
$Z\sim \bar{s}s$. 
As the strange PDF has a large impact on the  $W^\pm/Z$  measurements,
this observation could provide incisive information on the
individual $s$ and $\bar{s}$ distributions.

These points are further exemplified in Fig.~\ref{fig:rap_distr_str}
which displays $W^\pm$ production for both 2 and 5 flavors
as a function of lepton rapidity $y_{\ell^\pm}$.
For large $y_{\ell^\pm}$, (small lead $x_2$) the CT10 proton result
separates from the collective nuclear results; presumably, this
is due to the nuclear shadowing at small $x_2$. 
Again, we note
that in this small $x_2$ region there are minimal experimental constraints
and the nPDFs come largely from extrapolation at higher $x_2$ values. 
Additionally, by comparing the 2 and 5 flavor results, we
clearly see the impact of the heavier flavors, predominantly the strange quark PDF.

Furthermore, different strange quark PDFs in the baseline PDFs compared in 
Figs.~\ref{fig:wpwmCrossSecCorrelations} and \ref{fig:wzCrossSecCorrelations}, 
make it challenging to distinguish nuclear effects from different strange 
quark distributions. 
Thus, we find that the extraction of the nuclear corrections is intimately intertwined with 
the extraction of the proton strange PDF, and we must be careful to separately distinguish each of these effects. 
Fortunately, the above observations can help us disentangle these two effects.

\section{Reweighting\label{sec:reweighting}}
\figDefHessGlue

In this section we perform a reweighting study to estimate the possible
impact of the $W^\pm /Z$  data on \ncteqfit lead PDFs. For
this purpose we will use only the \ppb data sets. 

We refrain from using
PbPb data as typically the agreement of these data with current nPDFs
is much better (in part due to the large uncertainties), so the impact in the reweighting analysis will be minimal. Secondly the factorization in lead-lead
collisions is not firmly established theoretically~\cite{Qiu:2003cg} such that the interpretation may be complicated.

\subsection{Basics of PDF reweighting}

In this section we summarize the PDF reweighting technique
and  provide formulas for our specific
implementation of this method. 
Additional details can be found in the literature~\cite{Giele:1998gw,Ball:2010gb,Ball:2011gg,Sato:2013ika,Paukkunen:2014zia}.

In preparation for the reweighting, we need to convert
the \ncteqfit set of Hessian error PDFs into a set of 
PDF replicas~\cite{Watt:2012tq,Armesto:2015lrg}
which serve as a representation of the underlying probability distribution.
The PDF replicas can be defined by a simple formula,%
    \footnote{A detailed discussion on the construction of replicas from Hessian
    PDF sets in the case of asymmetric errors can be found in ref.~\cite{Hou:2016sho}.}
\begin{equation}
f_{k}=f_{0}+\sum_{i=1}^{N}\frac{f_{i}^{(+)}-f_{i}^{(-)}}{2}R_{ki},\label{eq:rep-gen}
\end{equation}
where $f_{0}$ represents the best fit (central) PDF, $f_{i}^{(+)}$
and $f_{i}^{(-)}$ are the plus and minus error PDFs corresponding to
the eigenvector direction $i$, and $N$ is the number of eigenvectors
defining the Hessian error PDFs. Finally, $R_{ki}$ is a random number
from a Gaussian distribution centered at 0 with standard deviation
of 1, which is different for each replica~($k$) and each eigen-direction~($i$).

After producing the replicas, we can calculate the average and variance
of any PDF-dependent observable as moments of the probability distribution:
\begin{equation}
\begin{split}\left<\ord\right> & =\frac{1}{N_{\text{rep}}}\sum_{k=1}^{N_{\text{rep}}}\ord(f_{k}),\\
\delta\left<\ord\right> & =\sqrt{\frac{1}{N_{\text{rep}}}\sum_{k=1}^{N_{\text{rep}}}\left(\ord(f_{k})-\left<\ord\right>\right)^{2}}
\quad  .
\end{split}
\end{equation}
In particular, it can be done for the PDFs themselves; we should
be able to reproduce our central PDF $f_{0}$ by the average $\left<f\right>$, 
and the (68\% c.l.) Hessian error bands $\Delta f=\frac{1}{2}\sqrt{\sum_{i}^{N}(f_{i}^{(+)}-f_{i}^{(-)})^{2}}$
by the corresponding variance $\delta\left<f\right>$. Of course,
the precision at which we are able to reproduce Hessian central PDFs
and corresponding uncertainties depends on how well we reproduce the
underlying probability distribution, and this will depend on the number
of replicas, $N_{\text{rep}}$, we use. In the following we use $N_{\text{rep}}=10^{4}$
which allows for a very good reproduction of both central and error
PDFs (within $\sim~0.1\%$ or better). 

We note here that since the \ncteqfit error PDFs correspond to the 90\% confidence
level (c.l.) we need to convert the obtained uncertainties such that they
correspond to the 68\% c.l.\footnote{The 68\% c.l. is necessary to correspond with the
variance of the PDF set defined below.} 
The conversion is done using the following approximate
relation between the 68\% c.l. and 90\% c.l. Hessian 
uncertainties:
$\Delta^H_{90}\ord \approx 1.645\, \Delta^H_{68}\ord$.%

In Fig.~\ref{fig:rep-vs-hess} we perform the above exercise and determine 
if our procedure is self consistent. 
Specifically,  in Fig.~\ref{fig:rep-vs-hess-a} we display 
the central value and uncertainty bands for the 
original gluon PDF and those generated from the replicas; they are indistinguishable. 
Additionally, in Fig.~\ref{fig:rep-vs-hess-b} we demonstrate the convergence of the
average of replicas to the central Hessian PDF for $N_{rep}=\{10^2,10^3,10^4\}$.
For $N_{rep}=10^4$ the central gluon is reproduced to better than 1\% except at the  
highest $x$ values. This is certainly a sufficient  accuracy considering the size of the PDF errors. 
Even the $N_{rep}=10^2$ and  $N_{rep}=10^3$ replicas yield good results except
at larger $x$ ($\gsim 0.1$) where the PDFs are vanishing and the uncertainties are large. 
Since our computational cost will be mostly dictated by the number
of Hessian error PDFs, we will use $N_{rep}=10^4$  to get a better representation
of the underlying probability distribution. 

Having defined the replicas we can apply the reweighting technique
to estimate the importance of a given data set on our current PDFs. The
idea is based on Bayes theorem which states that the posterior distribution
representing the probability of a hypothesis (new probability distribution
representing the PDFs if we would perform a fit including the new data
set we are using in the reweighting) is a product of the prior probability
(PDFs without the new data set) and an appropriate likelihood function.
This allows us to assign a weight to each of the replicas generated earlier
according to eq.~\eqref{eq:rep-gen}.

In the context of Hessian PDFs using a global tolerance criterion the appropriate
weight definition is given by a modified Giele-Keller
expression~\cite{Giele:1998gw,Sato:2013ika,Paukkunen:2014zia,Armesto:2015lrg}%
    \footnote{In the context of Monte Carlo PDF sets a NNPDF weight definition
    should be used~\cite{Ball:2011gg}.}
\begin{equation}
w_{k}=\frac{e^{-\frac{1}{2}\chi_{k}^{2}/T}}
           {\frac{1}{N_{\text{rep}}}\sum_{i}^{N_{\text{rep}}}e^{-\frac{1}{2}\chi_{i}^{2}/T}},
\label{eq:GKweight}
\end{equation}
where $T$ is the tolerance criterion used when defining Hessian error
PDFs%
    \footnote{In the case of the \ncteqfit PDFs, the tolerance criterion is $T=35$ which
    corresponds to a 90\% c.l.,
    the detailed explanation on how it was defined can be found in appendix A of~\cite{Kovarik:2015cma}.
    The tolerance factor used in this analysis corresponds to the
    68\% c.l. which we obtain by rescaling the above:  $T \approx 35/1.645^2 \sim 13$.} 
and $\chi_{k}^{2}$ represents the $\chi^{2}$ of the data sets considered
in the reweighting procedure for a given replica $k$. 
The \ppb $W^\pm$ and $Z$ data  do not provide correlated errors
(the published errors are a sum of statistical and systematic errors in quadrature)%
    \footnote{In our analysis we also add the normalization errors in quadrature to
    the statistical and systematic ones.}
so it is sufficient for our analysis
to use a basic definition of the $\chi^{2}$ function given by
\begin{equation}
\chi_{k}^{2}=\sum_{j}^{N_{\text{data}}}\frac{(D_{j}-T_{j}^{k})^{2}}{\sigma_{j}^{2}},
\end{equation}
where index $j$ runs over all data points in the data set(s),
$N_{\text{data}}$ is the total number of data points, 
$D_{j}$ is the experimental measurement at point $j$, 
$\sigma_{j}$ is the corresponding experimental uncertainty,
and $T_{j}^{k}$ is the
corresponding theoretical prediction calculated with PDFs given by
replica $k$.

With the above prescription we can now calculate the weights needed for the  
reweighting procedure. 
The  expectation value and variance of any PDF-dependent observable can now
be computed in terms of weighted sums:
\begin{equation}
\begin{split}\left<\ord\right>_{\text{new}} & =\frac{1}{N_{\text{rep}}}\sum_{k=1}^{N_{\text{rep}}}w_{k}\ord(f_{k}),\\
\delta\left<\ord\right>_{\text{new}} & =\sqrt{\frac{1}{N_{\text{rep}}}\sum_{k=1}^{N_{\text{rep}}}w_{k}\left(\ord(f_{k})-\left<\ord\right>_{\text{new}}\right)^{2}}\quad .
\end{split}
\label{eq:ave-rew}
\end{equation}

For our reweighting analysis we will only use  the \ppb data sets. 
Because the uncertainty of the nuclear PDFs dominates the proton PDFs, 
it is sufficient to only vary the lead PDFs.
Consequently,  the \ppb cross sections
are linear in the lead  uncertainties, and we can compute the reweighting 
by evaluating cross sections only on the Hessian error PDFs (32+1 in case
of \ncteqfit) instead of the individual replicas 
($N_{\text{rep}}=10^{4}$)
\begin{equation}
\sigma_{k}=f^{\text{p}}\otimes\hat{\sigma}\otimes\left[f_{0}^{\text{Pb}}+\sum_{i}^{N}\frac{f_{i}^{\text{Pb}(+)}-f_{i}^{\text{Pb}(-)}}{2}R_{ki}\right].
\end{equation}
A similar decomposition can be used for pp or PbPb data
    to reduce the number of necessary evaluations. However, because 
    of the quadratic dependence on the PDFs, the reduction is smaller
    and does not necessarily
    lead to lower computational costs.

We will compare the $\chi^{2}$ for each 
experiment calculated with the initial PDFs (before reweighting) and
with the PDFs after the reweighting procedure; this will allow us 
to  estimate the impact of each individual data set. 
 We do this using the following
formula 
\begin{equation}
\chi^{2}=\sum_{j}^{N_{\text{data}}}\frac{\left(D_{j}-\left<T_{j}\right>\right)^{2}}{\sigma_{j}^{2}},
\end{equation}
where $\left<T_{j}\right>$ is a theory prediction calculated as an
average over the (reweighted or not-reweighted) replicas according to
eq.~\eqref{eq:ave-rew} (with or without weights).

Finally, the effectiveness of the reweighting procedure can be (qualitatively) estimated 
by computing the  \textit{effective number of replicas} defined as~\cite{Ball:2011gg}:
\begin{equation}
N_{\text{eff}}=\exp\left[\frac{1}{N_{\text{rep}}}\sum_{k=1}^{N_{\text{rep}}}w_{k}\ln(N_{\text{rep}/w_{k}})\right]\ .
\label{eq:neff}
\end{equation}
$N_{\text{eff}}$ provides a measure of how many of the replica sets are effectively 
contributing to the reweighting procedure.
By definition, $N_{\text{eff}}$ is restricted to be smaller than $N_{\text{rep}}$.
However, when $N_{\text{eff}}\ll N_{\text{rep}}$ it indicates that
there are many replicas whose new weight (after the reweighting procedure)
is sufficiently small that they provide a negligible contribution to the 
updated probability density. This typically happens when
the new data is not compatible with the data used in the original
fit, or if the new data introduces substantial new information; in both 
cases, the procedure becomes ineffective and a new global fit is recommended.

\subsection{Reweighting using CMS $W^{\pm}$ rapidity distributions\label{sec:rewCMSwpm}}
\figDefRWcmsW
\figDefTheoCMSw
\figDefTheoCMSWii

As an example, we consider the reweighting using the CMS $W^{\pm}$
production data from \ppb collisions~\cite{Khachatryan:2015hha}. In this
example we use rapidity distributions of charged leptons originating
from the decay of both $W^{+}$ and $W^{-}$ bosons with $N_{\text{r}ep}=10^{4}$
replicas leading to $N_{\text{eff}}=5913$.

In Fig.~\ref{fig:weights_cmsW} we display  
the distribution of the weights obtained from the
reweighting procedure. We  see 
that the magnitudes of the  weights are reasonable; 
they extend up to $\sim9$ with a peak at the lowest bin. 
It will be useful to compare this distribution with later 
results using  different observables and  data sets. 

In Fig.~\ref{fig:theo_after_before_cmsW} we show the comparison of the data
to theory  before and after the reweighting procedure.%
    \footnote{We note here the difference of PDF uncertainties compared
    to the plots presented in Sec.~\ref{sec:comp}; this is caused by the
    fact that now we use the 68\% c.l. errors whereas in Sec.~\ref{sec:comp}
    we have used the 90\% c.l. errors that are provided with the \ncteqfit PDFs.
    This holds for all plots in Sec.~\ref{sec:reweighting}.}
As expected, we see that after the reweighting procedure
the description of the data is improved.
This is true for both the $W^{+}$ (left figure) and $W^{-}$ (right figure) cases.
We can quantify  the improvement of the fit by examining the $\chi^{2}/N_{\text{data}}$
for the individual distributions.
For the $W^{+}$ case,  
the  $\chi^{2}/N_{\text{data}}$ is improved from $5.07$ 
before  reweighting to $3.23$ after reweighting. 
Similarly,  for  $W^{-}$ the $\chi^{2}/N_{\text{data}}$ is improved from 
$4.57$ to $3.44$. 
The amount of change due to the reweighting procedure
should be proportional to the
experimental uncertainties of the incorporated data;  this is 
the same as we would expect from a global fit. 
For  $W^{\pm}$ production investigated here, 
the uncertainties are quite substantial, 
and the effects are compounded by the lack of  correlated errors.

Finally, we show the effect of the reweighting on the PDFs themselves.
In Fig.~\ref{fig:pdfs_after_before_cmsW_10}, 
we display PDFs for the up quark and gluon at a scale of $Q=10$~GeV.
We can see that the reweighting has the largest effects in the low $x$ region, 
and this holds also for the other flavors  as well. Generally the effects at intermediate
and large $x$ values are limited, with the exception of the gluon which is poorly constrained 
and exhibits a substantial change for large~$x$.

In Figs.~\ref{fig:theo_after_before_cmsW} and~\ref{fig:pdfs_after_before_cmsW_10},
in addition to the reweighting results, we also show results calculated using the
Hessian profiling method~\cite{Paukkunen:2014zia}.
The Hessian profiling should agree precisely with our reweighting calculations, 
and this can serve as an independent cross-check of our results.
Indeed, in the figures we observe that the profiling exactly matches the reweighted results. 
In the following figures we will display only the reweighting results, but in all
presented cases we have checked that these two methods agree.

\vskip 0.2in
\subsection{Using Asymmetries instead of differential cross sections \label{sec:CMSwpmAsym}}
\figDefcmsAsym
\figDefcmsAsymii
\figDefcmsAsymiii

In this section we will re-investigate the reweighting analysis from the 
previous section employing the CMS $W^{\pm}$ production data. 
Instead of using rapidity distributions
(as in the previous section),
we will use two types of asymmetries which are constructed with the charged leptons.
The lepton charge asymmetry is
\begin{equation}
A_{\ell}(y_{\ell})=\frac{dN_{l^{+}}-dN_{l^{-}}}{dN_{l^{+}}+dN_{l^{-}}}\ ,\label{eq:chrgAsym}
\end{equation}
and is defined per bin in the rapidity of the charged lepton where
$N_{l^{\pm}}$ represents the corresponding number of observed events
in a given bin. For the purpose of the theory calculation, $N_{l^{\pm}}$
will be replaced by the corresponding cross-section in a given bin.

It is useful to consider the expression for the charge asymmetry at leading
order in the parton model assuming a diagonal CKM matrix:

 \begin{widetext} 
\begin{eqnarray}
A_\ell &=& \frac{u(x_1) \bar d(x_2) + \bar d(x_1) u(x_2) + c(x_1) \bar s(x_2) + \bar s(x_1) c(x_2)
- \bar u(x_1) d(x_2)- d(x_1) \bar u(x_2) - \bar c(x_1) s(x_2) - s(x_1) \bar c(x_2)}
{u(x_1) \bar d(x_2) + \bar d(x_1) u(x_2) + c(x_1) \bar s(x_2) + \bar s(x_1) c(x_2)
+ \bar u(x_1) d(x_2)+ d(x_1) \bar u(x_2) + \bar c(x_1) s(x_2) + s(x_1) \bar c(x_2)} \ .
\nonumber \\
\null
\end{eqnarray}
Here, the partons with momentum fraction $x_1$ are in the proton, and those with momentum fraction $x_2$ are
inside the lead.
At large negative rapidities (small $x_1$, large $x_2$), we have $f(x_1) = \bar f(x_1)$ for all parton flavors
($f=u,d,s,c$) and the expression for the asymmetry simplifies to the following form
\begin{equation}
A_\ell \to \frac{d(x_1) u_v(x_2) - u(x_1) d_v(x_2) - c(x_1) s_v(x_2) + s(x_1) c_v(x_2)}
{d(x_1) u_v(x_2) + u(x_1) d_v(x_2) + c(x_1) s_v(x_2) + s(x_1) c_v(x_2)}\, .
\end{equation}
Assuming $c_v(x_2) = c(x_2) - \bar{c}(x_2) = 0$ and  $s_v(x_2) = s(x_2) - \bar s(x_2) = 0$, 
as it is the case in all the existing nPDF sets, 
the expression further simplifies
\begin{equation}
A_\ell \to \frac{d(x_1) u_v(x_2) - u(x_1) d_v(x_2)}{d(x_1) u_v(x_2) + u(x_1) d_v(x_2)}
\simeq \frac{u_v(x_2) - d_v(x_2)}{u_v(x_2) + d_v(x_2)}\, .
\end{equation}
In the last equation, we have used the fact that the small $x_1$ up and down PDFs are very similar~\cite{Arleo:2015dba}.
Since the $d_v^{\rm Pb}(x_2) > u_v^{\rm Pb}(x_2)$, we expect the asymmetry to be negative at large negative rapidities.
One can also observe that the asymmetry calculated with either $n_f = 2$ or $n_f=5$ will be the same.
A non-zero strange asymmetry ($s(x_2) > \bar s(x_2)$) would lead to a decrease of the $A_\ell$ asymmetry, thereby
improving the description of the CMS data.

Conversely, at large positive rapidities (small $x_2$, large $x_1$), we have $f(x_2) = \bar f(x_2)$ 
for all parton flavors ($f=u,d,s,c$) and the expression for the asymmetry becomes
\begin{equation}
A_\ell \to \frac{u_v(x_1) d(x_2) - d_v(x_1) u(x_2) + c_v(x_1) s(x_2) - s_v(x_1) c(x_2)}
{u_v(x_1) d(x_2) + d_v(x_1) u(x_2) + c_v(x_1) s(x_2) + s_v(x_1) c(x_2)}\, .
\end{equation}
Again, assuming $c(x_1) = \bar c(x_1)$ and $s(x_1) = \bar s(x_1)$, this expression further simplifies to 
\begin{equation}
A_\ell \to \frac{u_v(x_1) d(x_2) - d_v(x_1) u(x_2)}{u_v(x_1) d(x_2) + d_v(x_1) u(x_2)}
\simeq \frac{u_v(x_1) - d_v(x_1)}{u_v(x_1) + d_v(x_1)}\, ,
\end{equation}
\end{widetext} 
where we have again used $u(x_2) \simeq d(x_2)$ at small $x_2$.
Since $u_v(x_1) > d_v(x_1)$ in the proton, we expect a positive asymmetry in the kinematic region of
large positive rapidities. Furthermore, the reweighting of the nuclear PDFs will have very little impact
on the charge asymmetry in this limit even if the precision of the data will increase in the future.

Another asymmetry used by CMS is the forward-backward asymmetry. 
This  is defined as a ratio of the number of events in the forward and
backward region in a given rapidity bin: 
\begin{equation}
A_{\text{FB}}^{\pm}(y_{\ell})=\frac{dN_{l^{\pm}}(+y_{\text{lab}})}{dN_{l^{\pm}}(-y_{\text{lab}})}\ .
\label{eq:FBasymPM}
\end{equation}
This asymmetry is defined separately for the $W^{+}$ and $W^{-}$ cases. 
It can also be  combined into a single quantity,  the forward-backward
asymmetry of charge-summed $W$ bosons: 
\begin{equation}
A_{\text{FB}}(y_{\ell})=
\frac{dN_{l^+}(+y_{\text{lab}})+dN_{l^-}(+y_{\text{lab}})}
     {dN_{l^+}(-y_{\text{lab}})+dN_{l^-}(-y_{\text{lab}})}
\ .\label{eq:FBasymSUM}
\end{equation}
This is the quantity we will use for our analysis in this section.

We now use  the asymmetries of Eqs.~\eqref{eq:chrgAsym}
and~\eqref{eq:FBasymSUM} to perform a reweighting of the \ncteqfit lead
PDFs. These asymmetries are just combinations of the rapidity distributions
used in Sec.~\ref{sec:rewCMSwpm}, and if both are employed at the
same time they should encode similar information to the rapidity
distributions themselves. 
In the literature it is sometimes argued
that the asymmetries are more sensitive to the PDFs and in turn are better
suited to performing PDF fits~\cite{Paukkunen:2010qg,Armesto:2015lrg,Khachatryan:2015hha}.
 We will empirically check this
statement by comparing reweighting predictions using rapidity distributions
and the above mentioned asymmetries.

In the following, we present  the results of the reweighting using the lepton charge
asymmetry and forward-backward asymmetry of charge-summed $W$ bosons.
In this case, the effective number of replicas is $N_{\text{eff}}=7382$.

The distribution of weights is displayed in Fig.~\ref{fig:weights_cmsWasym},
and we can see that compared to the reweighting using directly the rapidity
distributions (Fig.~\ref{fig:weights_cmsW}), the  weights are smaller
extending only to around $\sim2.7$ and more evenly distributed.

In Fig.~\ref{fig:theo_after_before_cmsWasym} we show a comparison
of data and theory before and after the reweighting procedure.
In the case of the charge asymmetry we do not see a large improvement, 
but this is not surprising as there is already good agreement between 
the data and theory before the reweighting. We note that the $\chi^{2}/N_{\text{data}}$
before the reweighting is $1.44$ and $1.27$ after the reweighting. 

In the case of the forward-backward asymmetry
the initial agreement between data and theory is not as good and 
the corresponding improvement is much larger;
$\chi^{2}/N_{\text{data}}$ changes from $4.03$ to $1.31$.

We now show the effect of the reweighting procedure on the PDFs.
In Fig.~\ref{fig:pdfs_after_before_cmsWasym_10}
we display the PDFs for the up quark and gluon at a scale of $Q=10$~GeV.
We can see that in both cases
the effect is limited to the low $x$ region and does not exceed few percent.
The results for other flavors are similar, and overall the asymmetries
with the current experimental uncertainties seem to have rather small
effect on the nPDFs.

In particular it seems that using asymmetry ratios 
yields a reduced impact, at least compared to the 
rapidity distributions of Sec.~\ref{sec:rewCMSwpm}.
This is possibly due to the fact that much of the information 
on the nuclear corrections is lost  when constructing the ratios.
However, asymmetries can be still useful to explore the very forward and
backward regions of the rapidity distributions (corresponding to higher/lower
$x$ values) where experimental uncertainties are typically large but
can cancel in the ratios.

\figDefwAll
\figDeffull
\figDefallCmsW
\figDefallAtlasw
\figDefallud
\figDefallubdb
\figDefallG
\subsection{Including all the data sets\label{sec:alldataset}}
\nobreak
Due to large experimental uncertainties, the effect of individual
data sets presented in Sec.~\ref{sec:comp} on the lead PDFs is rather
limited. The largest constraint is obtained from the 
CMS $W^{\pm}$ data~\cite{Khachatryan:2015hha}
(Secs.~\ref{sec:rewCMSwpm}),
and from (preliminary) ATLAS $W^{\pm}$ data~\cite{AtlasWpPb}. 
In order to
maximize the effects on the PDFs, we now employ all proton-lead
data sets from Tab.~\ref{tab:LHCdatasets}
 to perform the reweighting of the
\ncteqfit lead PDFs. 
Note that we use  both the rapidity distributions
and the asymmetries; although this can be regarded as ``double counting'',
it is a common practice in proton PDF analyses, e.g.~\cite{Dulat:2015mca}.

As the impact of the reweighting on the theory predictions for ALICE
$W^\pm$ production data~\cite{Senosi:2015omk}, LHCb $Z$ data~\cite{Aaij:2014pvu}
and both ATLAS~\cite{Aad:2015gta} and CMS~\cite{Khachatryan:2015pzs}
$Z$ production data is very small, we will not show the corresponding
comparisons of theory predictions before and after the reweighting.
We do note that in the majority of these cases the $\chi^{2}/N_{\text{data}}$
has improved indicating that the data sets are compatible, cf. Fig.~\ref{fig:chisqr_full}.
However,  the initial $\chi^{2}$ for these data sets was already very small
which reflects the large experimental uncertainties of these data sets
and their limited constraining power on the nPDFs.

We start by examining  the distribution of weights of the new replicas
which is displayed in Fig.~\ref{fig:weights_all}. We  see that
the distribution is steeply falling in a similar manner to the 
one from Fig.~\ref{fig:weights_cmsW}
obtained using only CMS $W^{\pm}$ rapidity distributions,
but it extends to higher values of $\sim17$.
These results are not very surprising as 
the CMS $W^{\pm}$  data set is the one introducing the most constraints. 
We also note that the reweighting procedure results in the effective number
of replicas $N_{\mathrm{eff}}=3603$ which is around 40\% of the number of initial replicas.
This suggests that the reweighting procedure should still yield reliable results.

Now we turn to the comparison of data with the theory predictions
before and after the reweighting procedure. In Fig.~\ref{fig:theo_after_before_all_cmsW}
we show the  predictions for the CMS $W^{\pm}$ data~\cite{Khachatryan:2015hha}, 
and in Fig.~\ref{fig:theo_after_before_all_atlasW} 
we show  the corresponding predictions for 
the ATLAS $W^{\pm}$
data~\cite{AtlasWpPb}. We can see that in both cases we observe
an improvement in the data description that is confirmed by the corresponding
$\chi^{2}/N_{\text{data}}$ values (see figures). 
The $\chi^{2}$ values tell us also that the largest effect comes from the CMS data
which has  smaller errors and for which the initial description
(before the reweighting) was worse than in the ATLAS case. 

Furthermore, comparing the values of $\chi^{2}/N_{\text{data}}$ for the CMS
$W^{\pm}$ data after the reweighting using all data sets and using only CMS
data (Sec.~\ref{sec:rewCMSwpm}) we see further improvement of $\chi^{2}/N_{\text{data}}$
when more data is included. This shows that the different data sets are compatible
with each other and that they pull the results in the same direction.

In addition, we show in Fig.~\ref{fig:chisqr_full} the $\chi^{2}/N_{\text{data}}$
before and after the reweighting for each of the experiments, 
as well as the $\chi^{2}/N_{\text{data}}$ combining all 102 data points
from the different experiments. 
This highlights the fact that the 
CMS $W^{\pm}$ measurement yields the largest impact on the PDFs 
out of all the considered data sets.

Finally, in Figs.~\ref{fig:pdfs_after_before_all_ud}-\ref{fig:pdfs_after_before_all_g}
we present the effects of the reweighting on the $\{u,d,\bar{u},\bar{d},g, s \}$
distributions in lead for a scale $Q=10$ GeV.
The effects are similar  when looking at different scales.
From the figures we can see that changes in the PDFs are
generally affecting the low-$x$ distributions,
and to a lesser extent the moderate to high-$x$ distributions.

When considering the ratios of PDFs, the effects of the
reweighting appear to be quite substantial at large $x$, 
especially for the gluon; however, as is evident from looking at the 
plots of the PDFs directly, they are approaching zero
at large $x$ so the impact for physical observables is minimal.

Furthermore, when interpreting the results of the reweighting analysis
it is important to remember that this method can only estimate the
effects a given data set might have on the PDFs; it is not equivalent
to a full fit.
For example, a reweighting analysis cannot be used to explore new
parameters or other dimensions that are not already spanned by the
original PDF uncertainty basis.
In particular, this study has shown us that the strange quark PDF can
play an important role in the LHC pPb production of $W/Z$. As our
current $s(x)$ is parameterized proportional to $\bar{u}(x)+\bar{d}(x)$,
this restricts our ability to vary the strange PDF
independently;%
    \footnote{This point was explored in more detail in ref.~\cite{Kusina:2017bsq}.}
hence, an independent fit (in progress) is needed to
better the impact of this data on the nPDFs.

\subsection{Comparison with EPPS16\label{sec:epps}}
%
\begin{figure*}[!htb]
\centering{}
\subfloat[$u$ PDF]{
\includegraphics[width=0.48\textwidth]{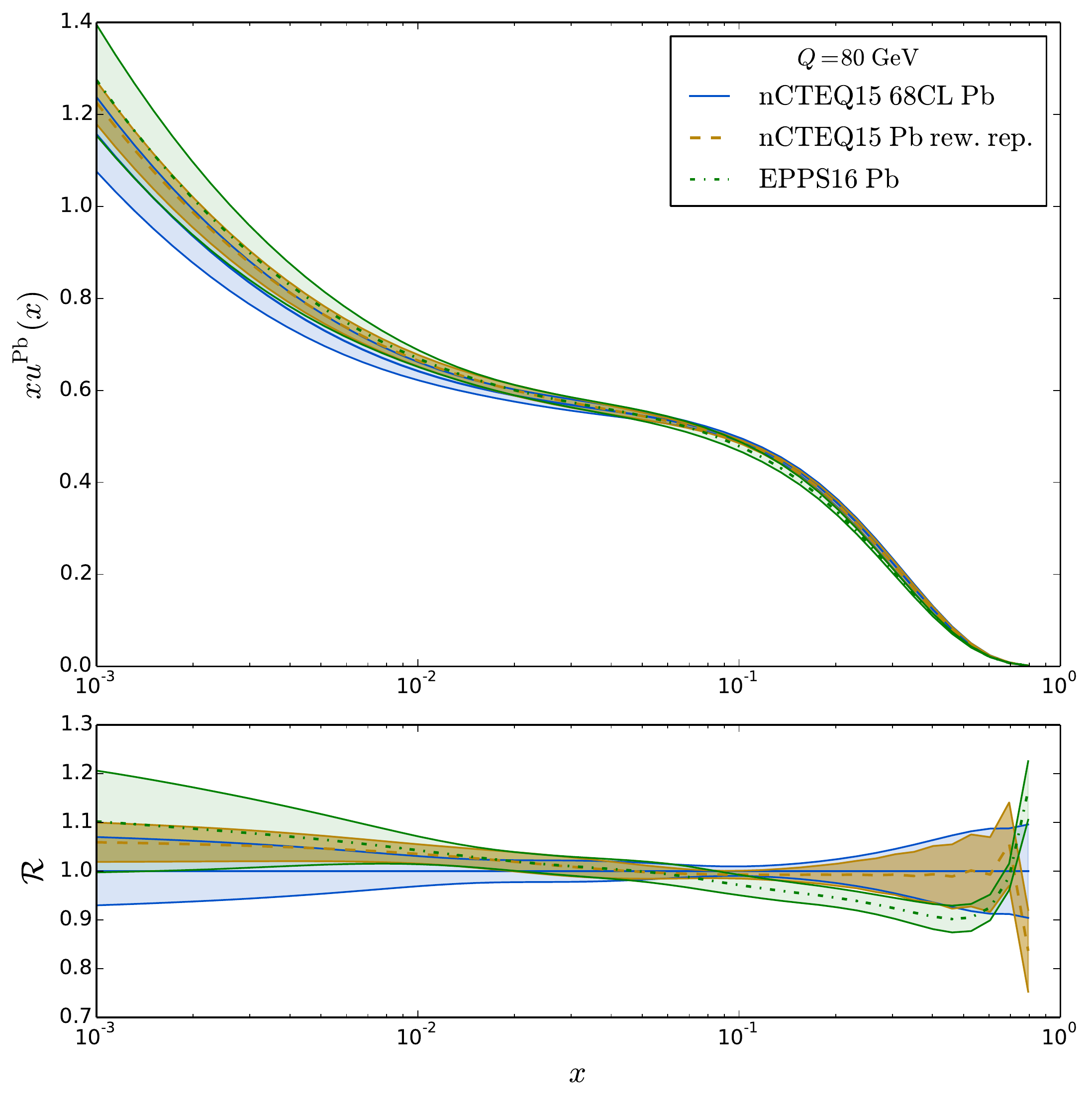}}
\subfloat[$d$ PDF]{
\hfil
\includegraphics[width=0.48\textwidth]{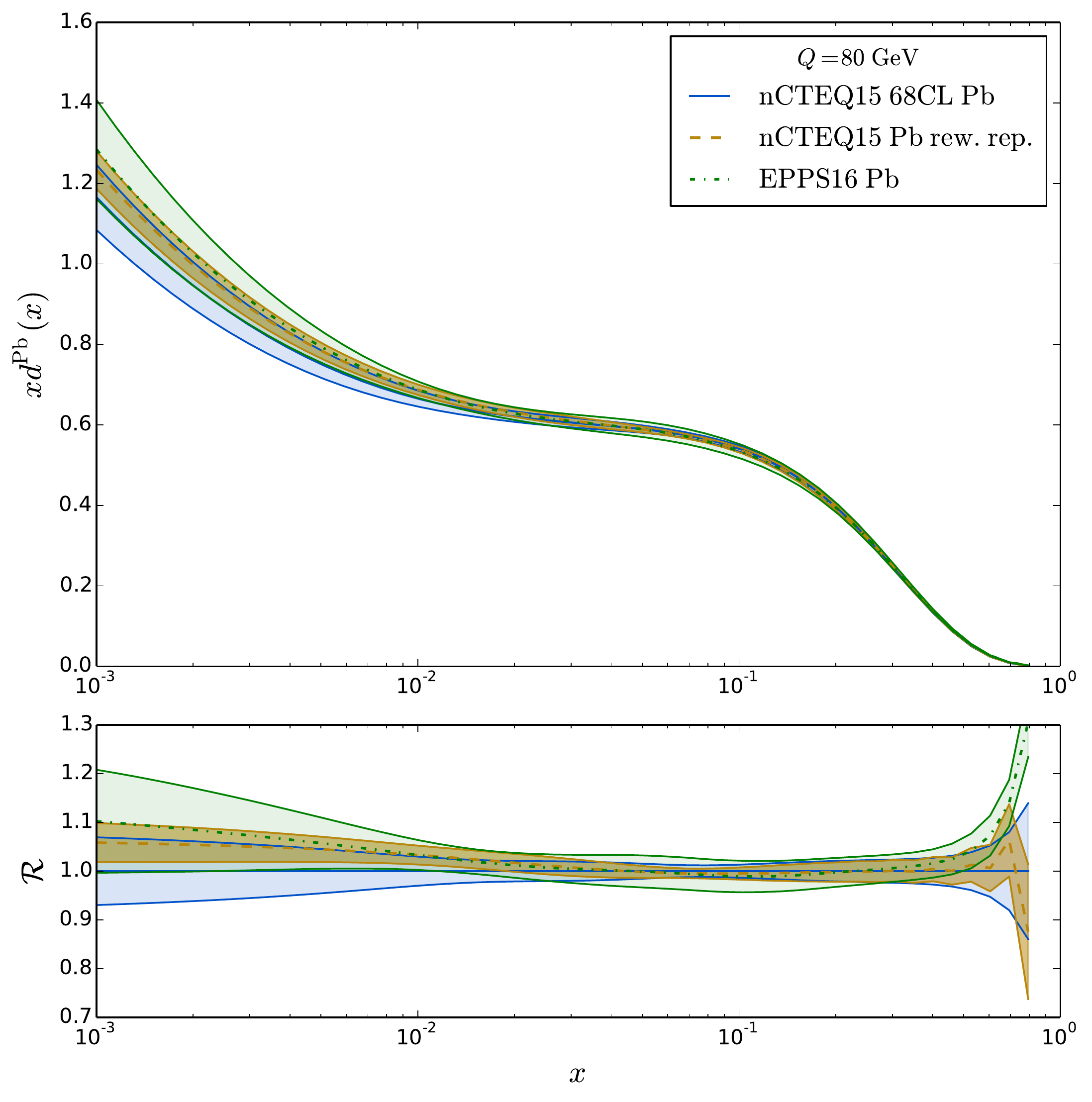}}
\\
\subfloat[$\bar{u}$ PDF]{
\includegraphics[width=0.48\textwidth]{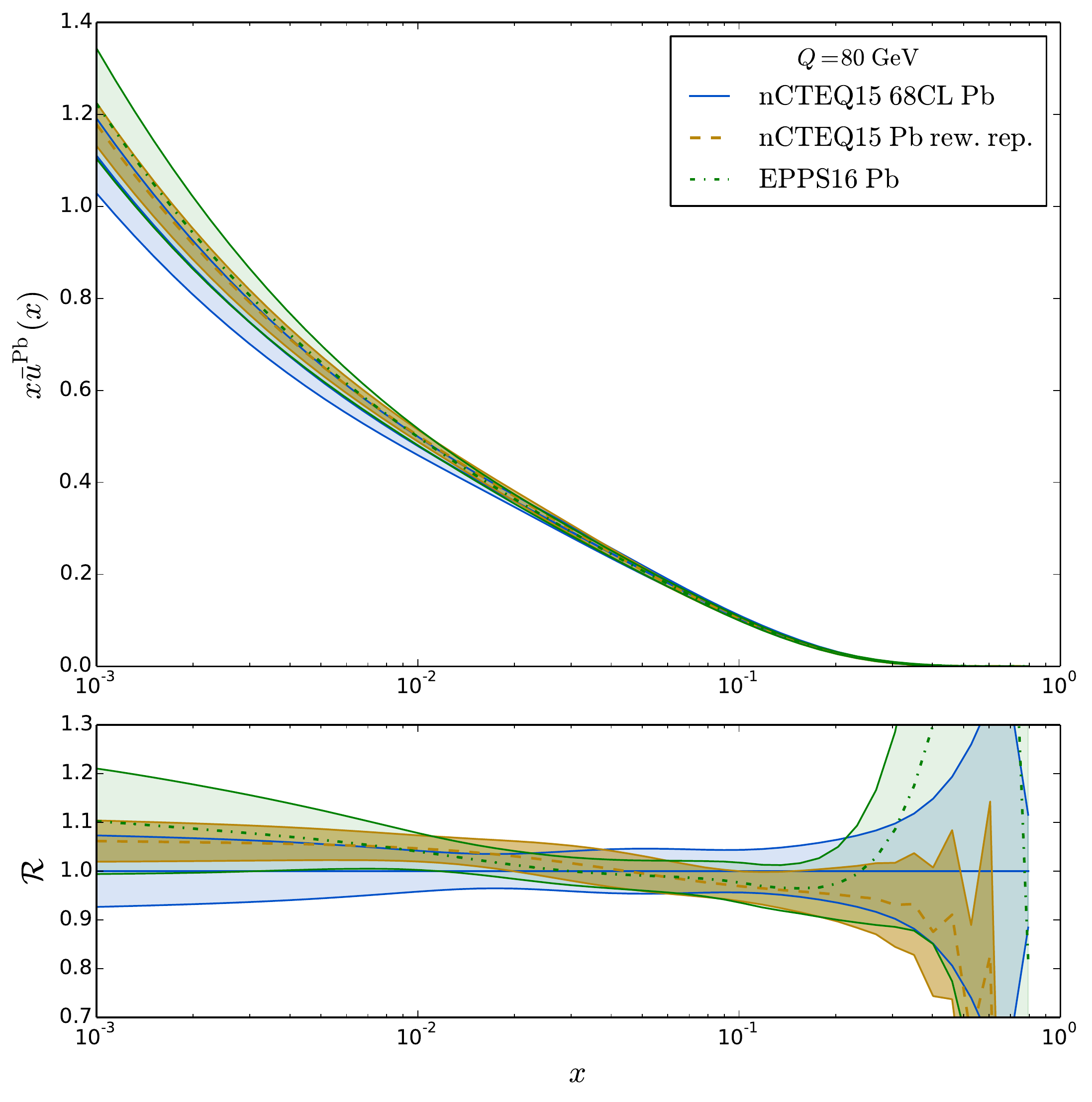}}
\subfloat[$\bar{d}$ PDF]{
\hfil
\includegraphics[width=0.48\textwidth]{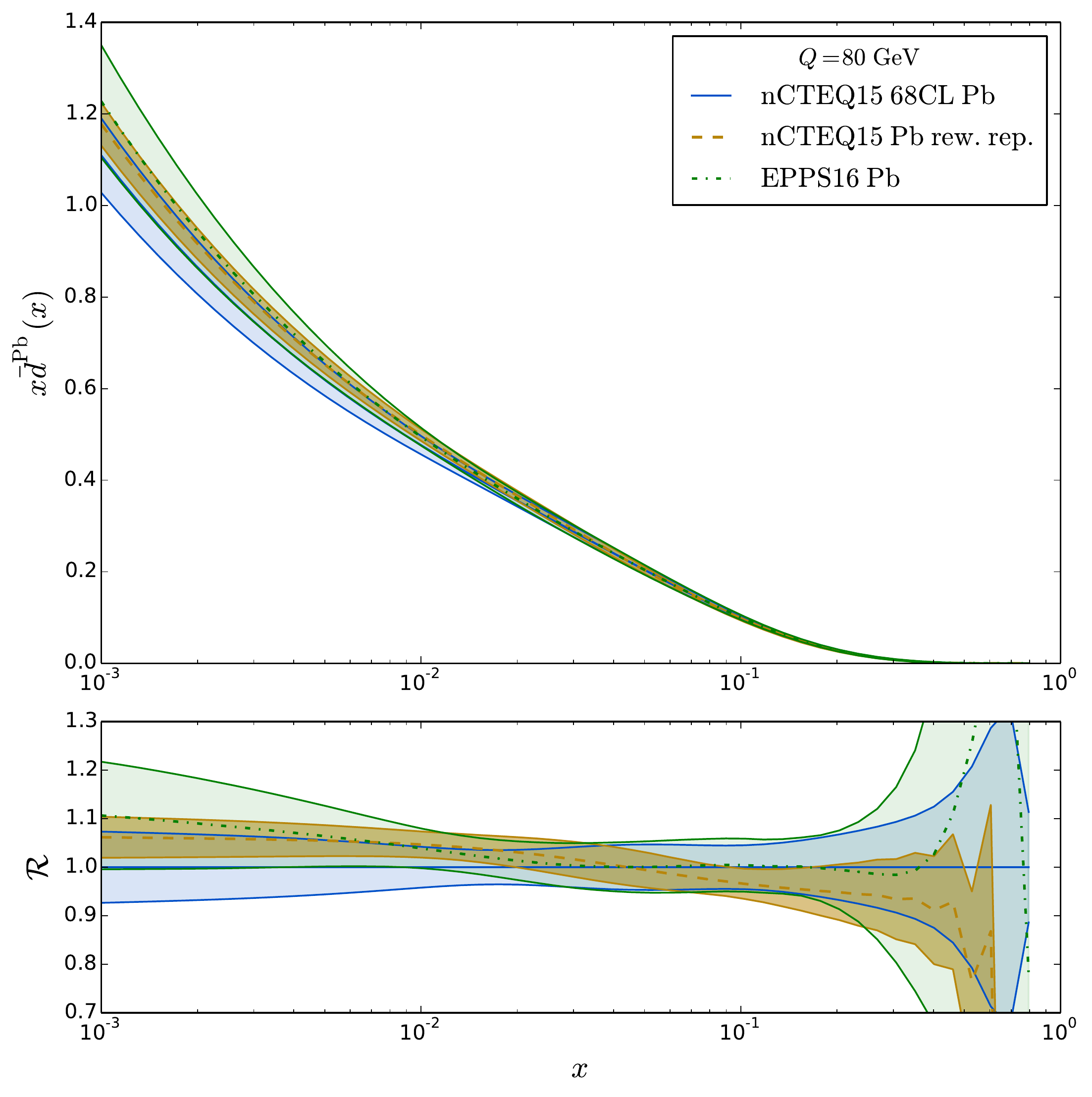}}
\caption{Comparison of the nCTEQ15 PDFs before and after the reweighting using all pPb data sets
with the EPPS16 PDFs including LHC data. The EPPS16 error bands include only the nuclear errors
(unlike what is provided in LHAPDF where also the proton baseline errors are included) and they
are calculated using symmetric formula.}
\label{fig:pdfs_after_before_EPPS16}
\end{figure*}
\begin{figure*}[!htb]
\centering{}
\subfloat[$g$ PDF]{
\includegraphics[width=0.48\textwidth]{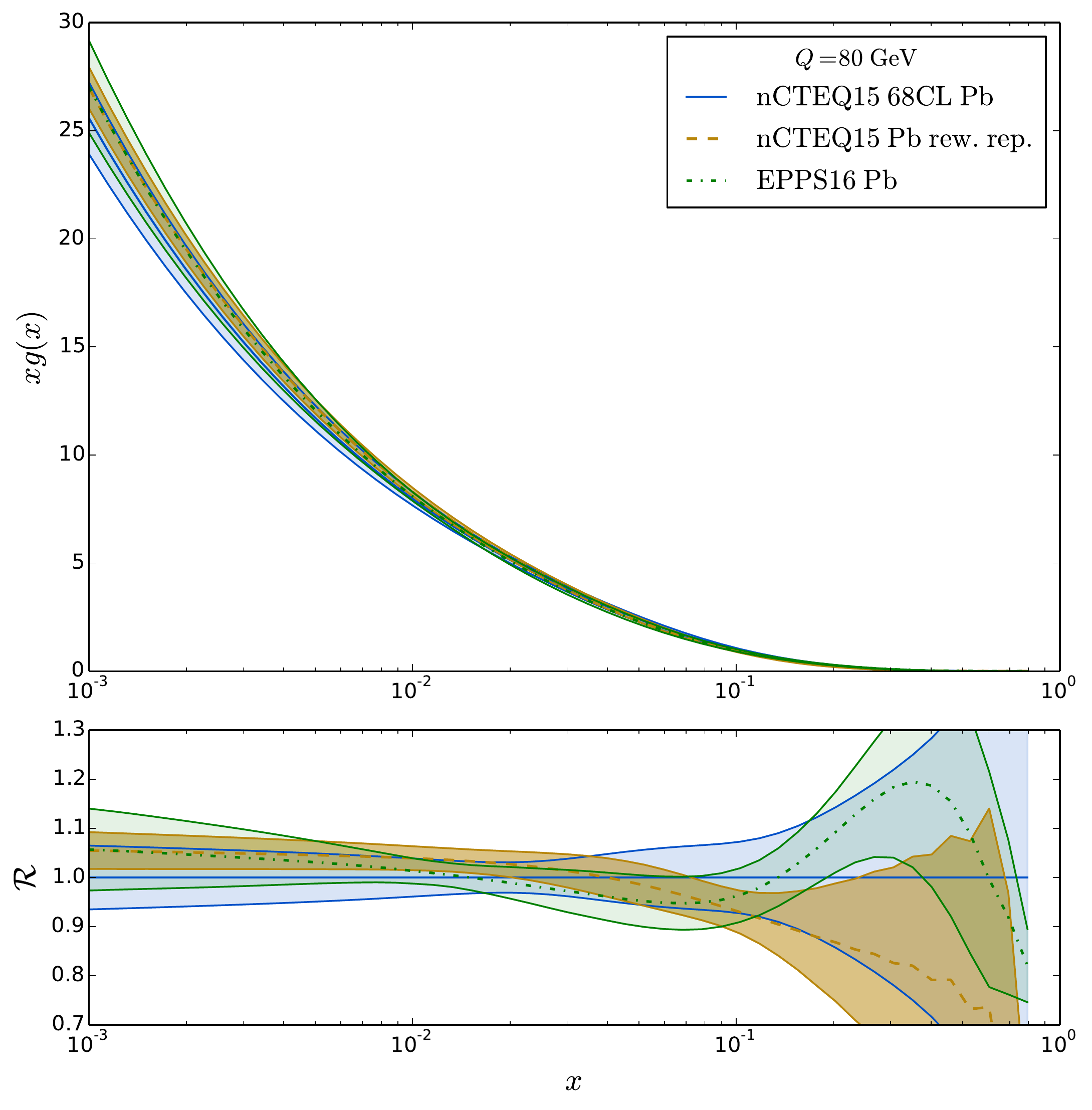}}
\subfloat[$s$ PDF]{
\hfil
\includegraphics[width=0.48\textwidth]{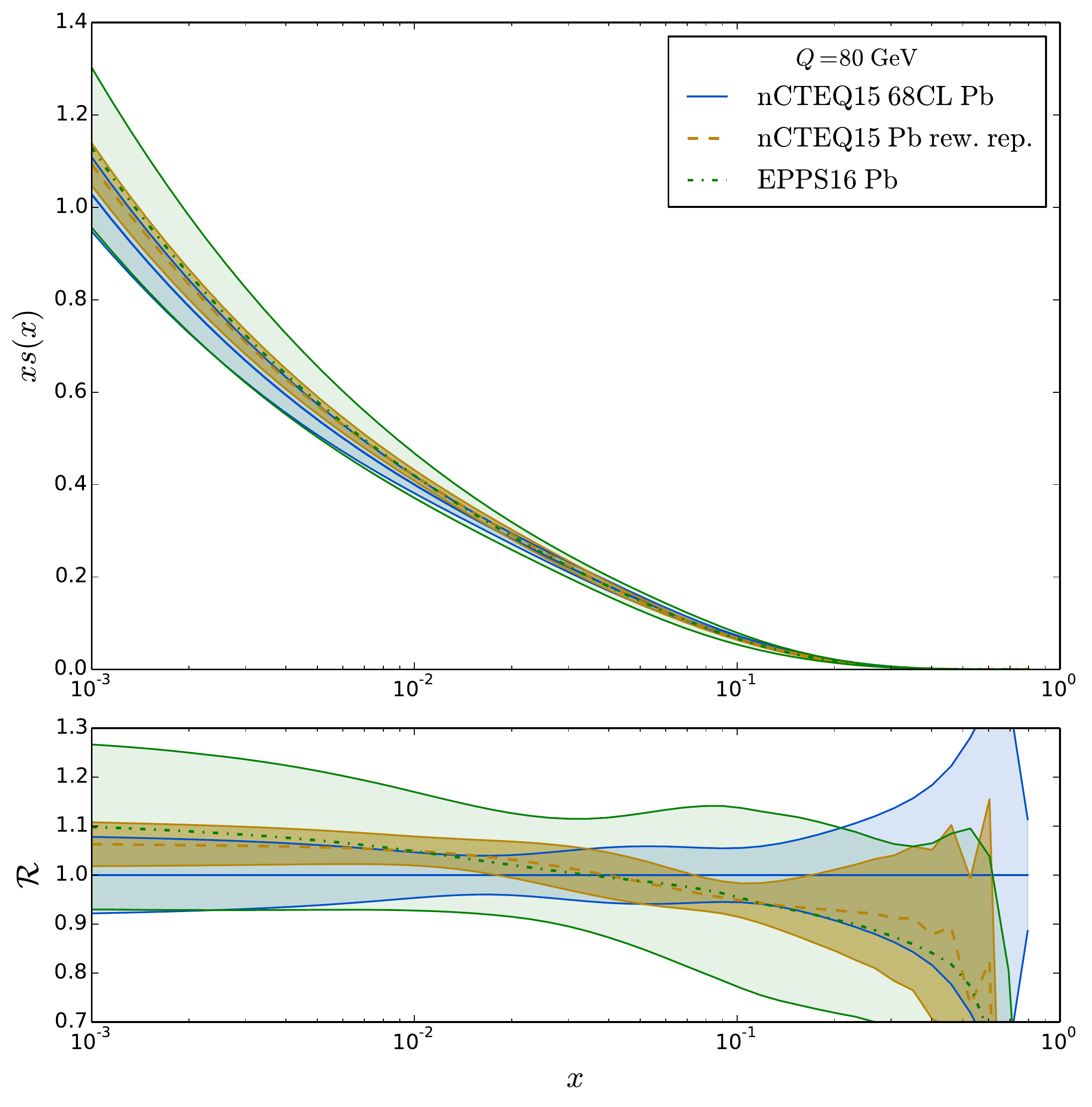}}
\caption{Continuation of Fig.~\ref{fig:pdfs_after_before_EPPS16}.}
\label{fig:pdfs_after_before_EPPS16_2}
\end{figure*}
%
During the course of our analysis, a new global fit including LHC data  (EPPS16~\cite{Eskola:2016oht})
has been released. This gives us an opportunity to compare the results of our reweighting
study with these new PDFs. We note here that this is a qualitative comparison as the
data sets used in these two studies are different.
Another important  difference is that the EPPS16 fit has more parameters to describe  the sea-quark PDFs
as  compared to the nCTEQ15 analysis; this provides  EPPS16 additional flexibility to accommodate all the considered data. 
As mentioned earlier, our reweighting of nCTEQ15  cannot compensate for our more restrictive parametrization, 
so this must be considered when evaluating these comparisons.

In Figs.~\ref{fig:pdfs_after_before_EPPS16} and~\ref{fig:pdfs_after_before_EPPS16_2}
we present a comparison of $u$, $d$, $\bar{u}$, $\bar{d}$, $g$ and $s$ for the nCTEQ15 PDFs
before and after the reweighting, with the EPPS16 distributions at the scale of 80 GeV. 
There are a number  of trends which emerge. 
\begin{enumerate}

\item
In the low $x$ region, the reweighted nCTEQ15 PDFs approach  the EPPS16 distributions;
for  the  $g$ and $s$ PDFs,  the central values are very close. 
The effect of the reweighting appears  mostly in this region 
where (prior to the LHC data) there were minimal constraints  on the PDFs. 
Therefore, adding the LHC data is able to significantly adjust the PDFs in this region. 

\vspace{6pt}

\item
In the intermediate $x$ range ($\sim 3\times 10^{-2}$),  the central values of the  EPPS16
and both reweighted and initial nCTEQ15 PDFs coincide, and their uncertainty bands are also similar (except for the strange quark). 
This region was previously constrained by pre-LHC data,
and we observe minimal changes in this region. 

\vspace{6pt}

\item
On the contrary, where $x$ is large, the differences are more important
with no consistent pattern. 
This is a challenging region as the absolute value of the PDFs is small,
and the nCTEQ15 parameterization may not be sufficiently flexible
to accommodate the  new data. 
Additionally, the inclusion of certain data sets in the EPPS16 analysis
(such as the CHORUS $\nu$-Pb data~\cite{Onengut:2005kv}) can have a significant impact.

\end{enumerate}

Finally, we also see that the  EPPS16 PDFs have consistently larger uncertainty bands (especially at low $x$).
As the nCTEQ15 uncertainty bands in this region are essentially extrapolated  from larger $x$ results, 
the EPPS16 uncertainties are probably a more realistic assessment. 
The issue of PDF parameterization is a perennial challenge  for  the nuclear PDFs as 
there is   less data and more degrees of freedom as compared to the proton PDFs. 
The common solution is to impose assumptions on the nPDF parameters, or to limit the
flexibility of the parameterization, and thereby underestimate the uncertainty. 
These issues highlight the importance of including this new 
LHC data in the nPDF analyses as they not only will help determine the central fits, 
but  also provide for more reliable error estimation.

\section{Conclusions\label{sec:conclusions}}

We have presented a comprehensive study of vector boson production
($W^{\pm},Z$) from lead collisions at the LHC.
This LHC lead data is of particular interest for a number of reasons.
\begin{enumerate}

\item Comparisons with LHC proton data can determine nuclear corrections
for large $A$ values; this is a kinematic $\{x,Q^2\}$ range very different
from nuclear corrections provided by fixed-target measurements.

\vspace{6pt}

 \item The $W^{\pm},Z$ lead data are
sensitive to the heavier quark flavors (especially the strange PDF),
so this provides important information on the nuclear flavor
decomposition.

\vspace{6pt}

\item Improved information on the nuclear corrections from the LHC lead
data can also help reduce proton PDF uncertainties as fixed-target nuclear data
is essential for distinguishing the individual flavors. 

\end{enumerate}

Predictions from the recent \ncteqfit nPDFs are generally compatible
with the LHC experimental data; however, this is partially due to the
large uncertainties from both the nuclear corrections and the data.
We do see suggestive trends (for example $W^\pm$ production in \ppb at
large $y_{\ell^+}$) which may impose influential constraints on the
nPDF fits as the experimental uncertainties are reduced.
Intriguingly, the large rapidity $W/Z$ data seem to prefer nuclear PDFs
with no or delayed shadowing at small $x$, similar to what has been observed
in $\nu$-Fe DIS.
This observation was validated by our reweighting study that
demonstrated the impact of the $W/Z$ pPb data on nPDFs.

The uncertainties of the currently available data are relatively large, 
and correlated errors are not yet available.
Fortunately, we can look
forward to more data (with improved statistics) in the near future as
additional heavy ion runs are scheduled.

While the above reweighting technique provides a powerful
method to quickly assess the impact of new data, there are limitations.
For example, the reweighting method cannot introduce or explore  new degrees of freedom.
Thus, if the original fit imposes artificial constraints (such as linking
the strange PDF to the up and down sea distributions),
this limitation persists
for the reweighted PDF~\cite{Kusina:2017bsq}.

Most importantly, our correlation study (Sec.~\ref{sec:corr}) demonstrated the
importance of the strange distribution for the vector boson ($W/Z$)
production at the LHC,
possibly even pointing to a nuclear strangeness asymmetry ($s(x)>\bar{s}(x)$).
The comparison of the 2 flavor and 5 flavor
results illustrates how flavor decomposition and nuclear corrections
can become entangled.  Therefore, it is imperative to separately
control the strange PDF and the nuclear correction factor if we are to
obtain unambiguous results.
The investigations performed in this paper provide a foundation for
improving our determination of the PDFs in lead, especially the
strange quark component.
Combining this information in a new nCTEQ fit across the full $A$
range can produce improved nPDFs, and thus yield improved 
nuclear correction factors.
These improved nuclear correction factors, together with the LHC $W/Z$
production data for $pp$, can refine our knowledge of the strange PDF
in the proton.
\section*{Acknowledgments\label{sec:acks}}

The authors would like to thank 
C.~Bertulani, 
M.~Botje, 
C.~Keppel, 
S.~Kumano, 
J.G.~Morf{\'i}n, 
P.~Nadolsky, 
\\
J.F.~Owens,
F.~Petriello, 
R.~Pla\v cakyt\.e,
and 
V.~Radescu
for valuable discussions. We acknowledge the hospitality of CERN,
DESY, and Fermilab where a portion of this work was performed.
This work was also partially supported by the U.S.\ Department of Energy
under Grant No.\ DE-SC0010129 and
by the National Science Foundation
under Grant No.\ NSF PHY11-25915. 

%
\bibliographystyle{spphys}
\bibliography{refs}

\end{document}